\documentclass{aa}
\usepackage{graphicx, epsfig, fancyhdr, rotating, amsmath, epsf, txfonts, natbib, epstopdf, multirow}
\usepackage{subfigure, tabularx, appendix, dcolumn, threeparttable, longtable, makecell, multirow}
\usepackage{psfig}
\def \kms {km\,s$^{-1}$}
\usepackage{color, ulem}

\begin{document}

\title{Eruptions from quiet Sun coronal bright points. I.  Observations}

\author{Chauzhou Mou\inst{1}
\and Maria S. Madjarska\inst{2}
\and Klaus Galsgaard\inst{3}
\and Lidong Xia\inst{1}}

\institute{Shandong Provincial Key Laboratory of Optical Astronomy and Solar-Terrestrial Environment, Institute of Space Sciences, Shandong University, Weihai, 264209 Shandong, China
\and Max Planck Institute for Solar System Research, Justus-von-Liebig-Weg 3, 37077, G\"ottingen, Germany
\and Niels Bohr Institute, Geological Museum, {\O}stervoldgade 5-7, 1350 Copenhagen K, Denmark}

\date{Received date, accepted date}

\abstract
{Eruptions from coronal bright points (CBPs) are investigated in a two part study.}
{The present study aims to explore in full detail the morphological and dynamical evolution of these eruptions  in the context of the full lifetime evolution of CBPs. A follow-up study employs data-driven modelling based on a relaxation code to reproduce the time evolution of the magnetic field of these eruptive CBPs, and provide an insight on the possible causes for destabilisation and eruption.}
{Observations of the full lifetime of CBPs in data taken with the Atmospheric Imaging Assembly (AIA) on board the Solar Dynamics Observatory in four passbands, He~{\sc ii}~304~\AA, Fe~{\sc ix/x} 171~\AA, Fe~{\sc xii}~193~\AA, and Fe~{\sc xviii}~94~\AA\ are investigated for the occurrence of plasma ejections, micro-flaring, mini-filament eruptions and mini coronal mass ejections (mini-CMEs). Data from the Helioseismic Magnetic Imager are analysed to study the longitudinal photospheric magnetic field evolution associated with the CBPs and related eruptions.}
{First and foremost, our study shows that the majority (76\%) of quiet Sun CBPs (31 out of 42 CBPs) produce at least one eruption during their lifetime.  From 21 eruptions in 11 CBPs, 18 occur in average $\sim$17~hrs after the CBP formation for  an average lifetime of the CBPs in AIA~193~\AA\  of $\sim$21~hrs.   This time delay in the eruption occurrence coincides in each BP with the convergence and  cancellation phase of the CBP bipole evolution during which the CBPs become smaller until they fully disappear. The remaining three happen 4 -- 6~hrs after the CBP formation. In sixteen out of 21 eruptions  the magnetic convergence and cancellation involve the CBP main bipoles, while in three eruptions one of the BP magnetic fragments and a pre-existing fragment of opposite polarity converge and cancel. In one BP with two eruptions cancellation was not observed.   The CBP eruptions involve in most cases the expulsion of chromospheric material either as  elongated filamentary structure (mini-filament, MF) or a volume of cool material (cool plasma cloud, CPC), together with the CBP or higher overlying hot loops. Coronal waves were identified during three eruptions.
 A micro-flaring  is observed beneath all erupting MFs/CPCs. It remains uncertain whether the destabilised  MF causes the micro-flaring or the  destabilisation and eruption of the MF is triggered by reconnection beneath the filament. In most eruptions,  the cool erupting plasma  obscures partially or fully the micro-flare  until the erupting material  moves  away from the CBP.  From 21 eruptions 11 are found to produce mini-CMEs.  The dimming regions associated with the CMEs are found to be occupied by both the `dark' cool plasma and areas of weakened coronal emission caused by the depleted plasma density. }
{The present study demonstrates that  the evolution of small-scale loop structures in the quiet Sun  determined by their magnetic footpoint motions and/or ambient field topology, evolve into eruptive phase that triggers the ejection of cool and hot plasma in the corona. }


\maketitle

\section{Introduction}
\label{sect:intro}

Coronal bright points (CBPs) are one of the most typical small-scale phenomena in the solar corona. The earliest discovery of CBPs was in X-rays which initially  led to the introduction of the term ``X-ray BP'' \citep{vaiana1970a}. X-ray BPs were seen in rocket mission observations  as point-like structures with typical sizes ranging from 10\arcsec\ to 50\arcsec\ that form a bright core of $\sim$10\arcsec\ \citep{1974ApJ...189L..93G}. Their lifetimes differ when observed in different imaging channels, i.e. when observed at different temperatures. CBPs were found to have a lifetime in X-rays from 2 to 48 hours with an average lifespan of 8 hours \citep{1976SoPh...49...79G}. Later, using data from the Yohkoh X-ray telescope, \citet{1993AdSpR..13...27H} studied the lifetimes of 514 X-ray BPs  and established that coronal hole BPs (34\% of the total) have an average existence of 12~hrs 36~min, the QS BPs (60\%) 13~hrs and 11~hrs for active region BPs (6\%).  Studies based on extreme-ultraviolet (EUV) observations \citep[e.g.,][]{1981SoPh...69...77H,  2001SoPh..198..347Z} show that the lifetimes of CBPs in EUV range from a few minutes to a few days with an average value of 20~hrs \citep{2001SoPh..198..347Z}. For review of on the observational properties of CBPs and their modelling see Madjarska (in prep).

High-resolution spectroheliograms in Fe~{\sc xiv}~284~\AA\ taken with the extreme ultraviolet spectrograph on the Apollo Telescope Mount on board {\it Skylab} first revealed that CBPs consist of small-scale fast evolving loops \citep{1979SoPh...63..119S} which was later confirmed by the high-resolution TRACE data \citep[e.g.,][]{2004A&A...418..313U}. \citet{1990ApJ...352..333H} observed that the peaks of the emission in CBPs in six simultaneously recorded different spectral lines with various formation temperatures are not always co-spatial which made the authors conclude that CBPs are possibly composed of small-scale loops at different temperatures. From near limb spectroheliogram  observations of CBPs that cover chromospheric, transition region and coronal lines taken with the Harvard EUV experiment in 1973, \cite{1981SoPh...69...77H} found that the coronal emission in CBPs is located a few arcseconds above the CBP chromospheric emission sources which should be assumed if CBPs are formed by magnetic loops connecting bipolar regions.   Recent spectroscopic observations from the Extreme Ultraviolet Spectrometer (EIS) on board the Hinode satellite further confirmed this by finding that the transition region bipolar emission can be interpreted as coming from the CBP footpoints \citep{2011A&A...526A.134A} while high temperature coronal emission appears compact and probably comes from loop tops.

CBPs are always associated with photospheric magnetic bipolar features (MBFs) \citep[e.g.,][]{1971IAUS...43..397K, 1977SoPh...53..111G, 1993SoPh..144...15W, 2003A&A...398..775M,2016ApJ...818....9M}. \citet{1994ASPC...68..377H} analyzed simultaneous time sequence of longitudinal magnetic field data and X-ray observations, and found that two-thirds of all magnetic bipoles appear not to be related to X-ray BPs. The authors concluded that ``emergence and cancellation of magnetic flux in the photosphere is not in itself a necessary and sufficient condition for the occurrence of XBPs but rather the interaction and reconnection of magnetic field with the existing, overlying magnetic field configuration that results in the occurrence and variability of X-ray BPs''. A recent study by \citet{2016ApJ...818....9M} on the formation of  MBFs associated with CBPs using Helioseismic Magnetic Imager (HMI)  on board SDO data confirmed the earlier finding by, e.g., \citet{1993SoPh..144...15W} and \citet{1985AuJPh..38..875H}, that three processes are involved in the formation and evolution of CBPs: emergence, convergence and local coalescence of photospheric magnetic concentrations. This study also established that 50\%\ of 70 CBPs are formed by bipolar flux emergence (the so-called ephemeral regions).

\begin{table*}
\centering
\caption{General information on the QS BPs associated with eruptions.}
\begin{tabular}{ccccccccccc}
\hline\hline
Event &
Original &
Location &
\multicolumn{2}{c}{Bipole form.} &
Flux emergence &
\multicolumn{3}{c}{CBP} &
Delay$^a$ \\
\cline{4-5}\cline{7-9}
No. &
event &
x, y&
Pos &
Neg &
Date - Time &
Start date - time &
End date - time &
Lifetime &
(min) \\
&
No. &
(arcsec) &
&
&
yyyy/mm/dd (UT) &
yyyy/mm/dd (UT) &
yyyy/mm/dd (UT) &
&
 \\
\hline
%----------------------------------------------------------------
1 & 4 & 130, -169 &  A & A & 2011/01/02 16:55 & 2011/01/02 17:25 & 2011/01/03 07:37 & 14h 12m  & 30 \\
\hline
2 & 5 &  60, -134 &  A & A & 2011/01/01 06:30 & 2011/01/01 07:30 & 2011/01/02 00:40 & 17h 10m  & 60 \\
\hline
3 & 8 & -109, -209 &  A & A & 2011/01/01 19:20 & 2011/01/01 19:52 & 2011/01/02 16:40 & 20h 48m  & 32 \\
\hline
4 & 10 & -89, -69  &  B & B &    --   & 2011/01/01 14:46 & 2011/01/02 07:07 & 16h 21m  &--  \\
\hline
5 & 26 & 160, -144 &  C & C & -- & 2011/01/01 01:45 & 2011/01/02 00:50 & 23h 5m  & -- \\
\hline
6 & 41 & -359, 335 &  A & A & 2011/01/02 10:15 & 2011/01/02 11:20 & 2011/01/03 17:00 & 29h 40m  & 65 \\
\hline
7 & 50 & -379, 205 &  A & A & 2010/12/31 20:00 & 2010/12/31 20:29 & 2011/01/01 20:20 & 23h 51m  & 29 \\
\hline
8 & 53 & -429, 205 &  A + B & B & 2011/01/01 12:50 & 2011/01/01 14:50 & 2011/01/02 01:40 & 10h 50m  & -- \\
\hline
9 & 57 & -239, 310 &  A & A & 2011/01/02 11:30 & 2011/01/02 12:04 & 2011/01/03 14:50 & 26h 46m  & 34 \\
\hline
10 & 59 & -14, 305  &  B & A + B & --  & 2011/01/03 02:30 & 2011/01/03 20:40 & 18h 10m  & -- \\
\hline
11 & 61 & -254, 315 &  B & A + B & 2011/01/01 22:20 & 2011/01/01 20:40 & 2011/01/02 23:30 & 26h 50m  & -- \\
\hline
\multicolumn{10}{l}{Notes.} \\
\multicolumn{10}{l}{A: emergence; B: convergence; C: local coalescence} \\
\multicolumn{10}{l}{$^a$: Time delay between the flux emergence and the CBP formation}
\end{tabular}%
\label{tab:bp}%
\end{table*}

CBPs at all latitudes were found to exhibit flaring activity that presents itself as an intensity increase of a few orders of magnitude in a small area of the CBPs \citep{1974ApJ...189L..93G}. \citet{1977ApJ...218..286M} associated ``flaring'' in X-ray BPs with macrospicules observed in H$\alpha$, thus for the first time establishing a link between CBP micro-flaring and eruptive activity in CBPs. Later, \citet{2009A&A...495..319I} introduced the term mini coronal mass ejections (mini-CMEs) as small-scale eruptions from the quiet Sun which appear conspicuously similar to CMEs in on-disk 171~\AA\ EUV image sequences. To recall, CMEs are large-scale eruptive phenomena. In white light observations, a classic CME is composed of three parts:  a bright loop, moving ahead of a dark cavity, and a bright core which corresponds to an eruptive prominence. However, only 30\% of the CMEs are found to have this structure with some lacking a bright core due to the draining of the prominence material towards the solar surface or rather a general lack of any prominence material \citep{2011LRSP....8....1C}.
The mini-CMEs were found at the junctions of supergranulation cells and the majority of the events showed cool plasma ejections that appear as darkening as referred to by the authors (e.g., mini- filament eruption) and micro-flaring in the eruption source region. Several events showed coronal dimmings that were identified as wave-like features propagating from the eruption site. Before the term mini-CME was introduced, a statistical study on EUV coronal jets by \citet{2009SoPh..259...87N} of 79 jet events in polar coronal holes identified 37 Eiffel tower-type jet events, 12 lambda-type  and five micro-CME-type. The micro-CME-type jets earned their name from the great similarities with classical CMEs. The micro-CMEs  appeared as small loops  expanding from the solar surface.

In \citet{2010A&A...517L...7I} two events, one in the quiet Sun and another in an equatorial coronal hole, were observed simultaneously on the solar disk in the extreme ultraviolet imaging telescope (EUVI)/SECCHI 171~\AA\ images of STEREO-A, and at the limb in the 304~\AA\ images from EUVI/SECCHI on board STEREO-B. The comparison of the timing of the chromospheric eruption observed at the limb in the 304~\AA\ passband with the brightening and the dimming in the disk centre in 171~\AA\ revealed that a coronal dimming appears prior to the chromospheric eruption in both events. The authors suggested that the removal of the overlying coronal field is fundamental in triggering these phenomena similar to CMEs. \citet{2010ApJ...709..369P} also analysed mini-CME events in EUVI/SECCHI 171~\AA. Their results showed that the mini-CMEs are characterised by a smaller size and a shorter lifetime compared with their large-scale counterparts. Small-scale coronal waves  (CWs) and coronal dimmings are both observed in their events. The speeds of the coronal wave generated from mini-CMEs were approximately 10--20 times smaller than large-scale ones. Furthermore, they found that the small-scale coronal dimmings have two types: deep core dimming and a more widespread dimming, which is very similar to the dimming in large-scale CMEs. None of the above studies investigated mini-CME events in the context of CBPs.

 Small-scale filamentary (often appearing as small loops) eruptive phenomena found in H$\alpha$ images of the quiet Sun were explored by \citet{1986NASCP2442..369H} using data from the Big Bear Solar Observatory (BBSO). The features had an average length of 15\arcsec\ and were found to have an occurrence rate of 600 per 24~hrs with an average duration of the eruption of 24~min from an average lifetime of the mini-filaments of 70~hrs. The authors note that the eruptions usually proceed as an expansion of the mini-filament (MF) into an arch or a loop. All events were related to small flares. Importantly, the filaments were associated with canceling magnetic bipoles. The authors state that the MFs were either located at the bipole inversion line, or one or both footpoints of the MF  were rooted there.

A dedicated CBP study by \citet{2014ApJ...796...73H} used data from the Atmospheric Imaging Assembly (AIA) on board the Solar Dynamic Observatory (SDO) to study the eruptive behaviour of 30 CBPs in a coronal hole. One-quarter to one-third of all CBPs were found to have one or more mini-filament eruptions. The authors concluded that convergence and cancellation of magnetic dipoles associated with the CBPs are possibly responsible for the erupting  MFs confirming the conclusions of \citet{1986NASCP2442..369H}.

 By exploring the state-of-the-art capabilities of the present solar observations and modelling, we aim to obtain in full detail the nature and evolutionary characteristics of eruptions from quiet Sun CBPs, by analysing AIA/SDO images in four channels,  He~{\sc ii}~304~\AA, Fe~{\sc ix/x} 171~\AA, Fe~{\sc xii}~193~\AA,  and Fe~{\sc xviii} 94~\AA, together with longitudinal magnetograms from HMI/SDO. The derived observational properties are used in the second part of this study where a data-driven modelling based on a relaxation code is used to model the time evolution of the analysed CBPs. The article is organised as follows: In Sect.~\ref{sect:obs} we describe the observations and the methodology. The results  are presented in Sect.~\ref{sect:res}.  The discussion is given in Sect.~\ref{sect:sum} and the conclusions  in Sect.~\ref{sect:concl}.

\section{Observations and methodology}
\label{sect:obs}

The present study is a follow-up of the work by \citet{2016ApJ...818....9M} where the magnetic field evolution of 70  randomly selected CBPs identified in images taken by the AIA\,\citep{2012SoPh..275...17L} and the Helioseismic and Magnetic Imager\,\citep[HMI,][]{2012SoPh..275..207S} on board the SDO\,\citep{sdo2010} was studied. Here, we use AIA data taken in the three AIA channels, He~{\sc ii}~304~\AA, Fe~{\sc ix/x} 171~\AA, Fe~{\sc xii}~193~\AA\ and Fe~{\sc xviii}~94~\AA\ (hereafter AIA 304~\AA, AIA 171~\AA, AIA 193~\AA, and AIA~94~\AA) and longitudinal magnetograms from HMI. The AIA data have a 12\,s cadence and 0.6\arcsec\ $\times$ 0.6\arcsec\ pixel size. The HMI longitudinal magnetograms have a 45\,s cadence and 0.5\arcsec\ $\times$ 0.5\arcsec\ pixel size. The magnetic data were corrected for the projection effect in the image plane away from the disk centre. In order to improve the qualities of the AIA images, we binned each three AIA images. This is especially needed for  the 193~\AA\ channel which has a lower signal-to-noise ratio (private communication H. Morgan). Thus the time resolution of the AIA data was reduced to 36\,s. These data were used to produce the AIA images (e.g. Fig.~1, top panel) and the AIA time-slice images (e.g. Fig.~1, bottom panel). To ensure an alignment between the AIA three channel images and the HMI magnetograms, the AIA 1600~\AA\ images were first aligned with the HMI magnetograms. Then the alignment between the AIA 1600~\AA\ images and the other three channel images were checked, and where necessary offset corrections were applied. The precision of the alignment is $\approx$1\arcsec.

\section{Results}
\label{sect:res}

The first study on the data used here by \citet{2016ApJ...818....9M} (hereafter Paper~I) investigated how magnetic bipolar features associated with CBPs form and how the consequent magnetic cancellation occurs.  Magnetic cancellation was observed in all 70 CBPs  and occurred: (I) between an MBF and small weak magnetic features; (II) within an MBF with the two polarities moving toward each other from a large distance; and (III) within an MBF whose two main polarities emerge in the same place simultaneously (the so-called ephemeral regions).

Our new investigation was motivated by the study of \citet{2009A&A...495..319I} on mini-CMEs and the question on whether the omnipresent CBPs can be linked to and their evolution be responsible for these mini eruptions. We started by  selecting the CBPs that are located at solar heliographic coordinates within 430\arcsec\ from the disk centre to avoid  the projection effect. Out of 70 CBPs 42 were selected and analysed for any possible dynamic activities in the form of mass ejections (mini-CMEs, jets, micro-ejections, mini-filament eruptions, etc., see Section~\ref{sect:intro}). For this, we created image animations (including base image difference) using images taken in the AIA~193~\AA\ channel as well as HMI magnetogram animations. Once mass ejections were identified, we visually inspected them for the presence of micro-flaring, MF eruptions and/or dimmings during their full lifetime as described by \citet{2009A&A...495..319I}.

We found that out of 42 CBPs, 6 CBPs do not show eruptive activity, however in these cases no magnetic field convergence is observed. In 5 cases any of the signatures  of eruptive activity was too weak or uncertain  to be classified as eruption.  All remaining 31  CBPs produced one or several (some up to 4) eruptions during their lifetime. That brings the proportion of CBPs with identified eruptive behaviour to 76\%. We selected 11 CBPs for a detailed study and modelling, and hereafter  their most prominent eruptions are presented in full detail. Only CBPs located in the quiet Sun were studied. A CBP coronal hole dedicated study is to follow.

The next step into our investigation was to study the identified eruptive activity and associated phenomena by analysing time-slice plots in the three AIA 304~\AA, 171~\AA\ and 193~\AA\  channels.  For the purpose of the present study, the further analysis included only original (not base or running difference)  images that is required by the objectives of the present and follow-up study investigating the complexity of the observed phenomena solely visible in the original images.The time-slice plots were made from cuts through the CBPs that cross the micro-flaring region as well the cool and hot eruptive material, and the dimming region when present. The dimming is known to represent a localised density decrease caused by plasma evacuation as well as temperature decrease (due to adiabatic cooling) caused by the rising overlying loops (see Sect.~4 for further discussion). Most importantly the eruptive activity was studied in the context of the CBP formation and  full lifetime evolution of the magnetic field, transition-region and coronal emission of each CBP.
In the QS the AIA 304~\AA\ channel is dominated by the two He~{\sc ii} 303.786 and 303.786~\AA\ lines (log$T (K) \sim$ 4.7) \citep{2010A&A...521A..21O} but it also has a certain coronal emission component that is still poorly understood \citep{2003A&A...400..737A}. The 171~\AA\ channel is dominated by Fe~{\sc ix} 171.07~\AA\ (log$T (K) \sim$ 5.85) but in general the emission comes from the temperature range 5.7 $< logT (K) < 6.1$. The 193~\AA\ channel is known to be dominated by Fe~{\sc xii} lines (log$T (K) \sim$ 6.2) \citep{2010A&A...521A..21O}. The 94~\AA\ passband includes the Fe~{\sc xviii} 93.93~\AA\ line $ logT (K) \sim 6.85$ that emits only during flares  in addition to  the Fe~{\sc x} 94.01~\AA\ line $ logT (K) \sim 6.05$ \citep{2010A&A...521A..21O}. The CBP photospheric (longitudinal) magnetic field evolution is also inspected regarding the eruption process itself but also in the context of the general CBP evolution including formation, evolution and disappearance.

\begin{figure*}[!ht]
\centering
\includegraphics[scale=0.85]{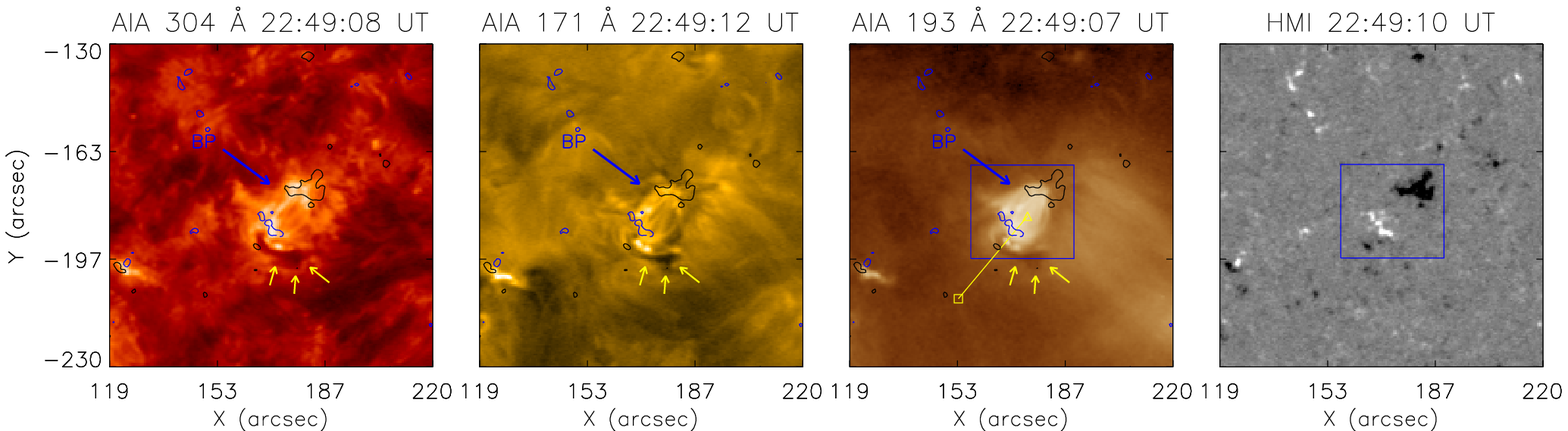}
\includegraphics[scale=0.85]{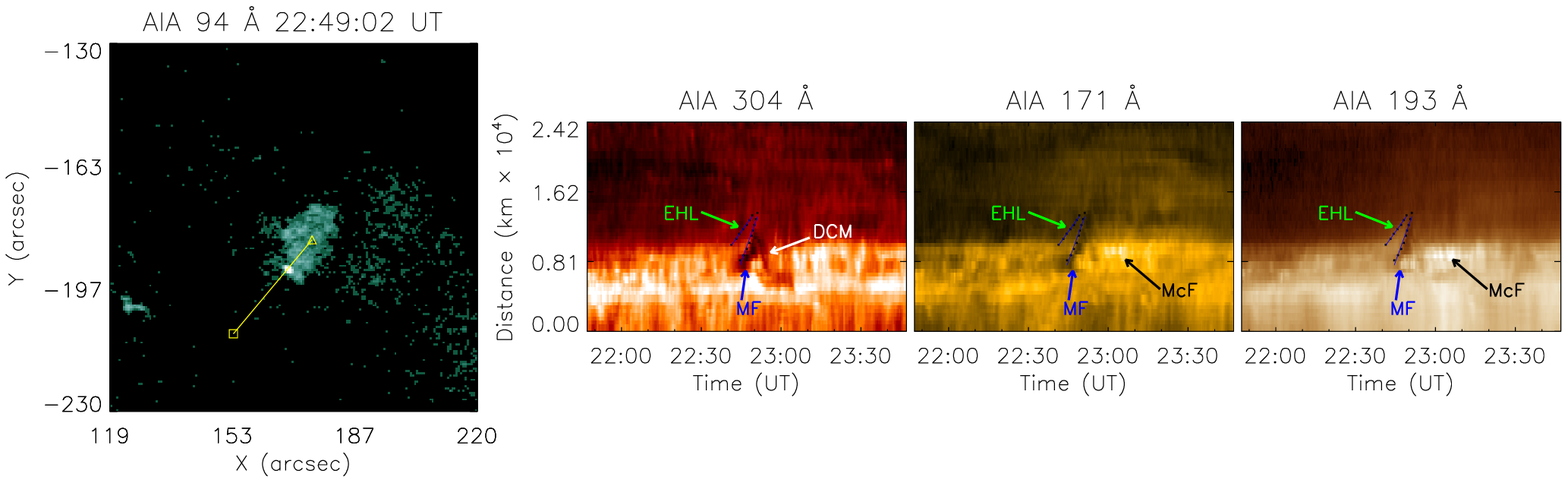}
\caption{BP1 ER1: Top row, from left to right: AIA 304~\AA, 171~\AA\ and 193~\AA\ images of first eruption BP1 and their corresponding HMI longitudinal magnetogram scaled from $-50$ to 50 ~G. The yellow arrows point at the erupting MF. The blue and black contours outline the positive and negative fluxes at $\pm$50~G. Bottom row, from left to right:   AIA 94~\AA\ image of BP1 during the first eruption showing the micro-flare. The left vertical line is a linear fit of visually selected points that outline the bright or dark front of the eruptive features. The following abbreviations are marked on the panels: BP -- bright point, EHL --  erupting hot loops, MF -- mini-filament, McF -- micro-flare, and DCM -- draining cool material. The yellow solid line indicates  the slice which has been extracted to produce the time-slice panels in the AIA 304~\AA, 171~\AA\ and 193~\AA\ channels. The bottom of the slice is marked with a triangle and the top with a square. 
An animation associated to the top row panels is available in the electronic edition.}
\label{fig:bp1_1}
\end{figure*}

\begin{figure}[!ht]
\centering
\includegraphics[scale=0.4]{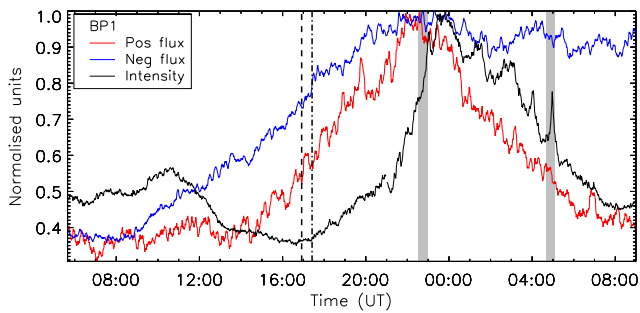}
\caption{BP1: Lightcurves of the normalized total magnetic positive (red solid line) and negative (blue solid line) flux from the region outlined with a box in the very right panel of the top row in Fig.~\ref{fig:bp1_1}. The solid black line shows the normalized total intensity in the AIA~193~\AA\ channel from the region outlined with a box in the third panel of the top row in Fig.~\ref{fig:bp1_1}. The grey areas show the eruption time, the dashed line the start of the flux emergence, and the dashed-dotted line, the time of the BP appearance in the AIA~193~\AA\ images.}
\label{fig:bp1_m}
\end{figure}

%%%%%%%%%%%%%%%%%%%%%%%%%%%%%%%%%%%%%%%%%%%%%%%%%%%%%%%%%%%%%%%%%%%%%%%%%%%%%%%%%%%%%%%%%%%%%%%%%%%%%%%%%%%%%%%%%%%%%%%%%%%%%%%%%%%%%%%%%%%%%%%%%%%%%%%%%%%%%%%%%%

The properties of the morphological and dynamical evolution of each CBP and related eruptions are given in Tables~\ref{tab:bp} and \ref{tab:erp}. Table~\ref{tab:bp} contains the general properties of the CBPs including the CBP  numeration from Paper~I, the solar disk location of each CBP, and the negative and positive polarity evolution taken from Paper~I. We added information on the starting time of the flux emergence associated with the CBP formation (information that is important for the modelling part of the present study) as well as the delay time between the flux emergence and the time when the CBP was first detected in AIA~193~\AA. We also included the start and end time of the CBP formation in AIA~193~\AA\ not only to have consistency in the present work but also because this channel response differs slightly from the AIA~211~\AA\ used in Paper~I. In Table~\ref{tab:erp} we provide the start and end times of each eruption, the eruption duration, the post-eruption change of the CBP and the feature(s) associated with the eruption. We also estimated the delay time between the CBP formation in AIA~193~\AA\ and the start of the eruptions which brings important information on the time the magnetic field system of each CBP requires to evolve into an eruptive state. Finally, we also included information on whether and when magnetic cancellation that is associated with the observed eruption has occurred. Here, we would like to remind the definition of the term magnetic cancellation (from Paper~I) -- a phenomenological description of a magnetic process that interprets the disappearance and/or decrease of opposite-sign magnetic features when they are in contact with each other observed in magnetograms. It is believed that magnetic cancellation is the observational evidence for either magnetic reconnection on photospheric level \citep[e.g., Ellermann bombs phenomena, see][]{2016A&A...590A.124R} or the submergence of small-scale loops \citep{1999SoPh..190...35H} that are being pulled into the convection zone by the magnetic tension following magnetic reconnection in the upper atmosphere also known as reconnection submergence \citep{1987smh..book.....P}.

Our investigation shows that eruptions from CBPs show large diversity even when originating from one and the same CBP. Homologous eruptions are also common. The term homologous defines a series of dynamic events that conform with the following criteria: the phenomenon must originate from the same region/location, it should have similar to identical appearance and display similar morphological evolution. Homologous phenomena are identified in the re-occurrence of solar flares \citep[e.g.,][]{1985AuJPh..38..875H,1938ZA.....16..276W, 1984AdSpR...4....5M}, coronal mass ejections \citep[e.g.,][]{2002ApJ...566L.117Z}, filaments/prominences \citep[e.g.,][]{2013ApJ...778L..29L, 2016NewA...48...66D} and jets \citep[e.g.,][]{2012ApJ...760..101W}. We selected the most prominent eruptions from 11 CBPs and report the important observational  properties  of each of these eruptions. This information is used to investigate but also constrain the data-driven modelling  of these dynamic phenomena. Hereafter, the CBPs and their corresponding eruptions will be referred to as, e.g., BP1 ER1 for the first CBP and its first eruption. The lightcurves off the CBP intensity and  total unsigned magnetic fluxes are given in the main text for BP1 and BP11, while the rest of the CBPs are shown in Fig.~\ref{fig:bps_m}.

\begin{table*}
\small
\centering
\caption{Eruption general properties.}
\begin{tabular}{cccccccc}
\hline\hline

%----------------------------------------------------------------
% First line
BP &
Start date - time &
End date - time &
Duration &
Change &
Eruptive &
Delay$^a$  &
 {Flux cancellation$^b$} \\
%----------------------------------------------------------------
% Scond line
No. &
yyyy/mm/dd (UT) &
yyyy/mm/dd (UT) &
(min) &
&
feature &
&
 $\pm$ time  \\
\hline
%----------------------------------------------------------------

\multirow{2}[0]{*}{1} & 2011/01/02 22:33 & 2011/01/02 23:00 & 27  & No    & EHL+MF  & 5h 8m  &   ongoing$^c$ weak$^d$   \\
      & 2011/01/03 04:42 & 2011/01/03 05:06 & 24 & Smaller & ELH+MF+CW+DIM/mCME & 11h 17m   &ongoing weak\\
\hline
\multirow{2}[0]{*}{2} & 2011/01/01 19:25 & 2011/01/01 19:45 & 20 & Smaller & EHL+MF+DIM& 11h 55m   &   -55m     \\
      & 2011/01/01 23:21 & 2011/01/01 23:37 & 16 & to Disappear & EHL+MF+DIM/mCME &  15h 51m   & -1h 31m \\
\hline
\multirow{2}[0]{*}{3} & 2011/01/02 09:23 & 2011/01/02 09:40 & 17 & No & EHL+CPC+DIM/mCME & 13h 31m & -- \\
      & 2011/01/02 13:12 & 2011/01/02 13:45 & 33 & Smaller & CPC+EHL & 17h 20m  & -55m      \\
\hline
\multirow{3}[0]{*}{4} & 2011/01/02 03:32 & 2011/01/02 03:52 & 20 & Smaller & EHL+MF & 12h 46m  & -1h 17m  \\
      & 2011/01/02 05:18 & 2011/01/02 05:49 & 31 & Smaller & EHL+MF  &  14h 32m   & ongoing$^c$ \\
      & 2011/01/02 06:24 & 2011/01/02 06:43 & 19 & to Disappear & CPC &  15h 38m   & ongoing \\
\hline
5     & 2011/01/01 21:30 & 2011/01/01 22:00 & 30 & Smaller & CPC+EHL & 19h 45m  &  $\pm$0  \\
\hline
6     & 2011/01/03 10:04 & 2011/01/03 10:23 & 19 & No & EHL+DIM/mCME & 22h 44m   & -8h 49m  \\
\hline
\multirow{4}[0]{*}{7} & 2011/01/01 00:32 & 2011/01/01 00:53 & 21 & No & EHL+MF+DIM+CW/mCME & 4h 3m   &-30m \\
      & 2011/01/01 15:30 & 2011/01/01 15:40 & 10 & Smaller & EHL+MF & 19h 1m &   ongoing weak   \\
      & 2011/01/01 16:17 & 2011/01/01 16:50 & 33 & No &  EHL+DIM+CW & 19h 48m   & ongoing \\
      & 2011/01/01 18:41 & 2011/01/01 18:54 & 13 & to Disappear & MF  & 22h 12m  &  ongoing    \\
\hline
\multirow{2}[0]{*}{8} & 2011/01/01 21:10 & 2011/01/01 21:50 & 40 & Smaller & EHL+DIM/mCME & 6h 20m   & -1h 25m  \\
      & 2011/01/01 23:42 & 2011/01/02 00:04 & 22 & Smaller & EHL+MF+DIM/mCME & 8h 52m   & ongoing \\
\hline
9    & 2011/01/03 13:40 & 2011/01/03 14:15 & 35 & to Disappear & EHL+CPC+DIM /mCME& 25h 36m  & -10h 53m  \\
\hline
10    & 2011/01/03 13:52 & 2011/01/03 14:15 & 23 & Smaller & EHL+CPC+DIM/mCME & 11h 22m  & -1h 12m \\
\hline
\multirow{2}[0]{*}{11} & 2011/01/02 09:10 & 2011/01/02 09:57 & 47  & Smaller & EHL+MF+DIM//mCME & 12h 30m  &ongoing weak \\
      & 2011/01/02 22:10 & 2011/01/02 22:50 & 40  & to Disappear & EHL+CPC+DIM//mCME & 25h 30m  & -2h 32m \\
\hline
\multicolumn{8}{l}Notes. \\
\multicolumn{8}{l}{EHL: eruptive hot loop(s), MF -- mini-filament, CPC -- cool plasma cloud, DIM -- dimming, CW -- coronal wave, mCME -- mini-CME} \\
\multicolumn{8}{l}{$^a$: Time delay between the eruption start and the CBP formation}\\
\multicolumn{8}{l}{$^b$: Magnetic cancellation related to the eruption}\\
\multicolumn{8}{l}{$^d$: The cancellation that started earlier and it is still ongoing}\\
\multicolumn{8}{l}{$^c$: Ongoing weak field ($\sim$10~G, at almost the noise level of HMI) cancellation.}

\end{tabular}%
\label{tab:erp}%
%\end{sidewaystable}%
\end{table*}

{\bf BP1--ER1} -- BP1 forms from bipolar flux emergence (ephemeral region, \citet{1973SoPh...32..389H}) that starts at 16:55~UT on 2011 January~2, while the CBP is first seen in AIA~193~\AA\ at $\sim$17:25~UT and disappears at $\sim$07:37~UT on January~3 with a total lifetime of 14~hrs 12~min (BP1\_193.mov). The first eruption of BP1 (Fig.~\ref{fig:bp1_1}, top row, and movie\_bp1\_er1.mov) occurs 5~hrs 8~min after the CBP formation and has a total duration of 27~min (Fig.~\ref{fig:bp1_m}). The start and end time of all eruptions are determined visually. At the beginning of the eruption a dark point-like feature appears near the southern footpoint of the CBP and later evolves into a rising loop-like phenomenon (Fig.~\ref{fig:bp1_1}). The feature is seen in absorption in all three AIA channels, 304~\AA, 171~\AA\ and 193~\AA, indicating chromospheric plasma temperatures ($<$10\,000~K), i.e. we observe an eruptive MF. The appearance of solar features as dark in coronal (171~\AA, 193~\AA\ etc) and He~{\sc ii}~304~\AA\ images is caused by the extinction of photons emitted from a hot coronal plasma in overlying neutral H~{\sc i} or He~{\sc i} atoms composing the cool plasma of, e.g., a filament or a spicule  \citep[illustrated in Fig.~10 of][]{1999ASPC..184..181R}. The loop-like MF appears to have at least one footpoint rooted at the inversion line between converging and cancelling weak ($\sim$10~G, close to the HMI noise level, \citet{2012SoPh..279..295L}) fluxes of the main positive and pre-existing negative flux fragments during the eruptive process.

The loop-like MF expands to $\sim$3~Mm away from the CBP as estimated from the time-slice plots in Fig.~\ref{fig:bp1_1}, bottom row (note that we use the term expansion rather than height as the measurements are influenced by a projection effect). This is not the real height of the eruption as the event propagates more towards the observer rather than in the plane-of-sky.  Whether this (or the rest of the eruptions studied) eruption propagates further in the upper corona is not certain given the bright QS background. A dedicated studied of  off-limb observed eruptions is needed to further investigate this. A point-like (a few pixels) brightening  (hereafter micro-flaring) appears at $\sim$22:47~UT (14~min after the first signature of the eruption, at the source region of the erupting MF and lasts until $\sim$23:09~UT. The micro-flaring is a signature of energy release where it has caused or is the result of the destabilisation of the MF. It occurs below the MF which obscures it for some time. Approximately half time through the eruption, the cool MF plasma is seen  draining back towards the source region of the eruption. The size of the CBP has no obvious change after the eruption but its brightness increases for a certain period of time which more or less coincides with the duration of the micro-flaring in the source region of the eruption. This eruptive phenomenon can be classified as a simple MF eruption that takes place at the edge of the CBP. The speed of the eruption derived from a linear fit in the time-slice image (Fig.~\ref{fig:bp1_1}, bottom row) is 6~\kms\ for the above lying loop and 10.4~\kms\ for the MF. Again, these numbers should be taken with caution given the projection effect. No coronal dimming is observed as a result of this eruption.

\begin{figure*}[!ht]
\centering
\includegraphics[scale=0.85]{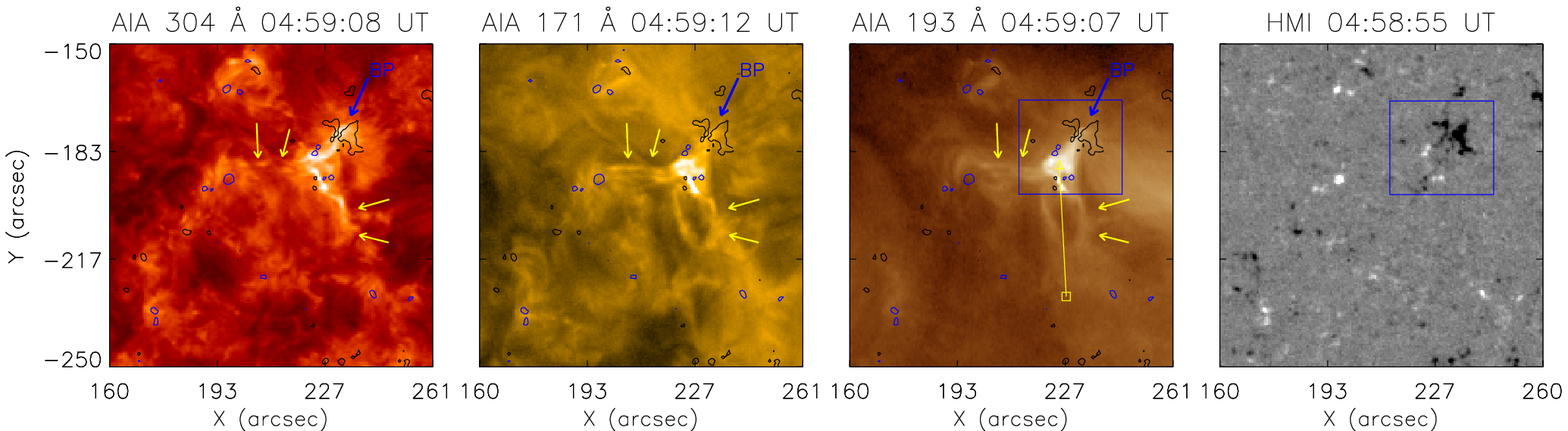}
\includegraphics[scale=0.85]{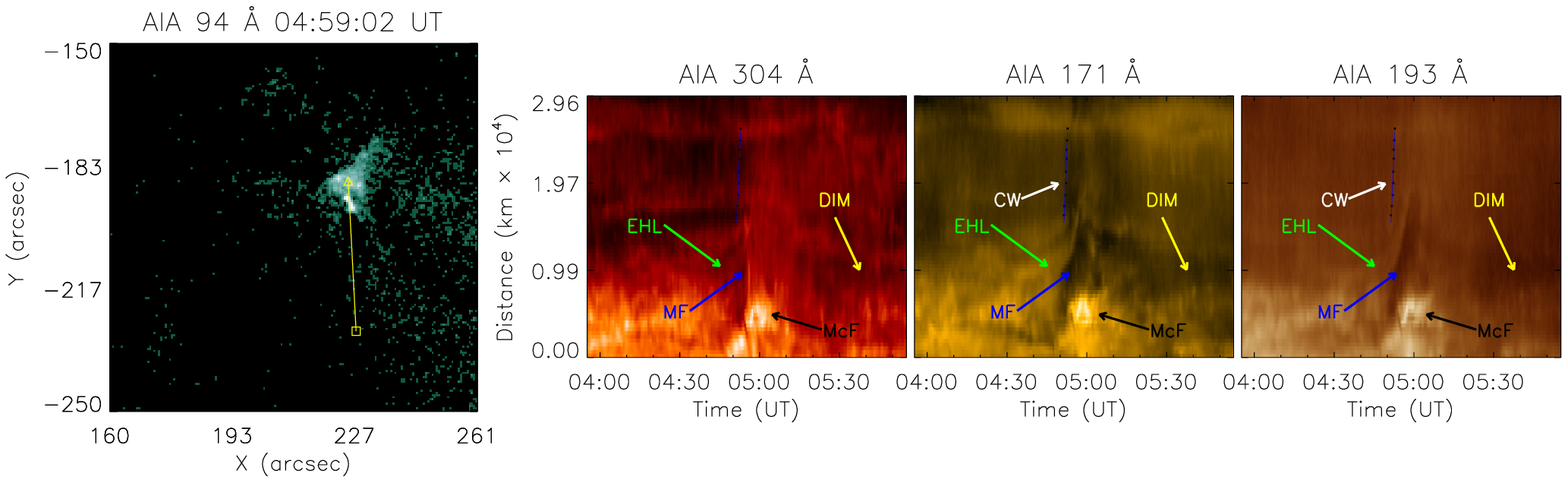}
\caption{The same as in Figure~\ref{fig:bp1_1} for  BP1 ER2. The following abbreviations are marked on the panels: BP -- bright point, EHL -- eruptive hot loops, MF -- mini-filament, McF -- micro-flare, DIM -- dimming, and CW -- coronal wave.  An animation associated to the top row panels is available in the electronic edition.}
\label{fig:bp1_2}
\end{figure*}

{\bf BP1--ER2} -- The onset of the  second eruption of BP1 (Fig.~\ref{fig:bp1_2}, top row, and  movie\_bp1\_er2.mov) is at $\sim$04:42~UT ending at $\sim$05:06~UT on January~3. This event is more complex with one part which is an erupting loop-like MF occurring at the same location as in the first eruption, and another (starting at $\sim$04:54~UT, i.e. 12~min later) that is a set of bright loops expanding eastward from the CBP (no cool material) (movie\_bp1\_er2.mov). The MF propagates southwards with footpoints that appear to be located between the main positive polarity of the CBP and the remnant of the negative pre-existing polarity that is associated with the first eruption. The pre-existing negative flux increases shortly before the eruption following its coalescence with another small-scale fragment of negative flux. The loops' eruption (eastward oriented) is most probably triggered by the dynamic changes in the CBP initiated by the erupting MF. The BP disappears around two and a half hours after this eruption.

A wave is seen moving ahead of the erupting MF and overlying loops, and it is recorded in the AIA 171~\AA\ and 193~\AA\ images. A base difference image is shown in the left panel Fig.~\ref{fig:base_diff}. The source region of the eruption brightens for almost 20~min. The dark erupting MF together with the above expending loop could be best seen in the 171~\AA\ and 193~\AA\ channels of the time-slice image (Fig.~\ref{fig:bp1_2}, bottom row). The MF ejection takes place at the outer edge of the CBP, and thus it does not disrupt significantly the  CBP, apart from a few erupted loops. The wave speed is estimated at $\approx$120~\kms. Similarly to the first eruption, magnetic cancellation of weak fluxes of one of the CBP main polarities and a pre-existing opposite flux is observed during this eruption. A dimming region expands to up to 25~Mm away from the CBP, i.e. we observed a mini-CME. The dimming is best seen in the AIA 171~\AA. A micro-flare becomes visible between the footpoints of the erupting MF at $\sim$04:54~UT, i.e. 12~min after the start of the eruption. Again, the MF obscures the micro-flare for some time.

Notably, the two eruptions start 5~hrs 8~min and 11~hrs 17~min after the CBP formation.  The two MF eruptions associated with this CBP strongly resemble a typical failed/confined filament/prominence  eruption \citep[e.g.][]{2003ApJ...595L.135J, 2007SoPh..245..287G,2012A&A...540A.127K}.
% \citep[e.g.,][]{}.

 %Bipolar convergence is the general evolution of the photospheric magnetic field that ends with the full cancellation of one of the BP main polarities \citep[e.g.,][]{1993SoPh..144...15W, 2003A&A...398..775M}.
%%%%%%%%%%%%%%%%%%%%%%%%%%%%%%%%%%%%%%%%%%%%%%%%%%%%%%%%%%%%%%%%%%%%%%%%%%%%%%%%%%%%%%%%%%%%%%%%%%%%%%%%%%%%%%%%%%%%%%%%%%%%%%%%%%%%%%%%%%%%%%%%%%%%%%%%%%%%%%%%%%

{\bf BP2--ER1} -- BP2 is first seen in AIA~193~\AA\ at $\sim$07:30~UT on January~1. It results from a bipolar flux emergence, i.e. ephemeral region, that first appears at $\sim$06:30~UT   (BP2\_193.mov, please see Sect.~\ref{details} for information where to find all movies). In all three AIA channels, 304, 171 and 193~\AA, an MF becomes visible at the west edge of the CBP at $\sim$18:50~UT. The first eruption of BP2 (Fig.~\ref{fig:bp2_1}, top row, and Fig.~\ref{fig:bps_m}) begins at $\sim$19:25~UT (i.e. 11~hrs 55~min after the CBP formation). A micro-flare is observed at $\sim$19:27~UT, i.e. 2~min after the eruption, partially obscured by the MF (Fig.~\ref{fig:bp2_1}, bottom row, and movie\_bp2\_er1.mov). Magnetic cancellation in the source region has started $\sim$55~min prior the eruption. The MF becomes more prominent (darker) while the brightening, partially covered by the MF, gets stronger and expands. Bright loops are also seen to grow forming a shamrock looking feature. The loops outline the edges of the ``shamrock leaves'' which themselves are dark with the brightening of the source region of the eruption representing the centre of the ``shamrock''. The dark region appears like a transient coronal hole, i.e. resembles a typical dimming during CMEs \citep{2000A&A...358.1097H}. We clearly see that part (the deeper/darker dimming) of the low intensity (dimming) region represents the erupting MF. The filament falls back towards the CBP around 20~min after the start of the eruption. The dimming region expands up to 18~Mm away from the CBP as determined from the time-slice images (Fig.~\ref{fig:bp2_1}, bottom row). To conclude, after a careful examination, we believe that this event represents a miniature replica of a classical CME.  The speed of the  loop expansion obtained from a linear fit in the time-slice image is $\sim$12~\kms.

%%%%%%%%%%%%%%%%%%%%%%%%%%%%%%%%%%%%%%%%%%%%%%%%%%%%%%%%%%%%%%%%%%%%%%%%%%%%%%%%%%%%%%%%%%%%%%%%%%%%%%%%%%%%%%%%%%%%%%%%%%%%%%%%%%%%%%%%%%%%%%%%%%%%%%%%%%%%%%%%%%

{\bf BP2--ER2}  -- The second eruption of BP2 starts at $\sim$23:21~UT and ends at $\sim$23:37~UT on January~1 (Fig.~\ref{fig:bp2_2}, top row, and Fig.~\ref{fig:bps_m}). We observe a MF appearing in the west of the CBP in all three AIA channels. A bright loop that encloses the MF could also be identified  moving ahead of it (movie\_bp2\_er1.mov). Approximately 10~min before the eruption a micro-flare appears above the polarity inversion line (PIL) where  the footpoints of the eruptive MF are located. A small, not so prominent, dimming region is seen in both the AIA~171~\AA\ and 193~\AA\ images. Therefore, this event can also be categorised as a mini-CME-like phenomenon. The CBP disappears $\sim$1~hr later. Magnetic cancellation between two main polarities starts $\sim$1~hr 31~min before the mini-CME. The time-slice image (Fig.~\ref{fig:bp2_2}, bottom row) shows the bright front (hot overlying loop eruption), the MF ejection and the dimming, as well the draining MF. The speed of the second eruption is similar to the first one, i.e. $\sim$11~\kms.

%%%%%%%%%%%%%%%%%%%%%%%%%%%%%%%%%%%%%%%%%%%%%%%%%%%%%%%%%%%%%%%%%%%%%%%%%%%%%%%%%%%%%%%%%%%%%%%%%%%%%%%%%%%%%%%%%%%%%%%%%%%%%%%%%%%%%%%%%%%%%%%%%%%%%%%%%%%%%%%%%%

{\bf BP3--ER1} -- BP3 appears in AIA~193~\AA\ at $\sim$19:52~UT on January~1 after a bipolar magnetic flux emergence  (Fig.~\ref{fig:bps_m} and BP3\_193.mov). The first eruption from BP3 (Fig.~\ref{fig:bp3_1}, top row, and Fig.~\ref{fig:bps_m}) begins at $\sim$09:23~UT (13~hrs 31~min after the BP3 formation). The eruption starts with a bright structure stretching northward from the CBP  evolving into a bundle of erupting CBP loops seen sidewise together with a  cool plasma cloud (CPC) (movie\_bp3\_er1.mov). At the same time a micro-flare is observed at the source region of the expanding loops at $\sim$09:26~UT. A small dimming region forms that is hardly distinguishable in the 304~\AA\ channel, and lasts for a short time in both coronal channels. The dimming also expands to a very small distance of no more than 10~Mm. The dimming region appears to result from the density depletion caused by  the evacuation of coronal material from the expanding loops.  A  cool material from the erupting MF is also present in this region. This event could be described as a mini-CME. There is no convergence of the main polarities during the eruption.  The loops slowly rise with a speed of 1~\kms\ for around 20~min while the eruption speed is ~14~\kms\ estimated from the linear fits in Fig.~\ref{fig:bp3_1}, bottom row.

%%%%%%%%%%%%%%%%%%%%%%%%%%%%%%%%%%%%%%%%%%%%%%%%%%%%%%%%%%%%%%%%%%%%%%%%%%%%%%%%%%%%%%%%%%%%%%%%%%%%%%%%%%%%%%%%%%%%%%%%%%%%%%%%%%%%%%%%%%%%%%%%%%%%%%%%%%%%%%%%%%

{\bf BP3--ER2} -- The second eruption of BP3 (Fig.~\ref{fig:bp3_2}, top row, and Fig.~\ref{fig:bps_m}) begins at $\sim$13:12~UT on January 2. We again observe  a CPC erupting northeast of the CBP. The eruption generates two dark round areas outlined by the expanding overlying loops (movie\_bp3\_er2.mov). The time-slice image (Fig.~\ref{fig:bp3_2}, bottom panels) reveals  that the cool material starts falling back to the solar surface as early as 10~min after the beginning of the eruption. One can also see that after $\sim$13:50~UT hot plasma rises above the dimming and moves towards the CBP which in the time-slice image (Fig.~\ref{fig:bp3_2}, bottom row) appears as falling back hot plasma. The CBP becomes around 30\% smaller after the eruption.
A convergence between two main polarities is observed during the eruption but no contact between them could be observed.  Cancellation at the PIL between weak magnetic fluxes (close to the noise level) is observed between 12:50 and 13:30~UT.  A micro-flare appears below the erupting CPC in the north of the CBP at $\sim$13:24~UT, i.e. 12~min after the start of the eruption as it has been obscured for some time from the CPC.  The micro-flare is too weak to be detected  in the AIA~94 channel. The speed of the eruption is 7.3~\kms.

%%%%%%%%%%%%%%%%%%%%%%%%%%%%%%%%%%%%%%%%%%%%%%%%%%%%%%%%%%%%%%%%%%%%%%%%%%%%%%%%%%%%%%%%%%%%%%%%%%%%%%%%%%%%%%%%%%%%%%%%%%%%%%%%%%%%%%%%%%%%%%%%%%%%%%%%%%%%%%%%%%

{\bf BP4--ER1} -- BP4 forms at $\sim$14:46~UT on January~1 following the convergence of bipolar fragments (Fig.~\ref{fig:bps_m} and BP3\_193.mov). The first eruption of BP4 (Fig.~\ref{fig:bp4_1}, top row, and Fig.~\ref{fig:bps_m}) starts at $\sim$03:32~UT (12~hrs 46~min after the BP4 formation). At the start of the eruption cool (what appears to be a MF) and hot (the CBP loops) plasma is ejected in northeast direction (movie\_bp4\_er1.mov). A micro-flare could be observed at $\sim$03:33~UT in the centre of the CBP which is also the source region of the erupting material. Magnetic cancellation between the two main polarities starts 1~hr 17~min before the eruption. The time-slice image (Fig.~\ref{fig:bp4_1}, bottom row) shows the propagation of the hot and cool material at a speed of $\sim$80~\kms, with a pre-eruptive rising phase of a few \kms. No dimming region is seen during this eruption. The MF expands not far from the CBP ($\sim$10~Mm), while the hot material spreads to at least $\sim$40~Mm away.

%%%%%%%%%%%%%%%%%%%%%%%%%%%%%%%%%%%%%%%%%%%%%%%%%%%%%%%%%%%%%%%%%%%%%%%%%%%%%%%%%%%%%%%%%%%%%%%%%%%%%%%%%%%%%%%%%%%%%%%%%%%%%%%%%%%%%%%%%%%%%%%%%%%%%%%%%%%%%%%%%%

{\bf BP4--ER2} -- The second eruption of BP4 starts at $\sim$05:18~UT with a slow expansion of some of the CBP loops (Fig.~\ref{fig:bp4_2}, bottom row), followed by a MF that appears above the CBP loops at $\sim$05:24~UT. Subsequently cool and hot plasma is ejected (Fig.~\ref{fig:bp4_2}, top row, and movie\_bp4\_er2.mov). The speed of the eruption estimated from the linear fit in the time-slice images (Fig.~\ref{fig:bp4_2}, bottom row) is $\sim$92~\kms. The MF starts draining back to the source region as soon as the eruption has started (see Fig.~\ref{fig:bp4_2}). Again, no dimming is seen during this event. A micro-flare appears in the centre of the CBP in the source region of the erupting material at $\sim$05:28~UT, again partially obscured by the erupting MF.

%%%%%%%%%%%%%%%%%%%%%%%%%%%%%%%%%%%%%%%%%%%%%%%%%%%%%%%%%%%%%%%%%%%%%%%%%%%%%%%%%%%%%%%%%%%%%%%%%%%%%%%%%%%%%%%%%%%%%%%%%%%%%%%%%%%%%%%%%%%%%%%%%%%%%%%%%%%%%%%%%%

{\bf BP4--ER3} -- The third and last eruption of BP4 (Fig.~\ref{fig:bp4_3}, top row, and movie\_bp4\_er3.mov) represents solely a MF or CPC eruption that has started at $\sim$06:24~UT, 24~min before the CBP end of life. The eruption starts with the MF appearing above the CBP and shortly after it is ejected. The MF is located at the PIL between the disappearing (cancelling) flux of the bipole. The MF appears to fall down but quite dispersed with no obvious traces in the time-slice image (Fig.~\ref{fig:bp4_3}, bottom row). The CBP disappears 5~min after the eruption. A micro-flare appears beneath the eruptive filament above the PIL of the cancelling magnetic flux at $\sim$06:33~UT, i.e. 9~min after the eruption has started.

The three consecutive eruptions of BP4 are an excellent example of homologous eruptions and their modelling will shade more light on the magnetic field evolution associated with such  phenomena. The micro-flares during ER2 and ER3 are weaker and are not registered in the AIA~94 channel. They originate from the same location in the CBP and are separated by 1~hr 46~min and 1~hr 6~min, between the ER1 and ER2, and ER2 and ER3, respectively.
%%%%%%%%%%%%%%%%%%%%%%%%%%%%%%%%%%%%%%%%%%%%%%%%%%%%%%%%%%%%%%%%%%%%%%%%%%%%%%%%%%%%%%%%%%%%%%%%%%%%%%%%%%%%%%%%%%%%%%%%%%%%%%%%%%%%%%%%%%%%%%%%%%%%%%%%%%%%%%%%%%

{\bf BP5} -- BP5 results from magnetic flux coalescence and it is first seen in AIA~193~\AA\ at $\sim$01:45~UT on January~1 (Fig.~\ref{fig:bps_m} and BP5\_193.mov). The eruption of BP5 starts at $\sim$21:30~UT, i.e. 19~hrs 45~min after the BP5 formation (Fig.~\ref{fig:bp5_1}, top row, and movie\_bp5\_er1.mov). A micro-flare appears at $\sim$21:32~UT in the east of the CBP where is also the eruption source region. It lasts for 43~min. During this eruption cool and hot material is ejected simultaneously. The two main magnetic polarities begin cancelling at the start of the eruption. From the time-slice image (timeslice\_bp5\_er1.eps) we estimated a rising phase speed of 1.4~\kms\ and an eruption speed of $\sim$13~\kms. No dimming is detected during this eruption. The hot and cool plasma falls back after expanding to up to $\sim$20~Mm away from the CBP.
%KG  expanding above!!!!

%%%%%%%%%%%%%%%%%%%%%%%%%%%%%%%%%%%%%%%%%%%%%%%%%%%%%%%%%%%%%%%%%%%%%%%%%%%%%%%%%%%%%%%%%%%%%%%%%%%%%%%%%%%%%%%%%%%%%%%%%%%%%%%%%%%%%%%%%%%%%%%%%%%%%%%%%%%%%%%%%%

\begin{figure*}[!ht]
\centering
\includegraphics[scale=0.85]{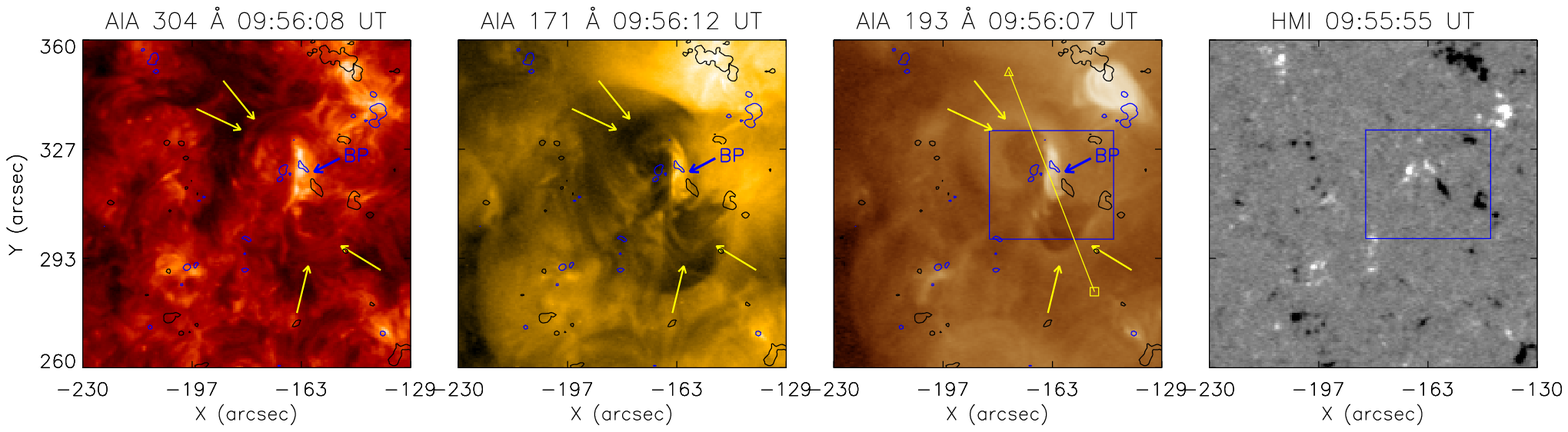}
\includegraphics[scale=0.85]{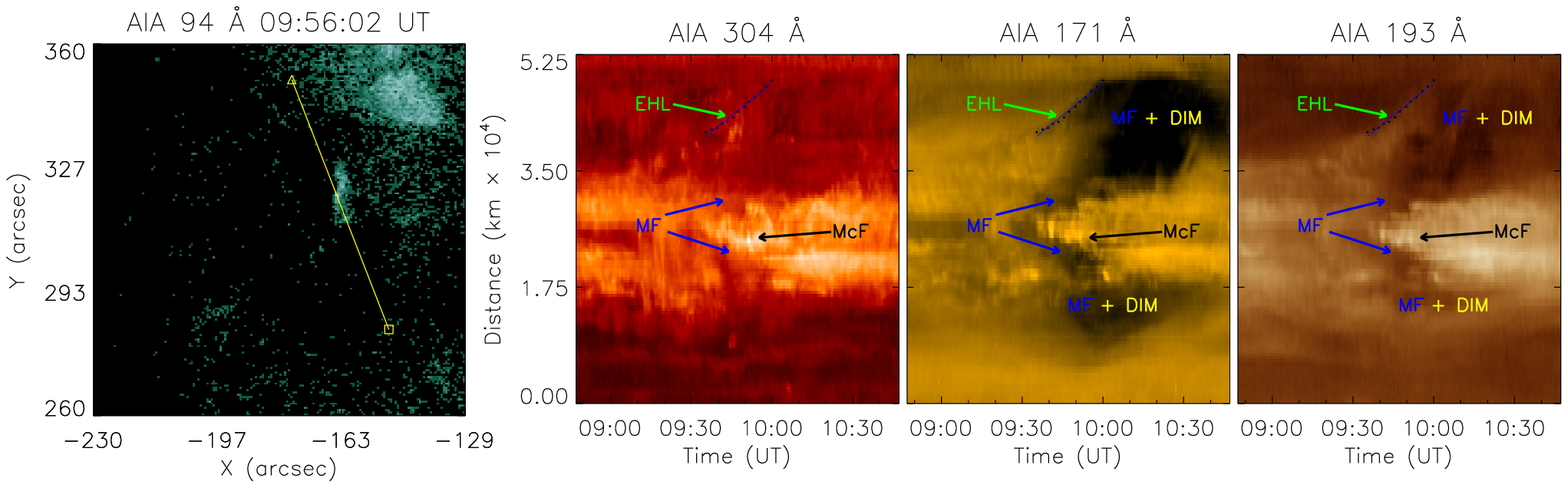}
\caption{The same as in Figure~\ref{fig:bp1_1} for the first eruption in BP11 ER1. The following abbreviations are marked on the panels: BP -- bright point, EHL -- eruptive hot loops, MF -- mini-filament, McF -- micro-flare, and  DIM -- dimming.  An animation associated to the top row panels is available in the electronic edition.}
\label{fig:bp11_1}
\end{figure*}

\begin{figure}[!ht]
\centering
\includegraphics[scale=0.4]{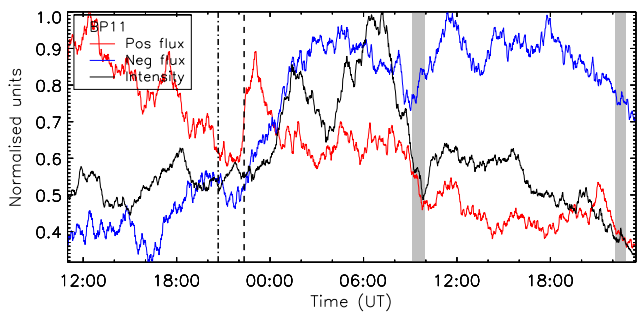}
\caption{The same as Figure~2 for BP11.}
\label{fig:bp11_m}
\end{figure}

\begin{figure*}[!ht]
\centering
\includegraphics[scale=0.85]{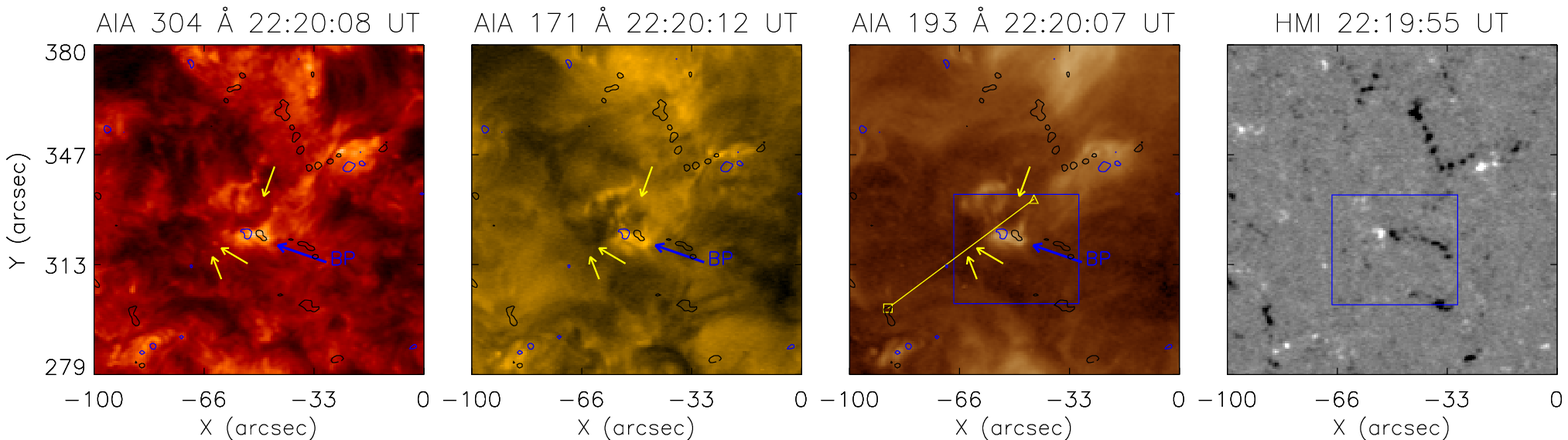}
\includegraphics[scale=0.85]{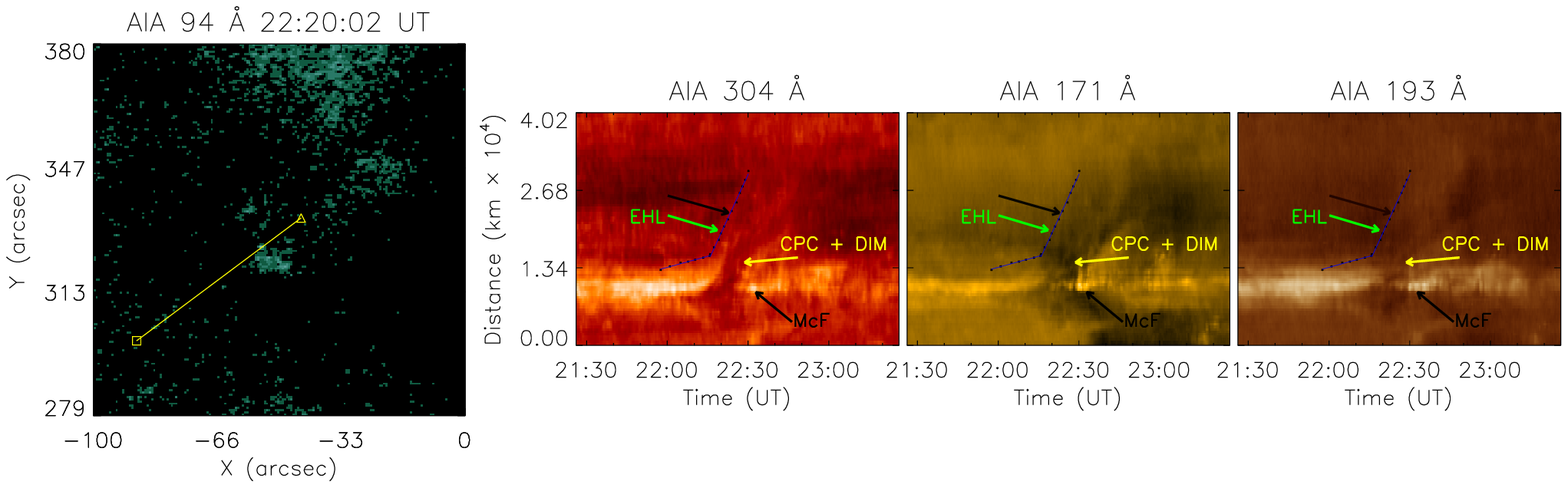}
\caption{The same as in Figure~\ref{fig:bp1_1} for BP11 ER2.  The following abbreviations are marked on the panels: BP -- bright point, EHL -- eruptive hot loops, MF -- mini-filament, McF -- micro-flare, and DIM -- dimming.  An animation associated to the top row panels is available in the electronic edition.}
\label{fig:bp11_2}
\end{figure*}

{\bf BP6} -- BP6 is associated with an ephemeral region that appears at $\sim$10:15~UT on January~2 and the emission from  BP6 is first registered around an hour later at $\sim$11:20~UT (Fig.~\ref{fig:bps_m}  and BP6\_193.mov). The eruption of BP6 begins $\sim$22~hrs 44~min  after the CBP appearance in EUV (Fig.~\ref{fig:bp6_1}, top row, and movie\_bp6\_er1.mov), i.e. close to the end of the lifetime of the CBP (at 10:04~UT on January~3). In the AIA~304 and 171~\AA\ time-slice images (Fig.~\ref{fig:bp6_1}, bottom row) we could distinguish cooler loop (no signature in AIA~193~\AA) oscillations with increasing amplitude, overlaying the CBP. The eruption involves the expulsion of both the higher overlying as well as the CBP loops. A dimming appears in the northeast of the CBP in the 171~\AA\ and 193~\AA\ images (Fig.~\ref{fig:bp6_1}, bottom row). A cool material may also be present judging from the appearance of a darkening, but we cannot be certain. No obvious micro-brightening was observed during this eruption, however, the whole CBP becomes brighter for almost 20~min after the eruption.  This brightening is too weak to be registered in the AIA~94 channel. A weak magnetic field at the PIL is seen interacting/cancelling which has been ongoing for already 8~hrs 49~min when the eruption took place. The eruption speed is estimated to ~13~\kms. This eruption could also qualify as a mini-CME.

%%%%%%%%%%%%%%%%%%%%%%%%%%%%%%%%%%%%%%%%%%%%%%%%%%%%%%%%%%%%%%%%%%%%%%%%%%%%%%%%%%%%%%%%%%%%%%%%%%%%%%%%%%%%%%%%%%%%%%%%%%%%%%%%%%%%%%%%%%%%%%%%%%%%%%%%%%%%%%%%%%

{\bf BP7--ER1} -- BP7 is an exceptional case of a double bipolar magnetic flux emergence (Fig.~\ref{fig:bps_m} and BP7\_193.mov). The first eruption of BP7 starts at $\sim$00:32~UT on January~1 (Fig.~\ref{fig:bp7_1}, top row, and movie\_bp7\_er1.mov), only $\sim$4~hrs after the BP formation. A MF first appears at $\sim$00:17~UT in the west of the CBP with a slow lift up of hot loops (movie\_bp7\_er1.mov). The MF lies at the PIL between the main positive polarity associated with the CBP and the nearby pre-existing negative polarity. At $\sim$00:29~UT, in the southern part of the MF a  micro-flare appears, becomes larger and longer while the filament turns darker and thicker, and shortly after it erupts. This is accompanied by the formation of a large dimming region, again best and longest seen in AIA~171~\AA. A shorter living dimming is observed in AIA~193~\AA\ (Fig.~\ref{fig:bp7_1}, bottom row). Magnetic cancellation between the main positive polarity of the CBP and the pre-existing negative polarities starts 30~min before the eruption. The magnetic cancellation is still ongoing during the following three eruptions but involves mostly weak field. A CW is observed with a speed of $\sim$87~\kms.  A base difference image is shown in the middle panel Fig.~\ref{fig:base_diff}. It is interesting to note that a MF that lies in the southern part of the CBP remains undisturbed during  this action. This eruption appears to carry all signatures of a mini-CME.

%%%%%%%%%%%%%%%%%%%%%%%%%%%%%%%%%%%%%%%%%%%%%%%%%%%%%%%%%%%%%%%%%%%%%%%%%%%%%%%%%%%%%%%%%%%%%%%%%%%%%%%%%%%%%%%%%%%%%%%%%%%%%%%%%%%%%%%%%%%%%%%%%%%%%%%%%%%%%%%%%%

{\bf BP7--ER2} -- The second eruption of BP7 begins at $\sim$15:30~UT and it is related to the main BP bipole evolution rather than its interaction with the pre-existing flux as during the first eruption (Fig.~\ref{fig:bp7_2}, top row, and movie\_bp7\_er2.mov). This event represents a MF embedded in the CBP erupting loops that is followed by the MF plasma draining  towards the source region of the eruption (Fig.~\ref{fig:bp7_2}). Oscillations of cooler overlying loops (not distinguishable in AIA~193~\AA) prior and during the eruption are clearly present. A dimming region forms more than 15~min after the start of the eruption. A micro-flare beneath the filament could be observed at $\sim$15:32~UT during the rise of the filament (Fig.~\ref{fig:bp7_2}, bottom row). The speed of the eruption is $\sim$26~\kms.

%%%%%%%%%%%%%%%%%%%%%%%%%%%%%%%%%%%%%%%%%%%%%%%%%%%%%%%%%%%%%%%%%%%%%%%%%%%%%%%%%%%%%%%%%%%%%%%%%%%%%%%%%%%%%%%%%%%%%%%%%%%%%%%%%%%%%%%%%%%%%%%%%%%%%%%%%%%%%%%%%%

{\bf BP7--ER3} -- The third eruption of BP7 happens at $\sim$16:17~UT (Fig.~\ref{fig:bp7_3}, top row, and movie\_bp7\_er3.mov). During the eruption, the CBP loops are ejected and a faint dimming region forms extending to at least 25~Mm. The lift-up of the loops is relatively slow ($\sim$3~\kms) and at 16:32~UT it accelerates rising with $\sim$10~\kms\ (Fig.~\ref{fig:bp7_3}, bottom row). No cool plasma is detected during this eruption. A micro-flare appears in the centre of the CBP which is also the location from where the dimming region starts to spread at $\sim$16:21~UT.  A base difference image is shown in the right panel Fig.~\ref{fig:base_diff} shows the CW related to this event.

%%%%%%%%%%%%%%%%%%%%%%%%%%%%%%%%%%%%%%%%%%%%%%%%%%%%%%%%%%%%%%%%%%%%%%%%%%%%%%%%%%%%%%%%%%%%%%%%%%%%%%%%%%%%%%%%%%%%%%%%%%%%%%%%%%%%%%%%%%%%%%%%%%%%%%%%%%%%%%%%%%

{\bf BP7--ER4} -- The fourth and last eruption of BP7 starts at $\sim$18:41~UT on January~1 (Fig.~\ref{fig:bp7_4}, top row,  and movie\_bp7\_er4.mov). A MF appears above the CBP at $\sim$18:06~UT and it lifts up 28~min later without triggering any changes in the CBP. No CBP loop ejection is observed. A micro-flare beneath the MF could be observed at 18:47~UT. During the eruption, the north footpoint of the MF rises and then the whole filament erupts (movie\_bp7\_er4.mov). The time-slice image (Fig.~\ref{fig:bp7_4}, bottom row) shows clearly the rising MF at a speed of $\sim$8~\kms.

%%%%%%%%%%%%%%%%%%%%%%%%%%%%%%%%%%%%%%%%%%%%%%%%%%%%%%%%%%%%%%%%%%%%%%%%%%%%%%%%%%%%%%%%%%%%%%%%%%%%%%%%%%%%%%%%%%%%%%%%%%%%%%%%%%%%%%%%%%%%%%%%%%%%%%%%%%%%%%%%%%

%%%%%%%%%%%%%%%%%%%%%%%%%%%%%%%%%%%%%%%%%%%%%%%%%%%%%%%%%%%%%%%%%%%%%%%%%%%%%%%%%%%%%%%%%%%%%%%%%%%%%%%%%%%%%%%%%%%%%%%%%%%%%%%%%%%%%%%%%%%%%%%%%%%%%%%%%%%%%%%%%%
{\bf BP8--ER1} -- BP8 forms from a newly emerged positive flux that converges with a pre-existing negative fragment (Fig.~\ref{fig:bps_m} and BP8\_193.mov). It first appears at $\sim$14:50~UT in AIA~193~\AA. The first eruption of BP8 starts at $\sim$21:10~UT on January~1 (Fig.~\ref{fig:bp8_1}, top row, and movie\_bp8\_er1.mov).  The positive flux decreases at the start of the eruption  but then increases so thus the negative flux (see the BP8 panel in Fig.~\ref{fig:bps_m}).
A micro-flare is observed at $\sim$21:05~UT in the east of the CBP which is also the  source region of the eruption followed by loops being ejected from the CBP. A dimming region expands not far from the CBP. The full extension of the eruption  cannot be determined due to the high background emission, but it reaches a minimum of 50~Mm and thus qualifies this event as a mini-CME. No filament eruption is detected during the event. Converging of the main polarities could be observed, leading to magnetic cancellation that starts at 1~hrs 25~min before the eruption. From the time-slice image (Fig.~\ref{fig:bp8_1}, bottom row) the speed of the eruption is estimated at $\sim$54~\kms.

{\bf BP8--ER2} -- The second eruption of BP8 starts at $\sim$23:42~UT (Fig.~\ref{fig:bp8_2}, top row). A micro-flare could be observed at $\sim$23:45~UT in the centre of the CBP, close to one of the footpoints of the eruptive MF. The CBP loops are ejected and expand  to a large distance of $\sim$40~Mm. The MF, after reaching a distance of almost 20~Mm, falls back. Magnetic cancellation is still ongoing in the same location as in  ER1. From the time-slice image (Fig.~\ref{fig:bp8_2}, bottom row) a $\sim$34~\kms\ speed of the hot material was estimated. We believe, this eruption also qualifies as a mini-CME despite the small dimming region.

%%%%%%%%%%%%%%%%%%%%%%%%%%%%%%%%%%%%%%%%%%%%%%%%%%%%%%%%%%%%%%%%%%%%%%%%%%%%%%%%%%%%%%%%%%%%%%%%%%%%%%%%%%%%%%%%%%%%%%%%%%%%%%%%%%%%%%%%%%%%%%%%%%%%%%%%%%%%%%%%%%

{\bf BP9} -- BP9  results from bipolar magnetic flux emergence that starts at $\sim$11:30~UT on 2011 January~2 (Fig.~\ref{fig:bps_m} and BP9\_193.mov). The eruption of BP9 begins at $\sim$13:40~UT on January~3 (Fig.~\ref{fig:bp9_1}, top row, and movie\_bp9\_er1.mov). A micro-flare could be observed at $\sim$13:42~UT in the south part of the CBP at the PIL between two main polarities where is also the source region of the erupting cool material, followed by the ejection of numerous loops,  and the appearance of a dimming region. The CBP disappears after the eruption. Magnetic flux cancellation has started long before the eruption, 10~hrs 53~min earlier. The dimming region extends to $\sim$23~Mm determined from the AIA~193~\AA\ image time-slice image (Fig.~\ref{fig:bp9_1}, bottom row). This eruption represents a mini-CME.

%%%%%%%%%%%%%%%%%%%%%%%%%%%%%%%%%%%%%%%%%%%%%%%%%%%%%%%%%%%%%%%%%%%%%%%%%%%%%%%%%%%%%%%%%%%%%%%%%%%%%%%%%%%%%%%%%%%%%%%%%%%%%%%%%%%%%%%%%%%%%%%%%%%%%%%%%%%%%%%%%%
%%%%%%%%%%%%%%%%%%%%%%%%%%%%%%%%%%%%%%%%%%%%%%%%%%%%%%%%%%%%%%%%%%%%%%%%%%%%%%%%%%%%%%%%%%%%%%%%%%%%%%%%%%%%%%%%%%%%%%%%%%%%%%%%%%%%%%%%%%%%%%%%%%%%%%%%%%%%%%%%%%

{\bf BP10} -- BP10 forms from the convergence of a pre-existing positive flux and the negative flux of a newly emerged bipole  (Fig.~\ref{fig:bps_m} and BP10\_193.mov). It first appears in AIA~193~\AA\ at $\sim$02:30~UT on 2011 January~3.  The eruption begins at $\sim$13:52~UT on January~3 (Fig.~\ref{fig:bp10_1}, top row, and movie\_bp10\_er1.mov). A micro-flare is observed at $\sim$13:58~UT beneath the erupting filament, i.e. the filaments is seen partially to obstruct it. A loop structure rises and expends together with  a dark material, a CPC, (best seen in AIA~171~\AA, movie\_bp10\_er1.mov). Magnetic cancellation began $\sim$1~hr 12~min prior to the eruption. From the linear fit of the expending loop in the time-slice images (Fig.~\ref{fig:bp10_1}, bottom row) a speed of $\sim$65~\kms\ is obtained. This event can also  be referred to as a mini-CME.

%%%%%%%%%%%%%%%%%%%%%%%%%%%%%%%%%%%%%%%%%%%%%%%%%%%%%%%%%%%%%%%%%%%%%%%%%%%%%%%%%%%%%%%%%%%%%%%%%%%%%%%%%%%%%%%%%%%%%%%%%%%%%%%%%%%%%%%%%%%%%%%%%%%%%%%%%%%%%%%%%%

{\bf BP11--ER1}  -- BP11 is the result from a newly emerged negative flux that converges with a pre-existing positive flux concentration (Fig.~\ref{fig:bp11_m} and  BP11\_193.mov). A few hours earlier ($\sim$00:30~UT) a filament appears above  the CBP.  The first eruption of BP11 starts at $\sim$09:10~UT on January~2 (Fig.~\ref{fig:bp11_1}, top row, and movie\_bp11\_er1.mov) with the filament and the CBP loops lifting up. A micro-flare appears at $\sim$09:35~UT at the source region of the eruption, followed at $\sim$09:44~UT by two dimming regions that expand north and south of the CBP. Both dimmings appear gradually, first in AIA~171~\AA, and later in AIA~193~\AA. Only the south dimming area moves further away from the CBP and expands to  $\sim$25~Mm from the source (micro-flare) region. The speed of the dimming expansion is $\sim$5~\kms (Fig.~\ref{fig:bp11_1}, bottom row) . A weak field that migrates from the main CBP polarities continuously cancels. As in earlier cases of weak flux cancellation, it is impossible to determine the precise time of the start of the cancellation as the flux involved is at the noise level of the HMI magnetograms. This event carries clear signatures of a mini-CME.

%%%%%%%%%%%%%%%%%%%%%%%%%%%%%%%%%%%%%%%%%%%%%%%%%%%%%%%%%%%%%%%%%%%%%%%%%%%%%%%%%%%%%%%%%%%%%%%%%%%%%%%%%%%%%%%%%%%%%%%%%%%%%%%%%%%%%%%%%%%%%%%%%%%%%%%%%%%%%%%%%%

{\bf BP11--ER2} --The second eruption of BP11 (Fig.~\ref{fig:bp11_2}, top row, and movie\_bp11\_er2.mov) starts at $\sim$22:10~UT on January~2. A micro-flare could  first be observed at $\sim$22:29~UT in the east of the CBP which is the source region of the erupting material. Cool material is seen erupting from the northeast of the CBP. The CBP disappears after the eruption. Magnetic cancellation between the CBP main polarities has started 2~hrs 32~min earlier.  From the time-slice image (Fig.~\ref{fig:bp11_2}, bottom row) we could see a bright front lifting up, best seen in AIA~304, and the formation of a dimming region in all three channels. The loop expansion is relatively slow at $\sim$2~\kms\ and at $\sim$22:17~UT it  accelerates  up to $\sim$17~\kms\ (Fig.~\ref{fig:bp11_2}, bottom row). The event is again a mini-CME.

%%%%%%%%%%%%%%%%%%%%%%%%%%%%%%%%%%%%%%%%%%%%%%%%%%%%%%%%%%%%%%%%%%%%%%%%%%%%%%%%%%%%%%%%%%%%%%%%%%%%%%%%%%%%%%%%%%%%%%%%%%%%%%%%%%%%%%%%%%%%%%%%%%%%%%%%%%%%%%%%%%

\section{Discussion}
\label{sect:sum}

Studying CBPs presents a unique opportunity to reach a better understanding of the most fundamental and outstanding questions in solar physics on how coronal plasma is heated to million degrees and what causes the eruptive behaviour in the solar atmosphere. Why CBPs? Alike active regions, CBPs are composed of coronal loops whose plasma is heated to millions of degrees. At the same time, CBPs are small, most of them have a very simple configuration of a set of loops  connecting well isolated bipolar magnetic fragments with  short lifetimes of $\le$24~hrs that permits  to study their full  lifetime. This together with the present observational and modelling capabilities, provide exceptional prospects in studying how fundamental physical processes operate in the solar atmosphere. In the following we summarize and discuss the obtained results.

First and foremost, the present study showed for the first time that the majority of quiet Sun CBPs produce at least one eruption during their lifetime. We selected 42 out of 70 randomly chosen CBPs investigated in Paper~I. The 42 CBPs were analysed searching for signatures of eruptive behaviour that includes at least two of the following features usually associated with solar eruptions:  micro-flaring, mini-filament (MF) eruption, loops' eruption, and/or corona dimming. We found that 76\% of the CBPs (31 out of 42) produced at least one eruption, 6 did not produce eruptions, and in 5 CBPs the identification of eruptions was inconclusive.  We selected 11 CBPs for detailed investigation of their most prominent eruptions. In the following, we summarise and discuss our main findings.

\subsection{Eruptions associated with CBPs}

In total 21 eruptions were analysed with 4 eruptions in one CBP (BP7), one CBP (BP4) with 3 eruptions, five  CBPs with 2, and four CBPs with only 1 eruption. None of the eruptions resembles a collimated flow, i.e. a jet, which is to expect because of the lack of open magnetic field in the surrounding of these CBPs in the quiet Sun.  Some of the multiple eruption cases appear to be homologous which will be further investigated in the second part of this study by modelling the magnetic field evolution associated with them and by comparing with  their observational properties described here.
Homologous eruptions require that the magnetic field structure on both local and global scales are comparable between the different eruptions. It is, therefore,  expected that the eruptions happen in a way where the original magnetic field structure is reconstructed as part of the process. This should involve for instance  magnetic field reconnection below the rising filament, recreating the previous magnetic field structure where a new filament can be formed.  It may happen both as part of the ongoing dynamical process, but also through the adding of new twist during the reconnection process taking place with the lift off of the previous eruption.

\subsection{Eruption timing and duration}

The 11 CBPs have lifetimes ranging from 10~hrs 50~min to 29~hrs 40~min, with an average value of 20~hrs 42~min. The time of the eruption after a CBP formation (either from emergence or occasional connectivity) provides important information on the time needed by a magnetic structure to evolve into a strongly sheared configuration that consequently erupts due to an instability or magnetic reconnection. We found that from 21 eruptions, 18 occur 9 to 25~hrs after the CBP formation, with an average delay of 16~hrs 40~min.  This time delay in the eruption occurrence coincides in each CBP with the convergence and later cancellation phase of the CBP bipoles during which the CBPs become smaller until they fully disappear.
In the remaining three cases the eruption occurred relatively soon after the CBP formation, i.e. 5~hrs 8~min (BP1, first eruption), 4~hrs 3~min (BP7, first eruption), and 6~hrs 20~min (BP8, first eruption) after the CBP appearance in the AIA~193~\AA\ channel. We come back  to these cases later. In 9 CBPs the last eruption occurred between 24~min and 2~hrs 36~min before the CBPs fully vanished. Two CBPs remain present for more than 6~hrs after their last eruption (BP6  and BP10). In both cases the bipole had a  magnetic flux strong enough to maintain the CBP for many hours after the eruption. The duration of the eruptions is in the range from 10 to 47~min with an average of 26~min.

\subsection{Magnetic flux convergence and cancellation}

The photospheric magnetic field evolution of the eruptions associated with CBPs was investigated in the context of the full lifetime evolution of the CBPs in order to find out how an eruptive state could be reached in what appears to be a simple magnetic bipolar configuration.   Six CBPs were formed from bipolar flux emergence, one from flux convergence, three from one polarity converging and one from both emerging and pre-existent flux, and one CBP resulted from the coalescence of pre-existing small-scale magnetic fragments. As already shown in Paper~I, more than 50\% of the CBPs are associated with ephemeral regions while the rest are related to a chance encounter of magnetic fluxes. \citet{1994ASPC...68..377H} established from X-ray and magnetic field observations that 20\% -- 30\%\ of CBPs relate to ephemeral regions. The present data hint towards more CBPs being formed by flux emergence which can be explained with both the higher resolution of the present data but also with their recent identification in lower temperature imaging data.   It is well known that the photospheric magnetic flux evolution of CBPs during the second half (even two-thirds of their lifetime)   is  associated with bipole convergence leading to magnetic flux cancellation towards the end of their lifetimes. The latter usually results in the full cancellation of one of the CBP main polarities \citep[e.g.,][]{1993SoPh..144...15W, 2003A&A...398..775M}. The 11 studied CBPs are no exception and all showed convergence followed by cancellation between the main polarities. In 6 out of 42 CBPs where no eruption is observed as either flux convergence and/or cancellation did not occur. The opposite flux fragments either moved apart which causes the CBP to fade away or  during the convergence one of the bipole fragments becomes too weak and fully disappears  while the distance between the two polarities is still too large leading to the CBP disappearance before cancellation has started.

Sixteen out of 21 eruptions occurred during the magnetic convergence and cancellation phase of the CBP bipoles.  As mentioned above, in  three CBPs the eruptions occurred soon after  the CBP formation (the flux emergence).  In these cases the magnetic flux of the CBP (one or both polarities) emerged close-by a stronger pre-existing magnetic fragment (e.g., BP1) or in a region seeded with numerous small-scale (covering just a few HMI pixels) magnetic concentrations (BP7 and BP8).  In all three cases the source of the eruption is found at the PIL between a pre-existing and newly emerging flux that converges and cancels.  In 5 eruptions (second or third of the homologous cases) the cancellation is ongoing and  has actually started long before the earlier eruptions in these CBPs (BP4, BP7 and BP8). In BP6, BP9 and BP11 cancellation has started more than 2 hours before the eruptions. In 4 eruptions only very weak fields (close to the noise level of HMI of $~10~G$) are involved in the cancellation that has started as soon as the magnetic flux has emerged. In two cases the cancellation was ongoing for more than 8~hrs before the eruption was triggered (BP9 and BP6).  Before and during one eruption cancellation was not observed (BP3 ER1).

The observed evanescence of bipolar flux referred to as cancellation appears to be related to the source of the eruptions in almost all CBPs. However, it remains unclear what physical processes take place leading to this observation. Several decades ago \citet{1986NASCP2442..369H} reported that cancelling magnetic flux results in the formation of MFs. The MFs erupted between 15 and 201~min later (in average of 70~min) and had an average duration of 26~min. The study is based on 61 MFs identified in H$\alpha$ images from the Big Bear Solar Observatory. \citet{1994ApJ...427..459P} incorporated this in the third and latest phase, the so-called cancellation phase, of their converging flux model of CBPs. Recently, \citet{2014ApJ...796...73H} reported that indeed, some jets in coronal hole CBPs are related to  MF eruptions that occurred during the convergence and cancellation of the CBPs bipoles. Several studies have identified magnetic cancellation as essentially responsible for the filament formation and eruption that triggers EUV jets both in the QS and coronal holes \citep{2000ApJ...530.1071W, 2003JKAS...36S..21L, 2015Natur.523..437S, 2016ApJ...821..100S, 2017ApJ...844..131P, 2018ApJ...853..189P}. Our impression from the provided example images in these articles is that all jets appear to occur in CBPs although this is not considered in any of these studies. The reported magnetic flux reduction during the flux convergence and cancellation  associated with the eruptions are estimated  at 20\% -- 60\%\ in \citet{2016ApJ...821..100S}, 10 -- 40\%\ in \citet{ 2017ApJ...844..131P}, and 20 -- 75\%\ in \citet{2018ApJ...853..189P}. With regard to our analysis, we believe that the flux decrease obtained in these studies is  related to the ``natural'' evolution of the CBPs associated with the eruptions, rather than with the eruptions themselves (see the magnetic flux lightcurves in the present study). The magnetic field reduction is possibly associated with the heating of the CBP plasma to coronal temperatures. How the coronal magnetic field evolves during the convergence and  cancellation phases that leads to the formation of MFs and their eruptions will be investigated in the follow-up study.

\subsection{Eruption phenomenology}

We identified a few phenomena related to CBP eruptions. They include a micro-flare seen as  a sudden micro-brightening starting in just a few AIA pixels, eruptive hot loop(s), a mini-filament (MF) or a cool plasma cloud eruption (CPC), a dimming (DIM), a coronal wave (CW) and a draining of the MF plasma. Among the 21 eruptions, MF or CPC ejections  are observed in 18 eruptions in 10 out of 11 CBPs.  Hot loop eruptions are detected in 19 cases and in 2 cases only a MF/CPC eruption is seen. Coronal dimming is identified in 11 cases. Coronal waves are clearly detected in 3 cases, the remaining would require dedicated study, i.e. image difference analysis, to investigate  their possible presence which could be a subject to a future dedicated study similar to the works by \citet{2015ApJ...812..173V,
2018arXiv180206152V}.

\subsection{Micro-flares}

Since the very first observations of CBPs, it has been  known that they exhibit brightness increases by several orders of magnitude on a time scale of minutes \citep{1974ApJ...189L..93G} which was named as flaring activity.   \citet{1974ApJ...189L..93G} concluded that 5 -- 10\%\ of all CBPs show this activity, and it is associated with CBPs at any latitudes. \cite{1979SoPh...63..113N} also noted that a significant number of CBPs brighten just before they disappear in X-rays. Here, micro-flaring is observed in 20 eruptions while in the eruption in BP6 the whole CBP is observed to brighten in relation to the eruption, but no micro-flaring can be detected. In 17 eruptions the micro-flares are registered in the AIA~94~\AA\ channel.  Magnetic flux cancellation is not observed in this CBP. In the case of BP6 and BP7 ER1 loop oscillations are followed by an eruption and brightening of the whole CBP. In all 19 cases the micro-flaring occurs at the polarity inversion line (PIL) where the cancelling process of opposite polarity fluxes takes place.
In 17 cases the brightening occurred beneath the erupting MF  or CPC. The flaring region, however, is often fully or partially ``obscured'' for a while  by the rising  cool material, and only after the MF/CPC moves further away from the CBP, the micro-flaring becomes visible. This is the reason why the micro-flarings are often observed a few minutes after an MF has started. In the case of BP2, the eruptions did not obscure the micro-flare region and it is registered   $\sim$5 and 10~min prior the MF eruption during ER1 and ER2, respectively.  Our analysis cannot, therefore, confirm with full confidence whether the filament destabilisation occurred before or was caused by the micro-flaring.   \citet{2015Natur.523..437S} reports  that first MFs lifts-off and then a jet bright point (JBP, i.e. micro-flaring) becomes visible at the PIL in the EUV and X-ray images, while \citet{2017ApJ...844..131P} report that JBPs usually appear together with  the lift-off of the MFs or shortly after. The micro-flaring in the wake of the MFs is also reported by \citet{2011ApJ...738L..20H} and \citet{2014ApJ...783...11A}. In  \citet{2018ApJ...853..189P} the timing of the JBPs is given but not the start time of the eruption/jet,  as it is also assumed  as the start of the eruption. \cite{2018ApJ...859....3M} report that the JBPs start  from 32~s to a few minutes after the MF is seen rising (see their Table~1). The possible obscuring effect of the rising MF is not considered in any of these studies.

\subsection{Mini-filaments and cool plasma cloud eruptions}

The eruptions from CBPs involve in most cases the expulsion of chromospheric material either as clearly seen filamentary structures or as a volume of cool material here referred  to as MF or as CPC. In two cases the eruptions involve only  MFs. These are the last eruptions of the BP4 and BP7. In the two cases they occurred at the end of the CBP lifetimes.  Mini-filament eruptions have been linked and considered as the main trigger of coronal jets in several studies  by \citet{2011ApJ...738L..20H}, \citet{2015Natur.523..437S}, \citet{2016ApJ...821..100S}, \citet{2017ApJ...844..131P}, \citet{2017ApJ...844...28S} and \cite{2018ApJ...853..189P}. None of these studies, however, investigated the eruptions in the context of CBPs and their evolution. As already mentioned, \citet{2014ApJ...796...73H} first reported that one-quarter to one-third of their CH CBPs (30 events)  were associated with mini-filament eruptions and suggested that ``CBPs are more complex in dynamical evolution and magnetic structure than previously thought''.

To determine the lifetimes of the filaments without using H$\alpha$ data appeared to be  an impossible task as our analysis concluded that  in several cases the filament became visible only during the eruption. In a few cases where MFs appear to have formed above the CBP loops (BP2 ER1, BP4 ER1, BP7 ER1 and ER4, and BP11 ER1), their formation times could eventually be determined but given the uncertainties with no available H$\alpha$ data, we cannot provide such information.  A forthcoming case study that is based on  both H$\alpha$ and AIA data will shade more light on the timing of the MF formation process and its follow-up eruption.

How do (mini)filaments form in first place? This has been debated for a long time. Observations of large CMEs with filament eruptions indicate that the convergence of opposite polarity fluxes towards the PIL is needed \citep{1997ApJ...486..534M}. This picture has been elaborated on in a cartoon like investigation by \citet{1990GMS....58..337M} that shows how the basic idea may work for a region with sheared magnetic field. This has also been investigated in a number of numerical experiments where different magnetic structures that have been either highly sheared \citep{1998ApJ...502L.181A} or directly converged to produce prominence like structures \citep{1999ApJ...510..444G}. 

\subsection{Mini-CMEs, waves  and coronal dimmings}

In total 11 eruptions qualify as mini-CMEs where at least two components of a classic CME are present: bright expanding loops ahead of a dark region, and eruptive MF or CPC. Coronal waves are distinguishable only in three cases. In only one CBP eruption the dimming is only associated with a coronal wave but no CME formation is detected. In 6 eruptions the dimming duration in the AIA~171~\AA\ channel exceeds the duration in the AIA~195~\AA\ channel. We find that the dimming region is clearly composed of two phenomena. One is the erupting filament material that  usually forms a darker core in the dimming region, \citet{2009A&A...495..319I} refer to it as darkening. When CPC erupts, it is usually more diffused and often hard to impossible to identify which part of the dimming region is filled with a cool plasma. The second dimming phenomenon is the region formed by the removal of the coronal plasma by the expending loops. \citet{2010ApJ...709..369P} who used an image difference techniques identified deep core dimming and dimming regions which are also typically observed in classic  CMEs \citep[][and the references therein]{2018arXiv180206152V}. The analysis of our data in all three channels showed the presence of one or both phenomena in the dimming regions. We find that the deep core dimming region with all its characteristics described in \citet{2010ApJ...709..369P} represents the eruptive MF, while the secondary dimming is the plasma-evacuated region. \citet{2010ApJ...709..369P} have also established that in most cases the micro-flaring occurred after the appearance of the micro-CME deep core dimming which conforms with our finding that the erupting MFs obscure  for some time the micro-flaring as mentioned above. In the cases when no mini-CME is observed, but only a filament eruption is detected, only a ``deep core dimming'' is identified. Studies of coronal dimming associated with classic CMEs have found that mass evacuation from the low corona is responsible for the reduction in intensity rather than temperature variations \citep[e.g.][]{2000A&A...358.1097H, 2003A&A...400.1071H}. Most recently a study by \cite{2018arXiv180206152V} on the plasma characteristics of six coronal dimming events using Differential Emission Measure (DEM) methods found that dimmings (secondary, for details see \citet{2018arXiv180206152V}) are related to a significant density decrease caused by mass evacuation (10--45\%) together with some temperature decrease. In all cases of dimmings (except the deep core dimming caused by the filament) we observed differences in the appearance of the dimming in the two AIA coronal channels, 171~\AA\ and 193~\AA. One possible explanation is that the observed eruptions are very localised phenomena in the low solar corona. While the 171~\AA\ channel samples best this temperature region, the 193~\AA\ channel records in addition a large amount of overlying (the localised CBP eruption region) background coronal emission that obscures the localised event (making it to appear brighter) and less distinctive. In some cases this can also be caused by a temperature effect as most of the plasma is heated to high temperatures only detectible in the AIA~193~\AA\ channel, while the AIA~171~\AA\ image appears dark simply because of the lack of emission (the ionization of Fe~{\sc ix} and {\sc x} to higher ionisation states) in this channel. Jets from CBPs have been detected in coronagraph and solar eclipse data  \citep[e.g.][]{2010SoPh..264..365P, 2018ApJ...860..142H} and are believed to contribute to the solar wind. An ejection from a QS CBP is reported by \citet{2005A&A...434..725M} who observed a sigmoid formation and ejection from a CBP located at the solar disk  centre. The eruption was indirectly linked to an interplanetary magnetic cloud. A statistical study of  white-light jets observed  in  SECCHI/COR1 by \citet{2010SoPh..264..365P} shows the relation of these jets to coronal hole and quiet Sun CBPs as well as EUV jets. \citet{2014ApJ...784..166Y} found a positive correlation between polar jets associated with X-ray BPs and high-speed responses in the interplanetary medium using LASCO C2 and STEREO SECCHI/COR2, as well as  Solar Mass Ejection Imager data. These studies indicate that CBPs can produce ejections escaping into the interplanetary space.

\citet{2009A&A...495..319I} estimated that 1400 mini-CMEs (or generally small-scale eruptions) occur per day over the whole Sun. A follow-up study by \citet{2012ApJ...746...12A} that uses automatic identification based on Zernike moments space-time slices  and a   support vector machine (SVM) classifier,  found 1217 events in  EUVI images and 2064 in AIA data.  A study by \citet{2005SoPh..228..285M} using EIT data finds $\sim$250~CBPs at any given time on the visible solar disk in the EIT 195~\AA\ channel (see their Figure~2), while a recent study by \cite{2015ApJ...807..175A} identifies 572 or 1144 for the whole Sun.  If we consider that 76\% of these CBPs produce at least one eruption during their lifetime of $\sim$24~hrs, the minimum number of CBP eruptions per day would be ~870.  Considering that  CBPs undergo a minimum of one eruption, this number can easily reach the number of eruptions detected by   \citet{2012ApJ...746...12A}. A dedicated large statistical survey of eruptions in CBPs is, however, required to provide more certainty on this matter.

\section{Conclusions}
\label{sect:concl}

 The observational evidence presented here indicates that the
majority of coronal bright points (CBPs) in the quiet Sun
produce mass ejections during their lifetime. The eruptions
usually occur late in the evolution of the CBPs during the
convergence and cancellation phase of the CBP bipole evolution
during which the CBPs become smaller until they
fully disappear. Eruptions can also occur earlier when CBP								
bipoles emerge in regions with a strong pre-existing flux. For
the majority of the eruptions magnetic convergence and cancellation
involve the CBP main bipoles, while in just three
eruptions one of the CBP magnetic fragments and a preexisting
fragment of opposite polarity converge and cancel.
The CBP eruptions appear as expulsions of chromospheric
material either as elongated filamentary structure (mini-filament,					
MF) or a volume of cool material (cool plasma
cloud, CPC), together with the CBP or higher overlying hot
loops. Coronal waves are occasionally identified during the
eruptions. A micro-flaring is observed beneath all erupting
MFs/CPCs (one case is inconclusive). It remains uncertain
whether the destabilised MF causes the micro-flaring
or the destabilisation and eruption of the MF is triggered
by reconnection beneath the filament. In most eruptions, the
cool erupting plasma obscures partially or fully the micro-flare
until the erupting material moves away from the CBP.
Around 50\% of the eruptions represent mini-CMEs. The
dimming regions associated with the CMEs are occupied by
both the `dark' cool plasma and areas of weakened coronal
emission caused by the depleted plasma density.

The present study demonstrates that the evolution of
small-scale loop structures in the quiet Sun determined by
their magnetic footpoint motions and/or ambient field topology					
evolve into an eruptive phase that triggers the ejection of					
cool and hot plasma in the corona. In our follow-up study we
model the time evolution of the magnetic field, using a potential
extrapolation of the first HMI magnetogram in the time
series. The 3D magnetic field structure is then advanced in
time assuming a relaxation approach where the photospheric
boundary conditions of the magnetic field is changed according			
to the observed HMI magnetograms. The evolution of
the magnetic fields in the CBP regions are then investigated
to learn more about the general magnetic field topologies and
the possibilities for building up structures that may  resemble				
the eruption configurations.

\begin{acknowledgements}
 We would like to thank the referee for the comments and suggestions. The authors thank very much Richard Harrison for the careful reading and comments on the manuscript.  C.M. and L.X. thank the National Natural Science Foundation of China (41474150 and 41627806). The AIA and HMI data are provided courtesy of NASA/SDO and corresponding  science teams. The AIA and HMI data have been retrieved using the Stanford University's Joint Science Operations Centre/Science Data Processing Facility.  M.M. and K.G.  thank the ISSI Bern for the support to the team``Observation-Driven Modelling of Solar Phenomena''.
\end{acknowledgements}

\bibliographystyle{aa}
%\bibliography{bibliography}

\begin{appendix}

\section{Details of rest of the  BPs}
\label{details}

Here we present the rest of the figures for the discussed eruptions (BP2 to BP10) to support our discussion and conclusions. Details on these eruptions are later related to the magnetic field modeling in the follow-up study.  All movies can be found in the following archive \url{https://doi.org/10.5281/zenodo.1300416}.

\begin{figure*}[!h]
\centering
\includegraphics[scale=0.85]{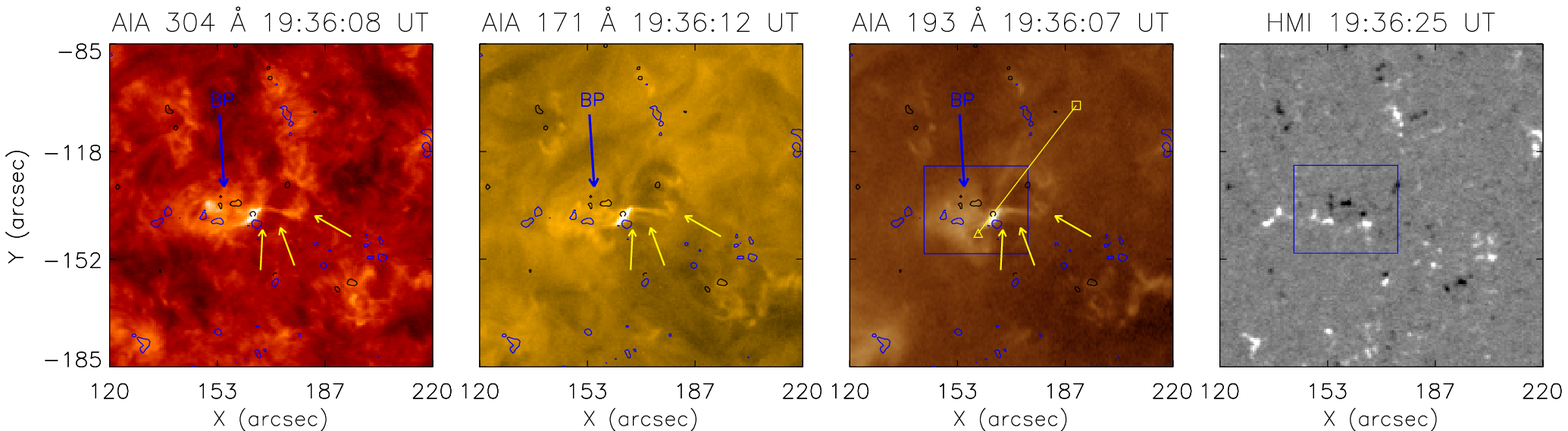}
\includegraphics[scale=0.85]{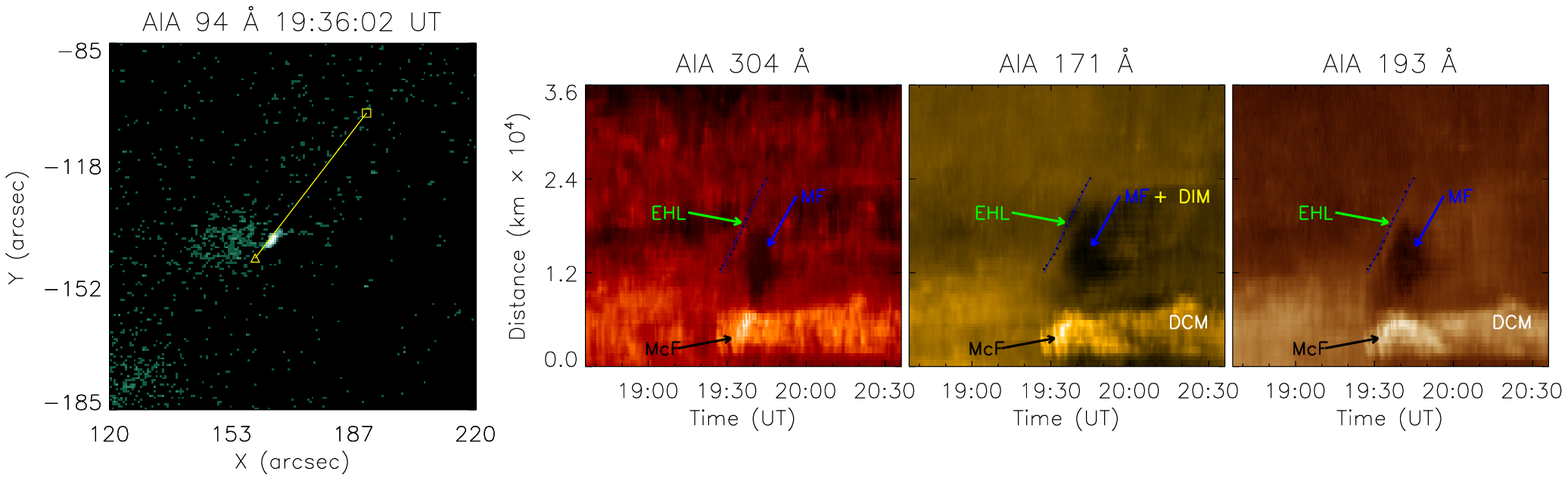}
\caption{BP2 ER1: Top row, from left to right: AIA 304~\AA, 171~\AA\ and 193~\AA\ images of first eruption BP1 and their corresponding HMI longitudinal magnetogram scaled from $-50$ to 50 ~G. The yellow arrows point at the erupting mini-filament. The blue and black contours outline the positive and negative fluxes at $\pm$50~G. Bottom row, from left to right: AIA 94~\AA\ image of BP2 during the first eruption showing the micro-flare.  The left vertical line is a linear fit of visually selected points that outline the bright and/or dark front of the eruptive features.  The following abbreviations are marked on the panels: BP -- bright point, EHL  -- eruptive hot loop(s), MF -- mini-filament, DIM -- dimming, DCM -- draining cool material, and McF -- micro-flare. The yellow solid line indicates  the pixel slice which has been extracted to produce the time-slice panels in the AIA 304~\AA, 171~\AA\ and 193~\AA\ channels. The bottom of the slice is marked with a triangle and the top with a square.}
\label{fig:bp2_1}
\end{figure*}

\begin{figure*}[!ht]
\centering
\includegraphics[scale=0.85]{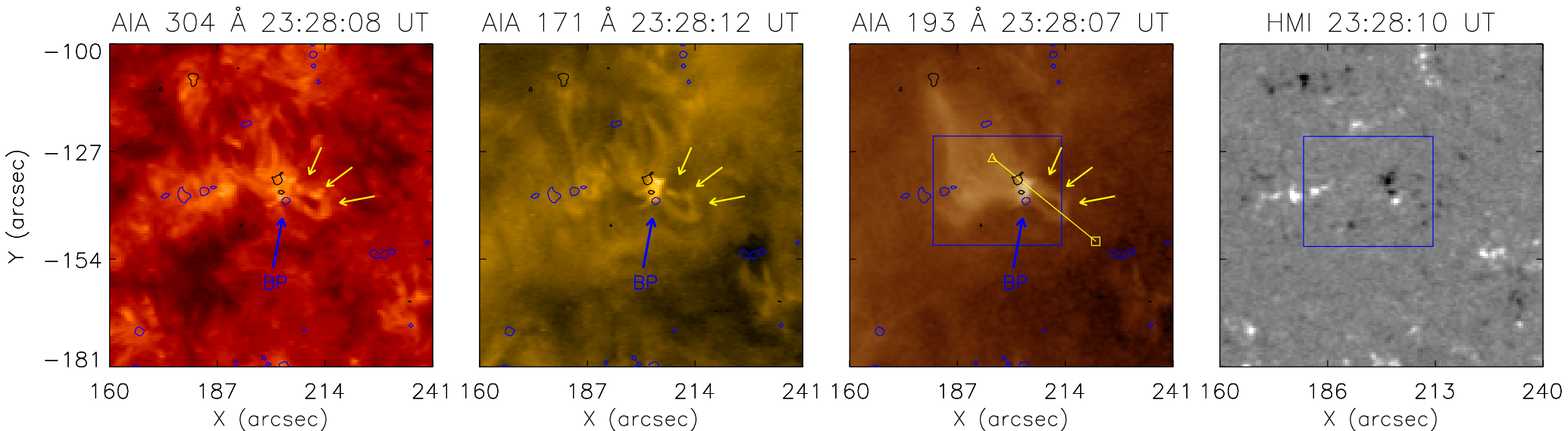}
\includegraphics[scale=0.85]{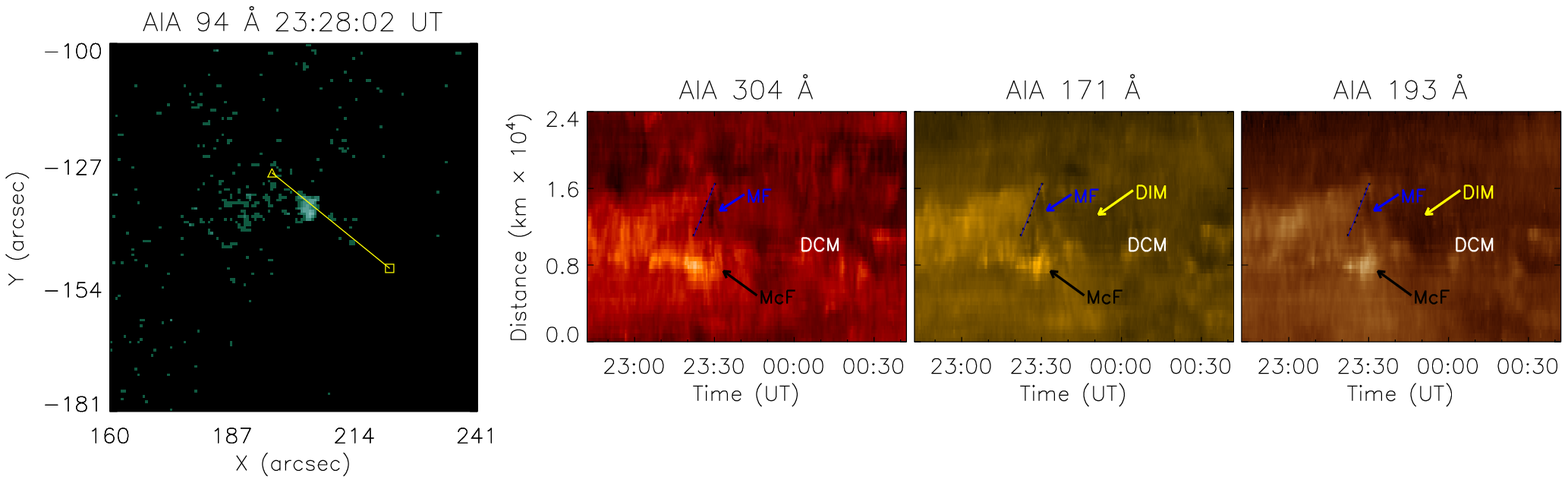}
\caption{The same as in Figure~\ref{fig:bp2_1} for BP2 ER2.  The following abbreviations are marked on the panels: BP -- bright point, HLE -- hot loop eruptions,  MF -- mini-filament,  DIM -- dimming, and McF -- micro-flare.}
\label{fig:bp2_2}
\end{figure*}

\begin{figure*}[!ht]
\centering
\includegraphics[scale=0.85]{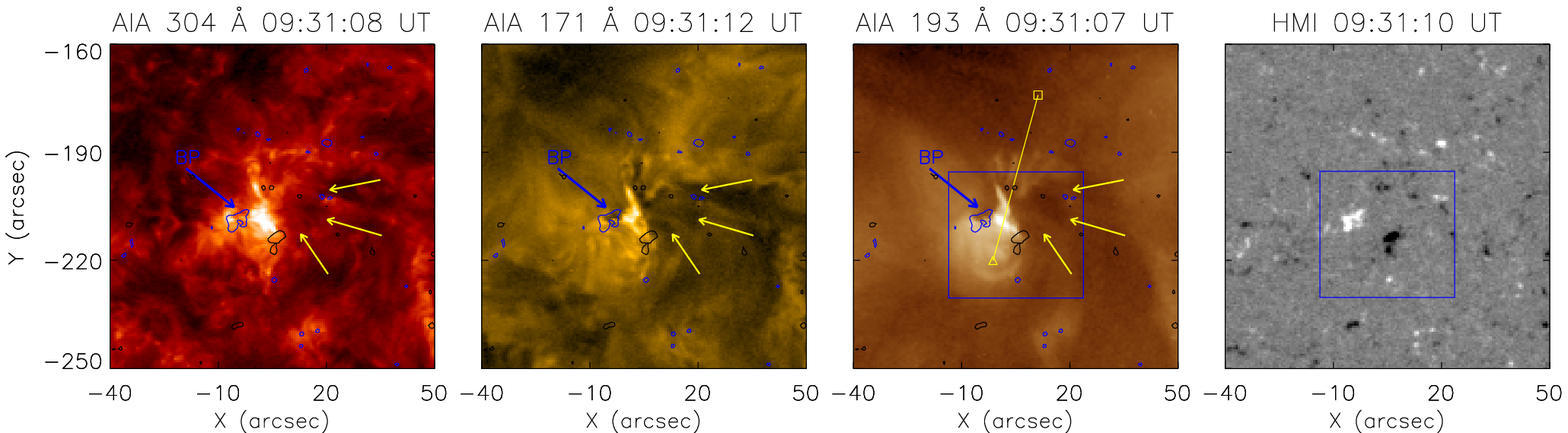}
\includegraphics[scale=0.85]{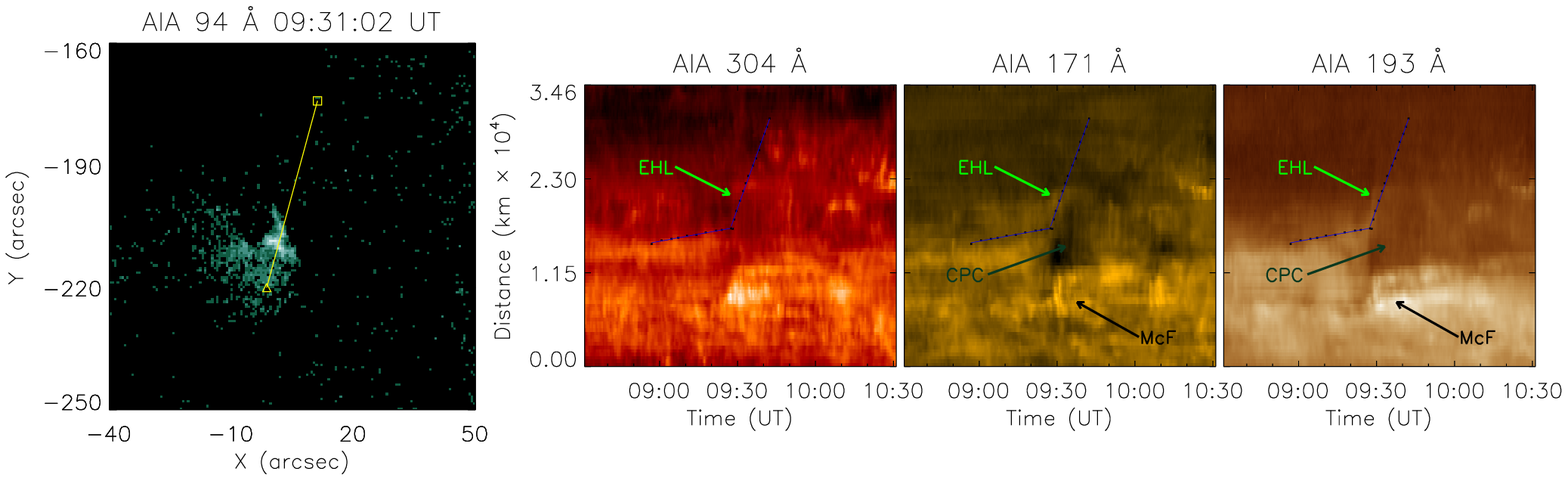}
\caption{The same as in Figure~\ref{fig:bp2_1} for BP3 ER1.  The following abbreviations are marked on the panels: BP -- bright point, HLE -- hot loop eruptions, CPC -- cool plasma cloud, DIM -- dimming, and McF -- micro-flare.}
\label{fig:bp3_1}
\end{figure*}

\begin{figure*}[!ht]
\centering
\includegraphics[scale=0.85]{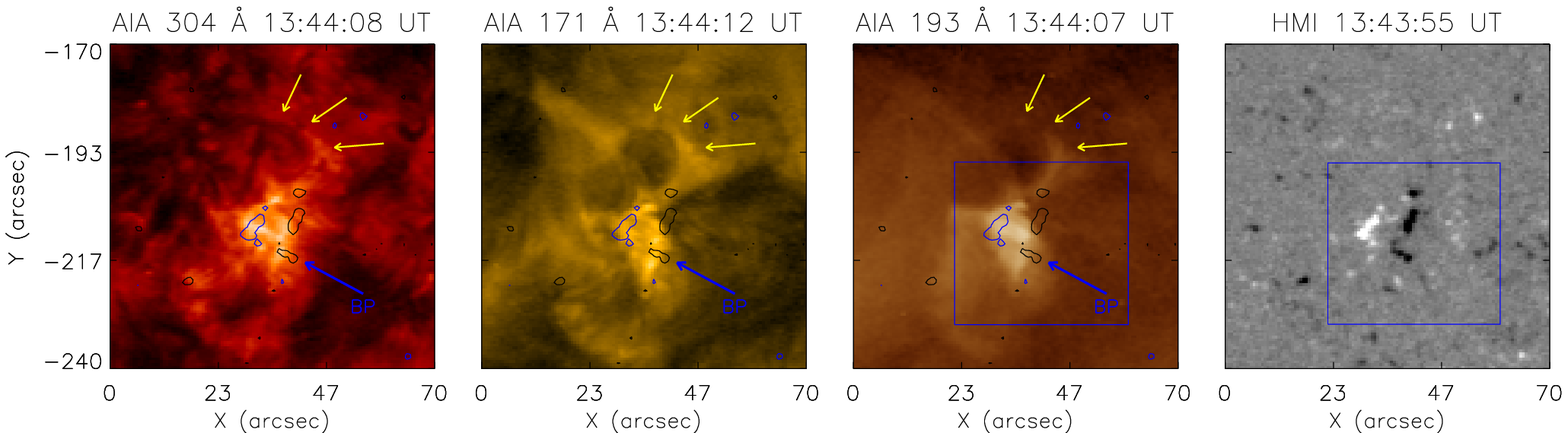}
\includegraphics[scale=0.85]{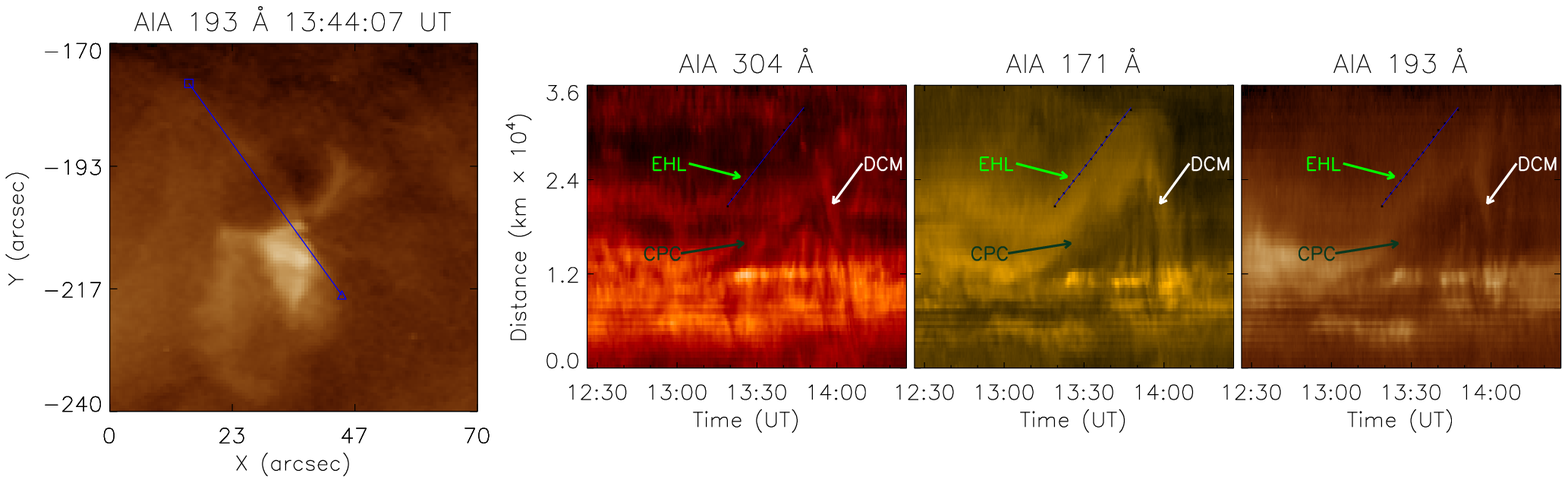}
\caption{The same as in Figure~\ref{fig:bp2_1} for the second eruption in BP3 ER2.  The following abbreviation are marked on the panels: BP -- bright point, EHL -- eruptive hot loop, CPC -- cool plasma cloud,  and DCM -- draining  cool material. }
\label{fig:bp3_2}
\end{figure*}

\begin{figure*}[!ht]
\centering
\includegraphics[scale=0.85]{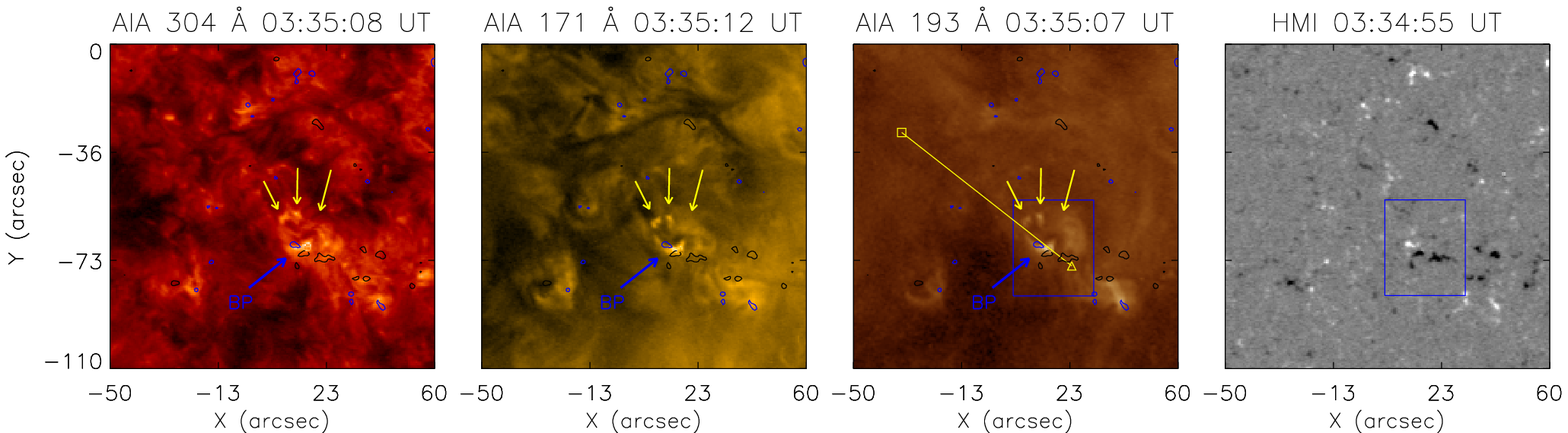}
\includegraphics[scale=0.85]{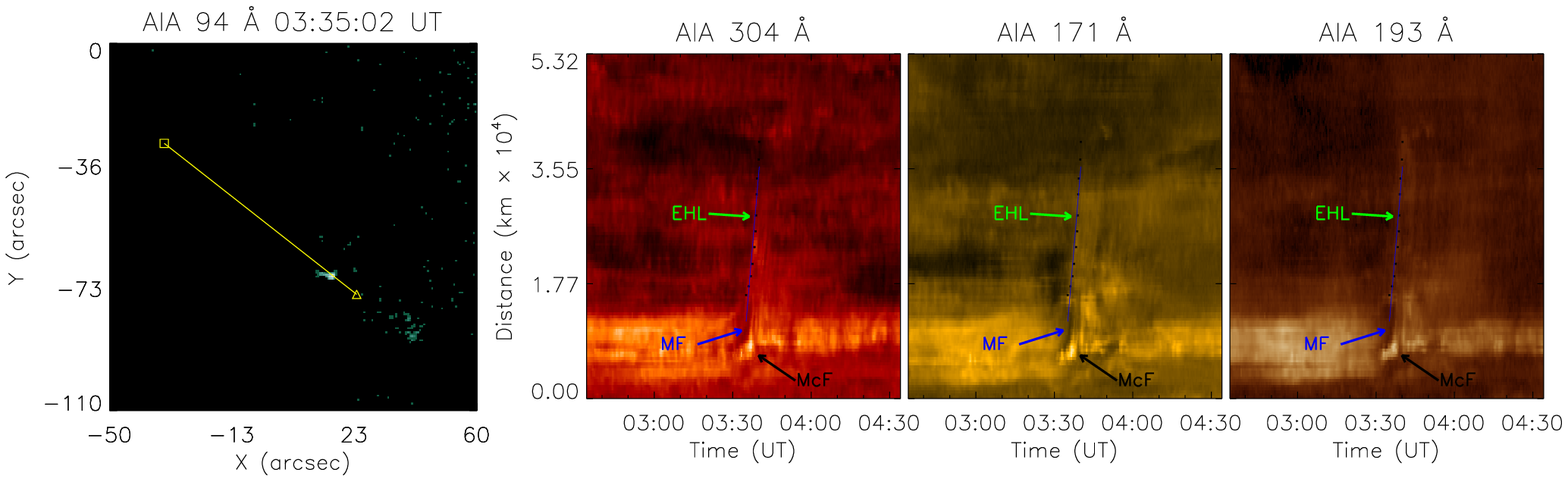}
\caption{The same as in Figure~\ref{fig:bp2_1} for BP4 ER1. The following abbreviations are marked on the panels: BP -- bright point, EHL -- eruptive hot loops, MF -- mini-filament, and McF -- micro-flare.}
\label{fig:bp4_1}
\end{figure*}

\begin{figure*}[!ht]
\centering
\includegraphics[scale=0.85]{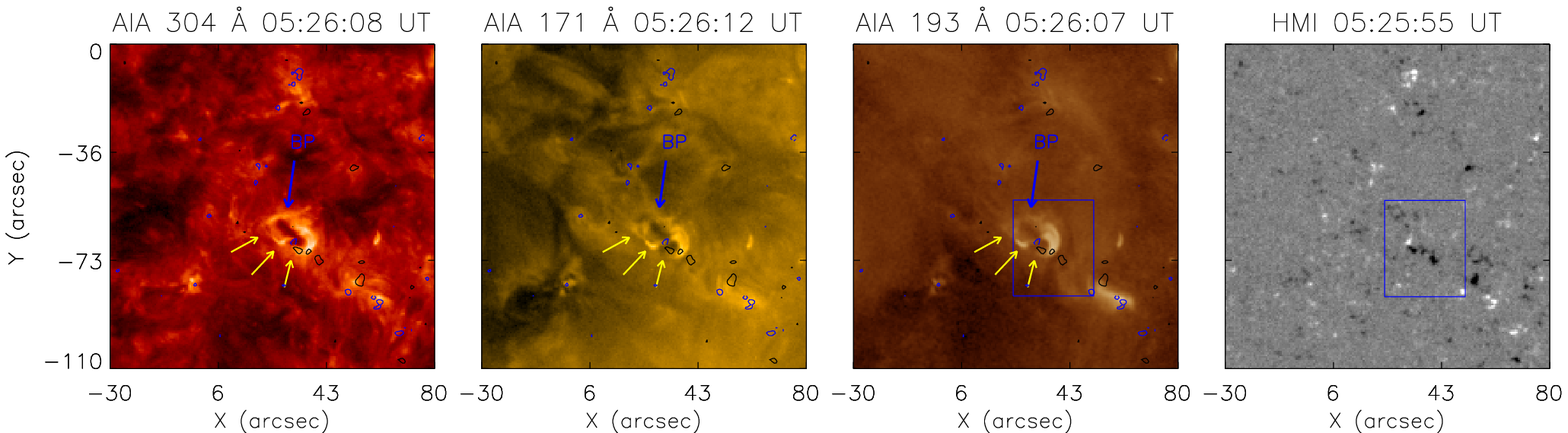}
\includegraphics[scale=0.85]{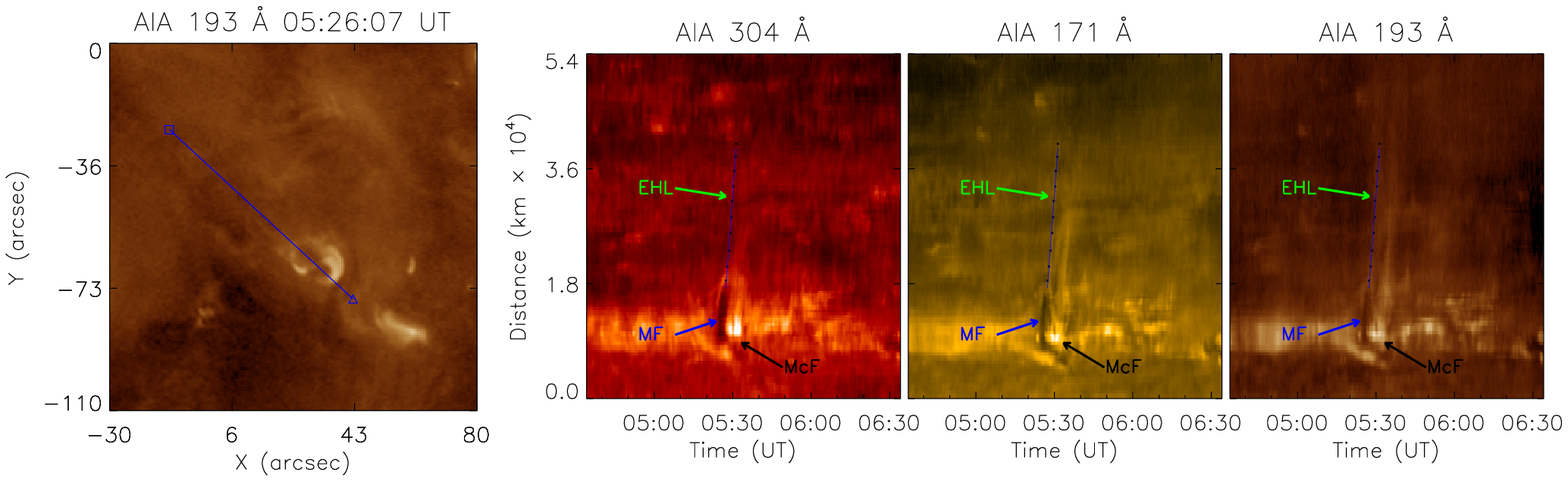}
\caption{The same as in Figure~\ref{fig:bp2_1} for BP4 ER2.    The following abbreviation are marked on the panels: BP -- bright point, EHL -- eruptive hot loop(s), MF -- mini-filament, and McF -- micro-flare.}
\label{fig:bp4_2}
\end{figure*}

\begin{figure*}[!ht]
\centering
\includegraphics[scale=0.85]{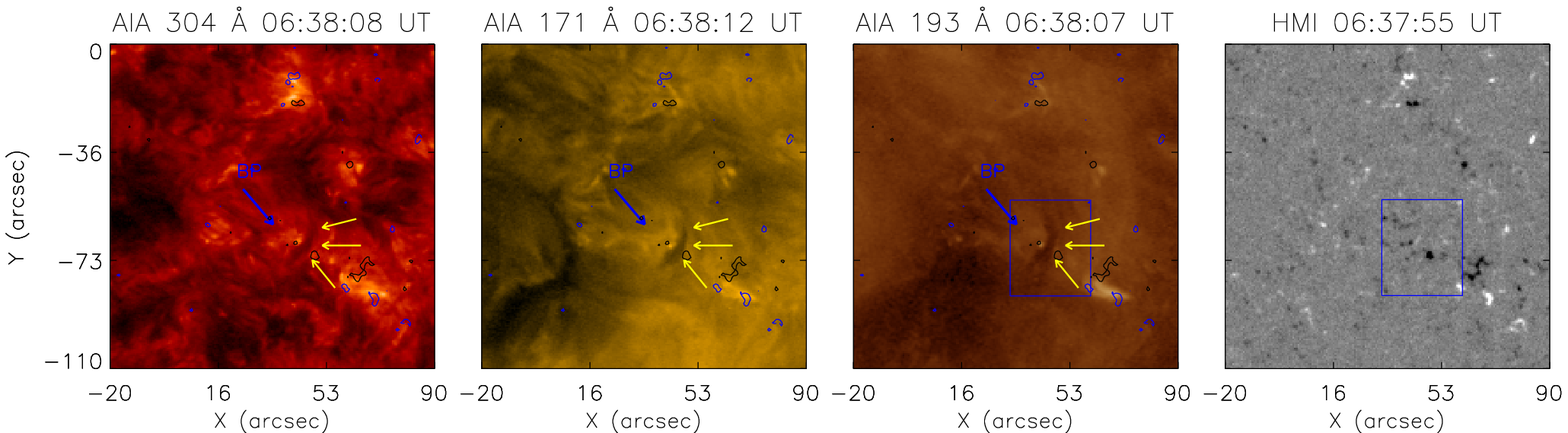}
\includegraphics[scale=0.85]{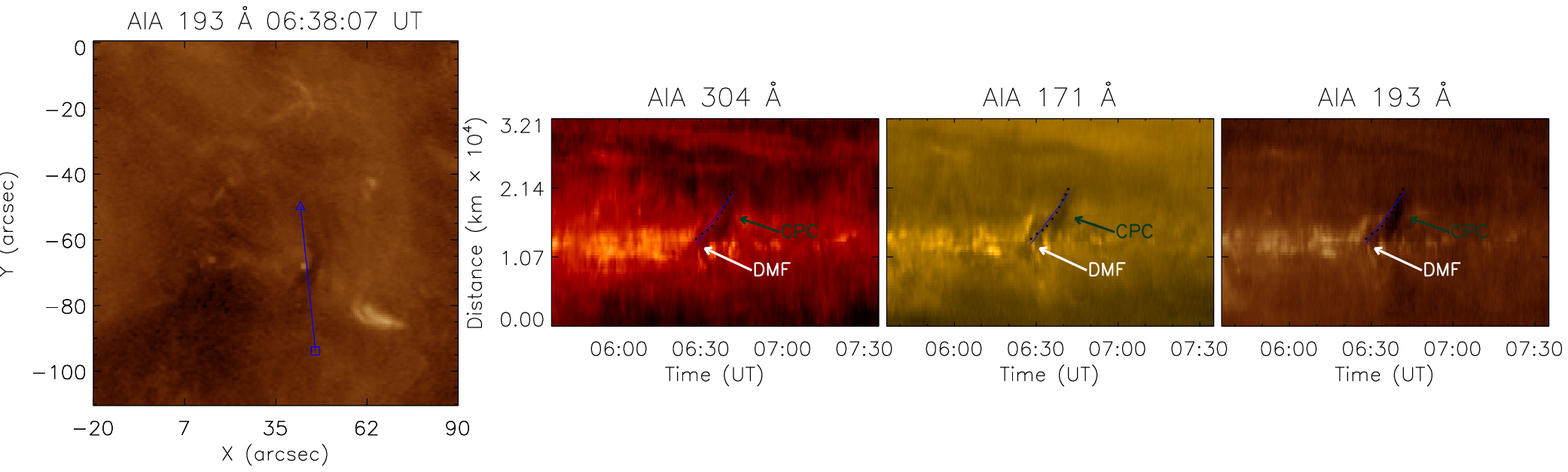}
\caption{The same as in Figure~\ref{fig:bp2_1} for BP4 ER3.   The following abbreviations are marked on the panels: BP -- bright point, EHL -- eruptive hot loops, MF -- mini-filament, and McF -- micro-flare. }
\label{fig:bp4_3}
\end{figure*}

\begin{figure*}[!ht]
\centering
\includegraphics[scale=0.85]{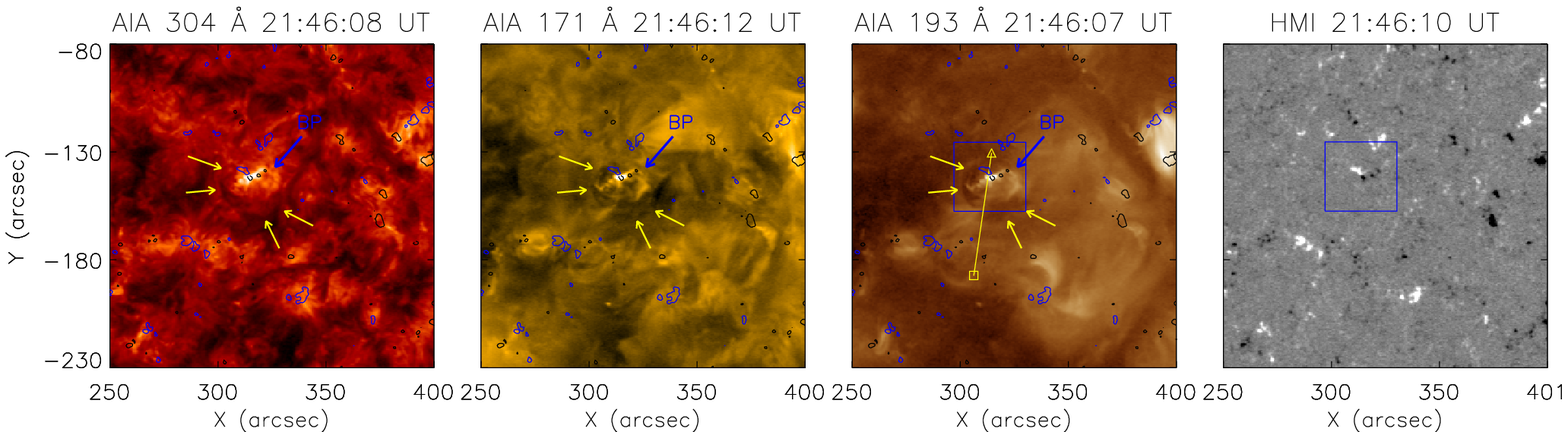}
\includegraphics[scale=0.85]{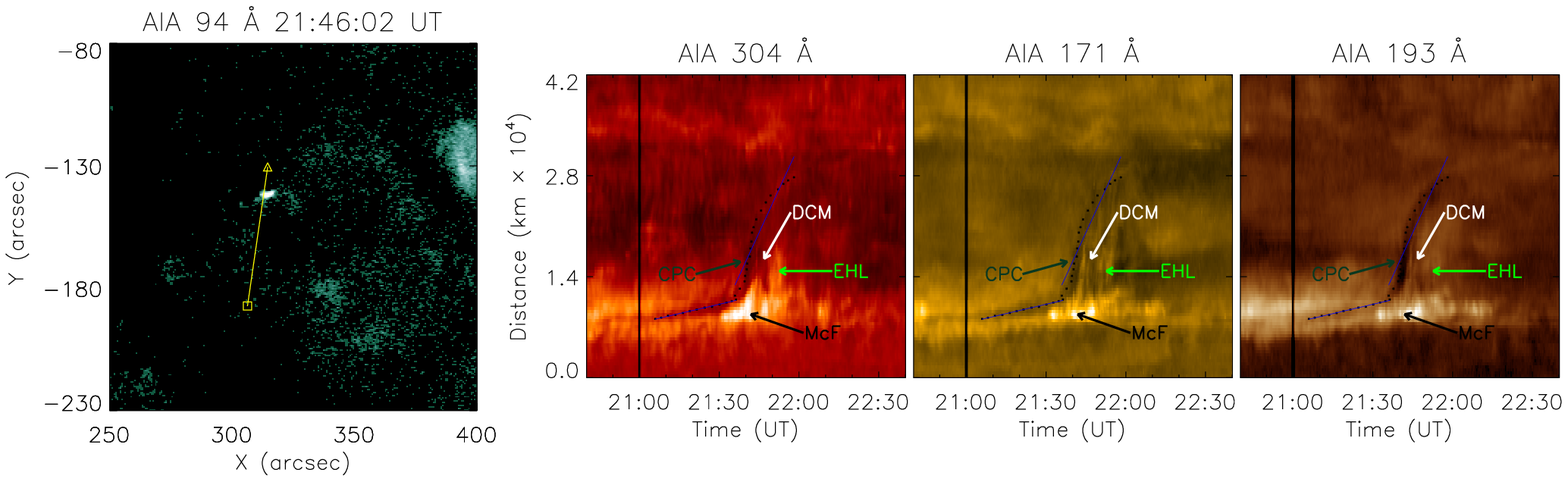}
\caption{The same as in Figure~\ref{fig:bp2_1} for BP5. The following abbreviations are marked on the panels: BP -- bright point, EHL -- eruptive hot loops, CPC -- cool plasma cloud, DCM -- draining cool material, and McF -- micro-flare}
\label{fig:bp5_1}
\end{figure*}

\begin{figure*}[!ht]
\centering
\includegraphics[scale=0.85]{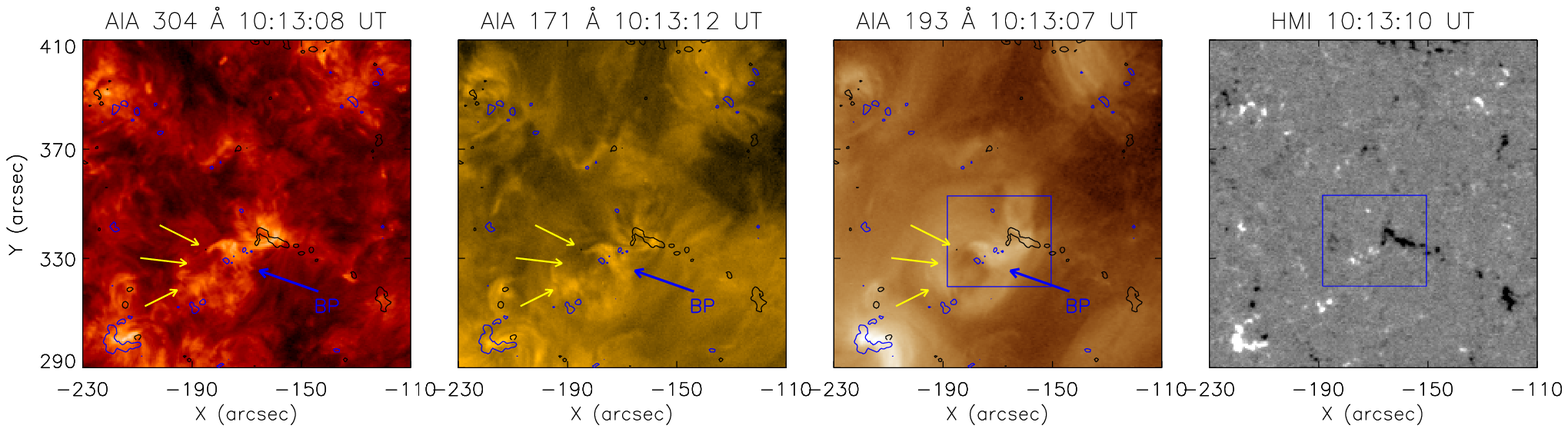}
\includegraphics[scale=0.85]{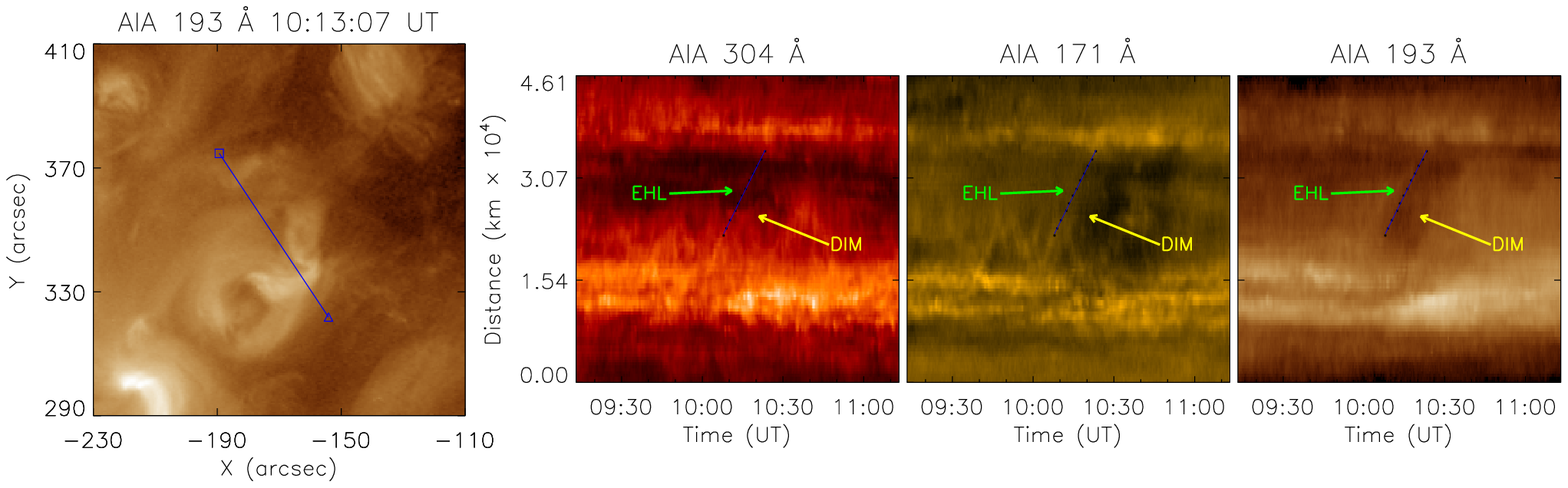}
\caption{The same as in Figure~\ref{fig:bp2_1} for BP6. The following abbreviations are marked on the panels: BP -- bright point, EHL -- eruptive hot loops, and DIM -- dimming.}
\label{fig:bp6_1}
\end{figure*}

\begin{figure*}[!ht]
\centering
\includegraphics[scale=0.85]{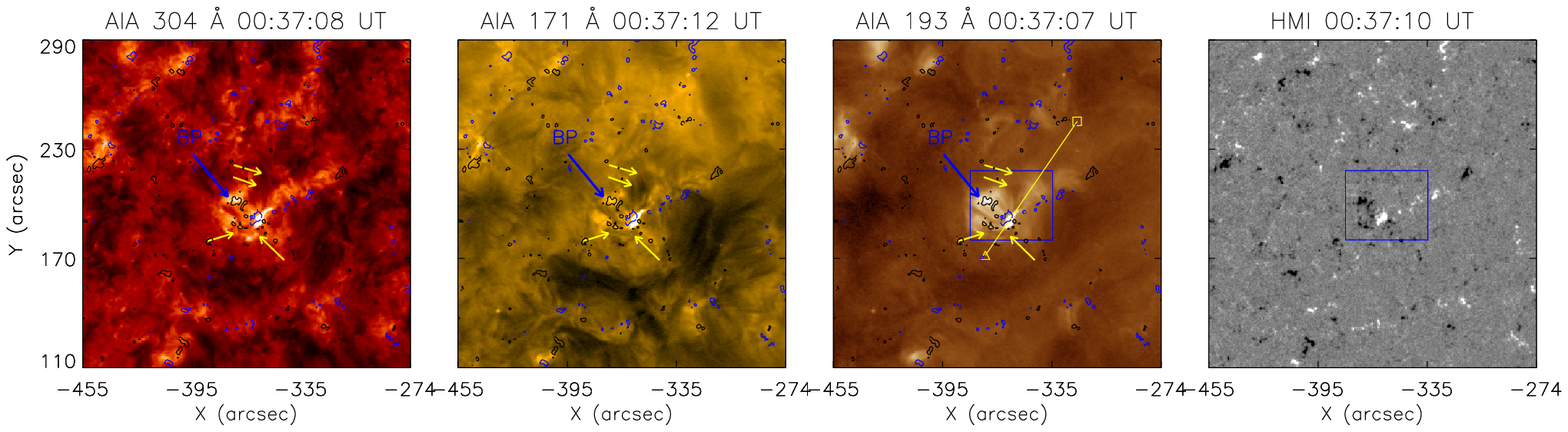}
\includegraphics[scale=0.85]{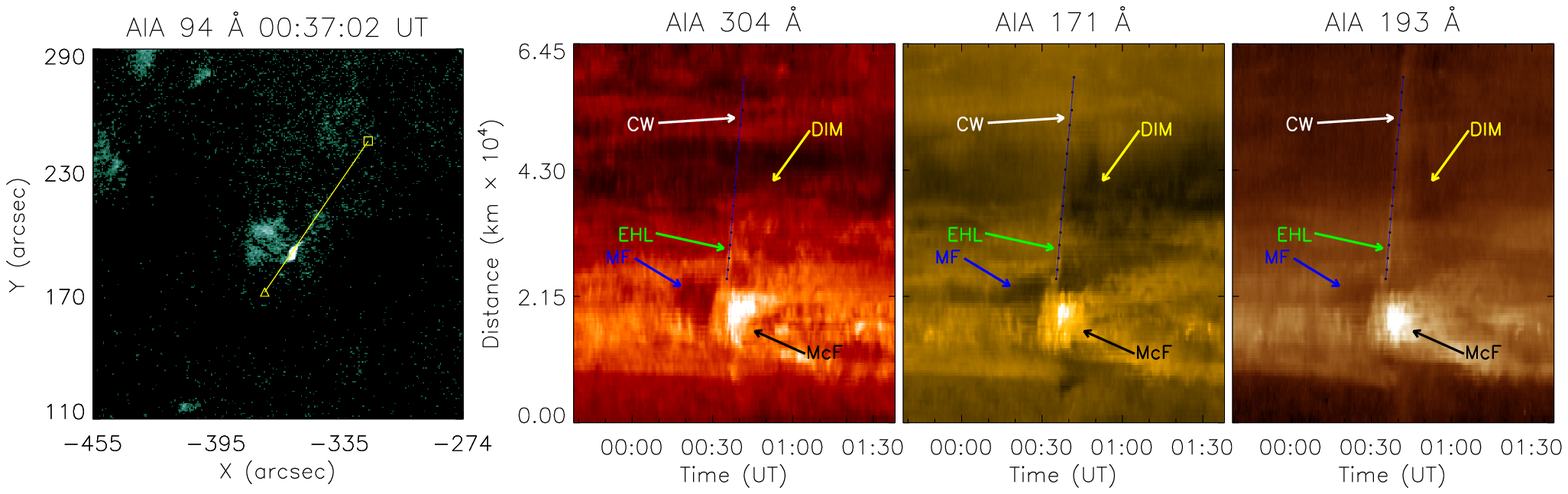}
\caption{The same as in Figure~\ref{fig:bp2_1} for BP7 ER1. The following abbreviations are marked on the panels: BP -- bright point, EHL -- eruptive hot loops, MF -- mini-filament, DIM -- dimming, and McF -- micro-flare.}
\label{fig:bp7_1}
\end{figure*}

\begin{figure*}[!ht]
\centering
\includegraphics[scale=0.85]{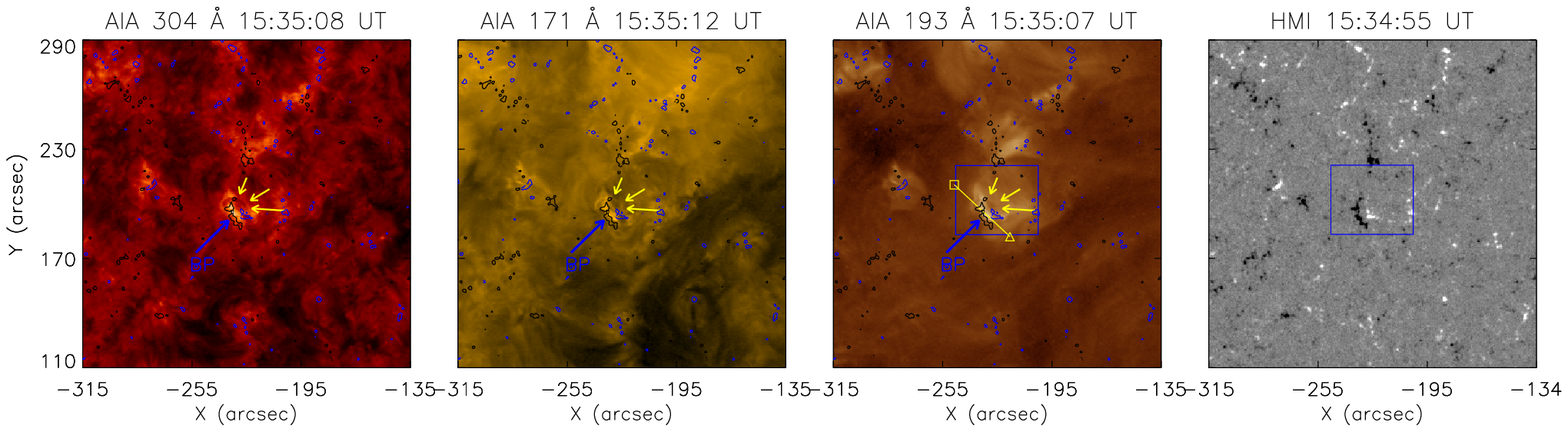}
\includegraphics[scale=0.85]{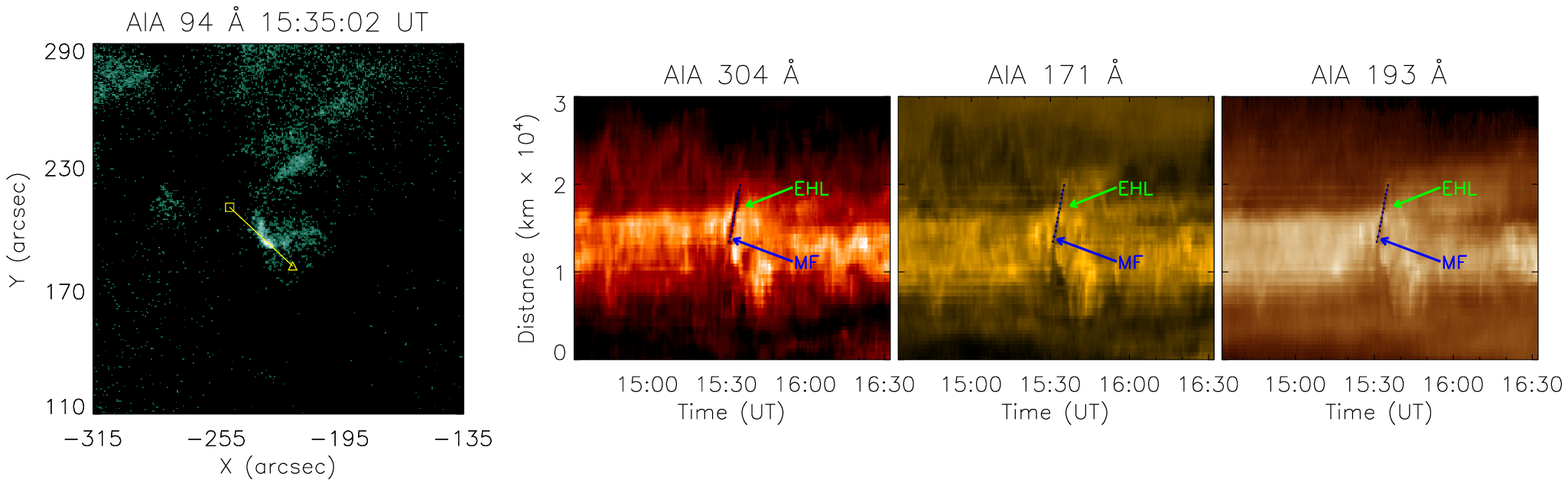}
\caption{The same as in Figure~\ref{fig:bp2_1} for for BP7 ER2. The following abbreviations are marked on the panels: BP -- bright point, EHL -- eruptive hot loops, MF -- mini-filament, and McF -- micro-flare.}
\label{fig:bp7_2}
\end{figure*}

\begin{figure*}[!ht]
\centering
\includegraphics[scale=0.85]{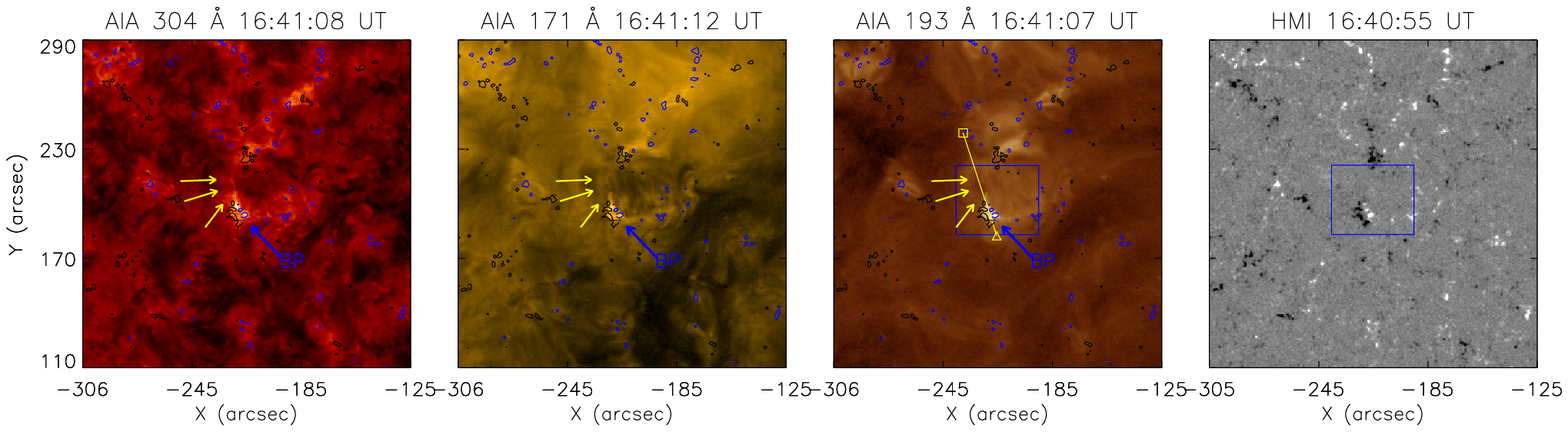}
\includegraphics[scale=0.85]{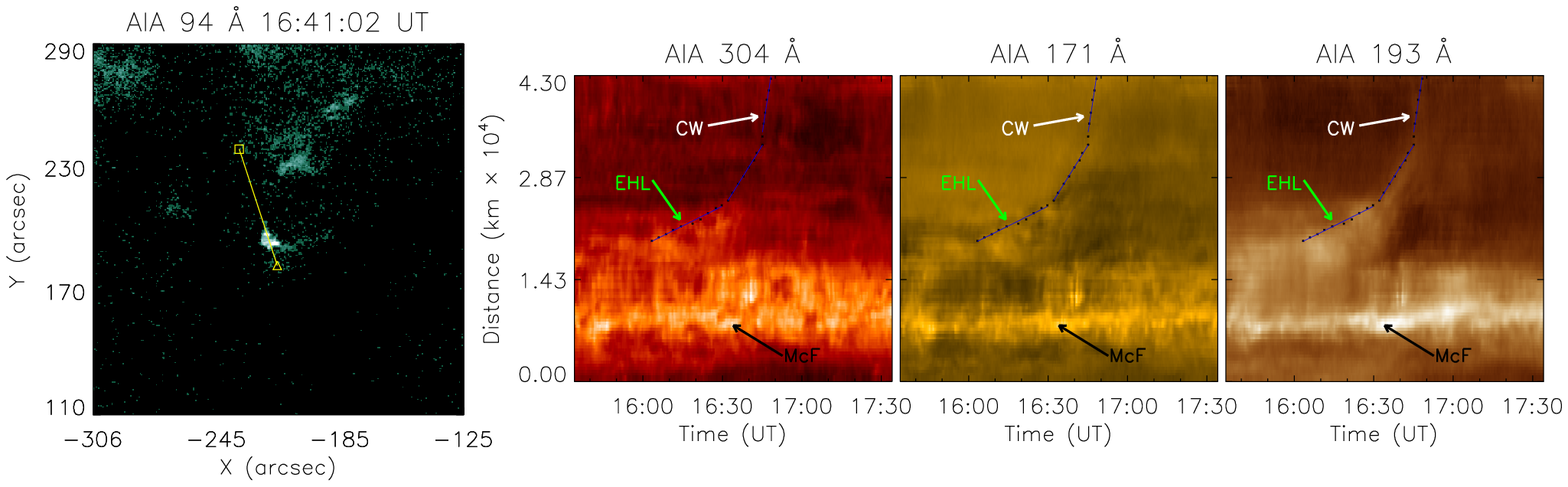}
\caption{The same as in Figure~\ref{fig:bp2_1} for for BP7 ER3. The following abbreviations are marked on the panels: BP -- bright point, EHL -- eruptive hot loops, CW - coronal wave, DIM -- dimming, and McF -- micro-flare.}
\label{fig:bp7_3}
\end{figure*}

\begin{figure*}[!ht]
\centering
\includegraphics[scale=0.85]{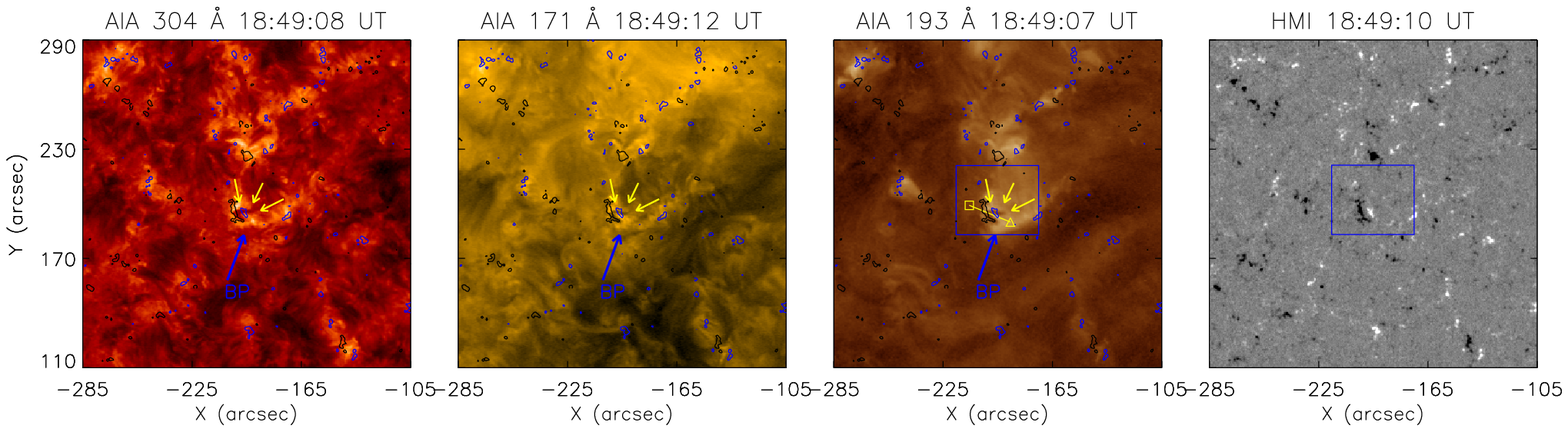}
\includegraphics[scale=0.85]{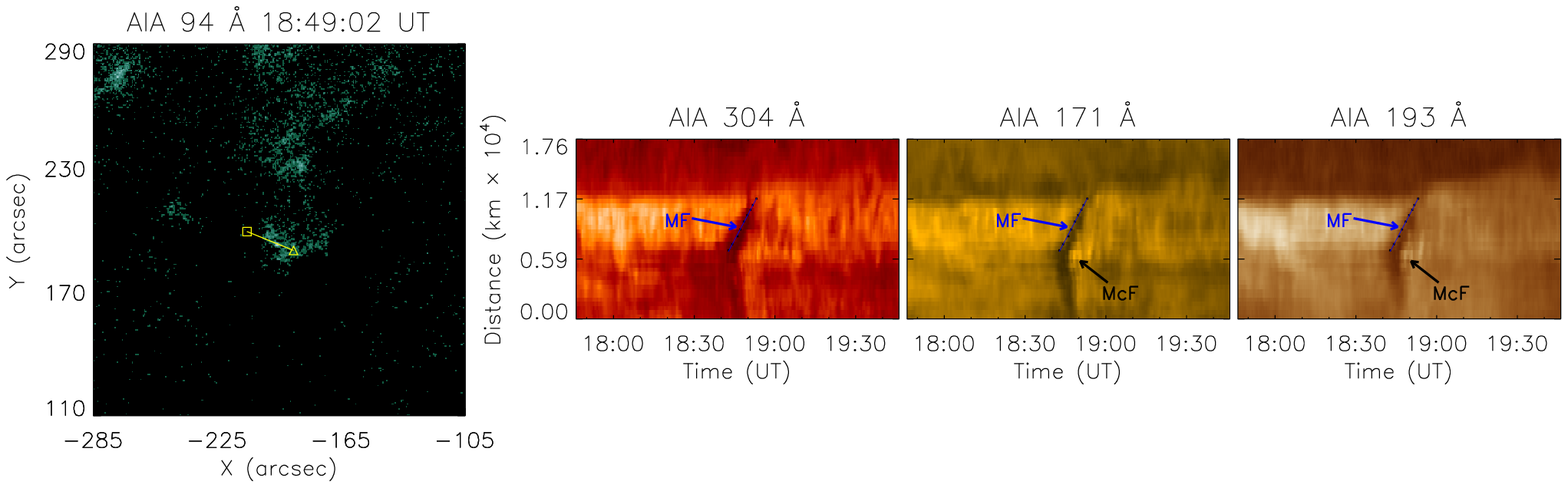}
\caption{The same as in Figure~\ref{fig:bp2_1} for for BP7 ER4. The following abbreviations are marked on the panels: BP -- bright point, MF -- mini-filament, and McF -- micro-flare.}
\label{fig:bp7_4}
\end{figure*}

\begin{figure*}[!ht]
\centering
\includegraphics[scale=0.85]{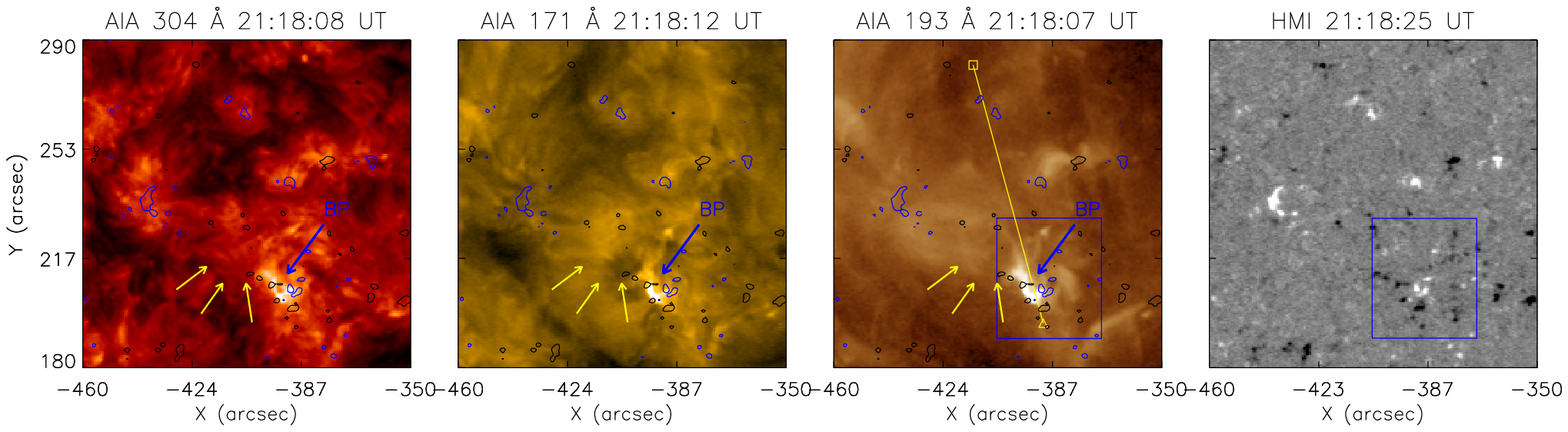}
\includegraphics[scale=0.85]{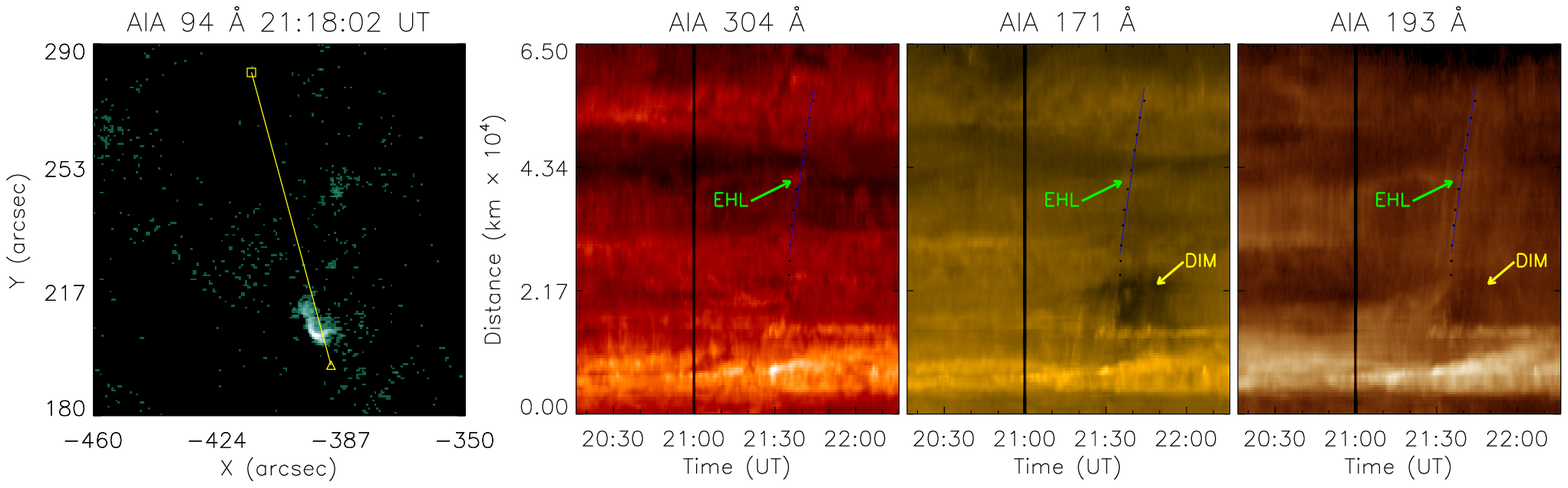}
\caption{The same as in Figure~\ref{fig:bp2_1} for for BP8 ER1. The following abbreviations are marked on the panels: BP -- bright point, EHL -- eruptive hot loops, DIM -- dimming, and  McF -- micro-flare.}
\label{fig:bp8_1}
\end{figure*}

\begin{figure*}[!ht]
\centering
\includegraphics[scale=0.85]{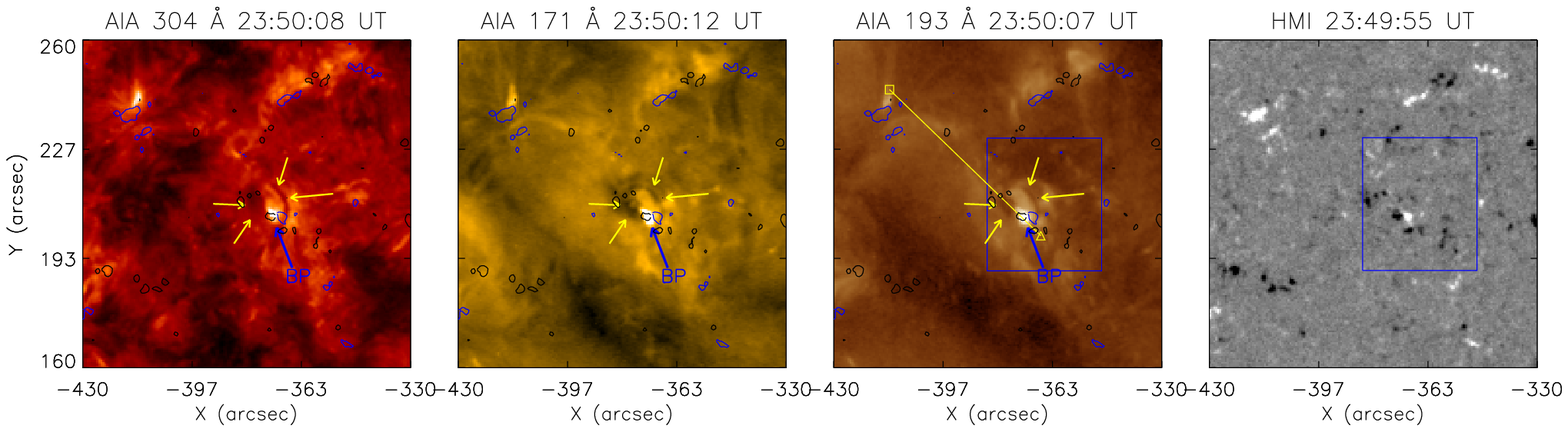}
\includegraphics[scale=0.85]{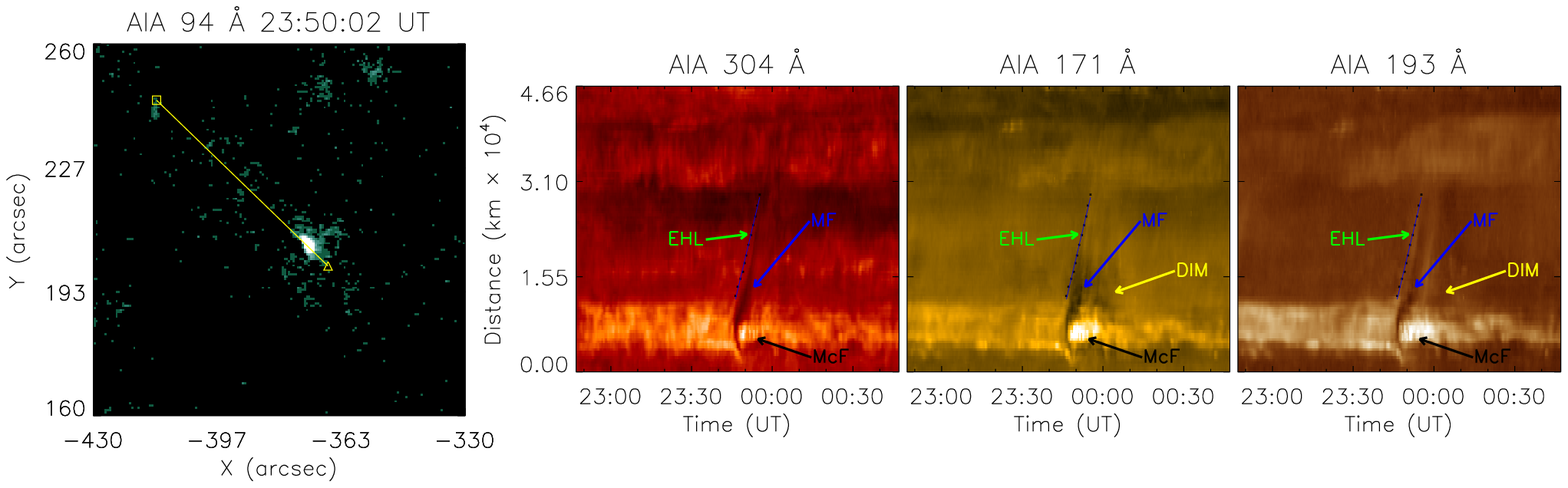}
\caption{The same as in Figure~\ref{fig:bp2_1} for BP8 ER2. The following abbreviations are marked on the panels: BP -- bright point, EHL -- eruptive hot loops, MF -- mini-filament, DIM -- dimming, and McF -- micro-flare.}
\label{fig:bp8_2}
\end{figure*}

\begin{figure*}[!ht]
\centering
\includegraphics[scale=0.85]{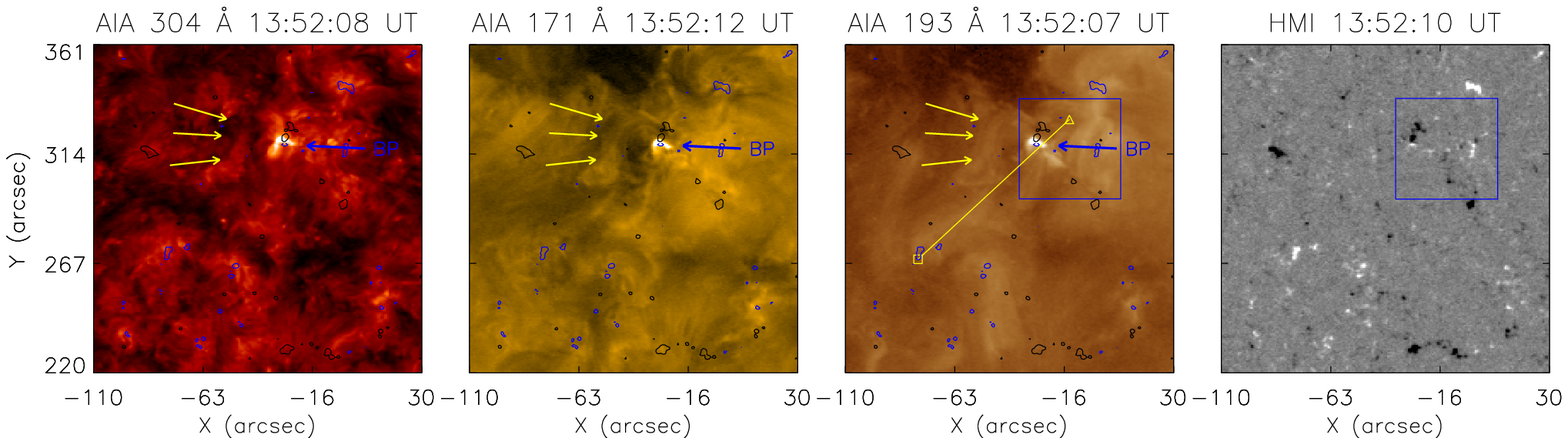}
\includegraphics[scale=0.85]{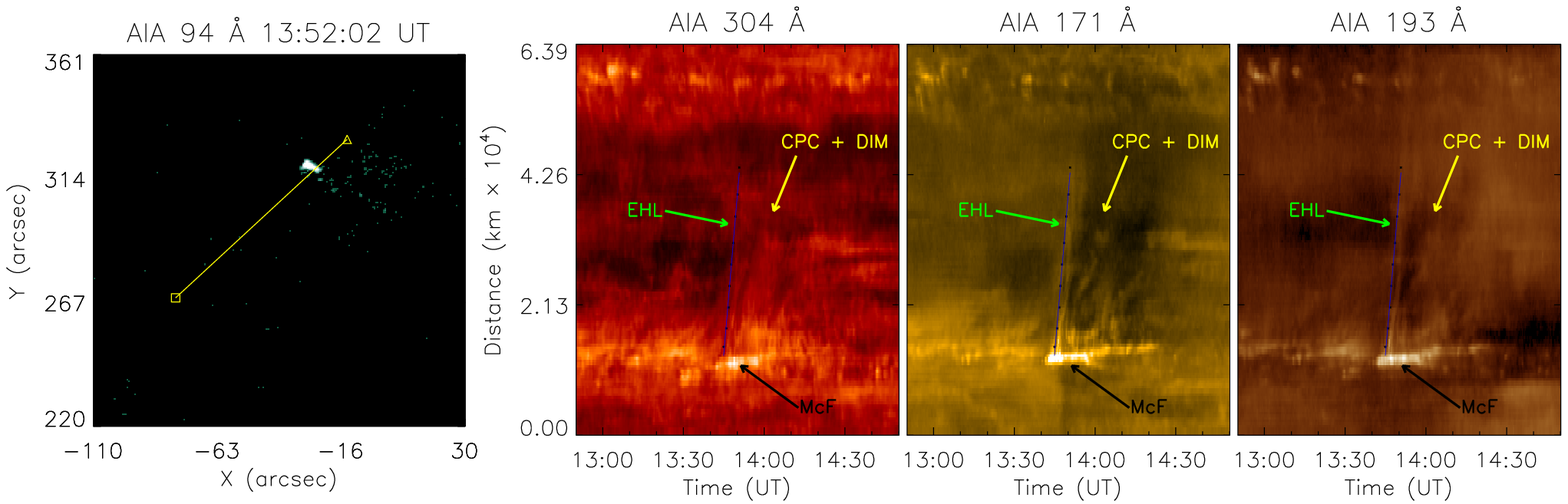}
\caption{The same as in Figure~\ref{fig:bp2_1} for BP9. The following abbreviations are marked on the panels: BP -- bright point, EHL -- eruptive hot loops, CPC -- cool plasma cloud, , DIM -- dimming, and McF -- micro-flare.}
\label{fig:bp9_1}
\end{figure*}

\begin{figure*}[!ht]
\centering
\includegraphics[scale=0.85]{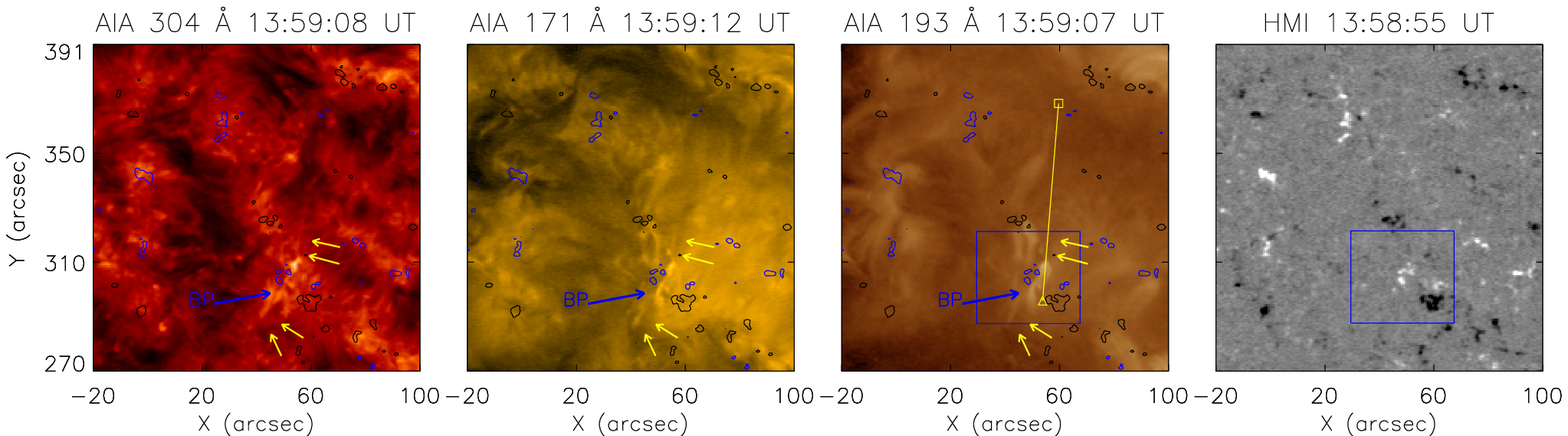}
\includegraphics[scale=0.85]{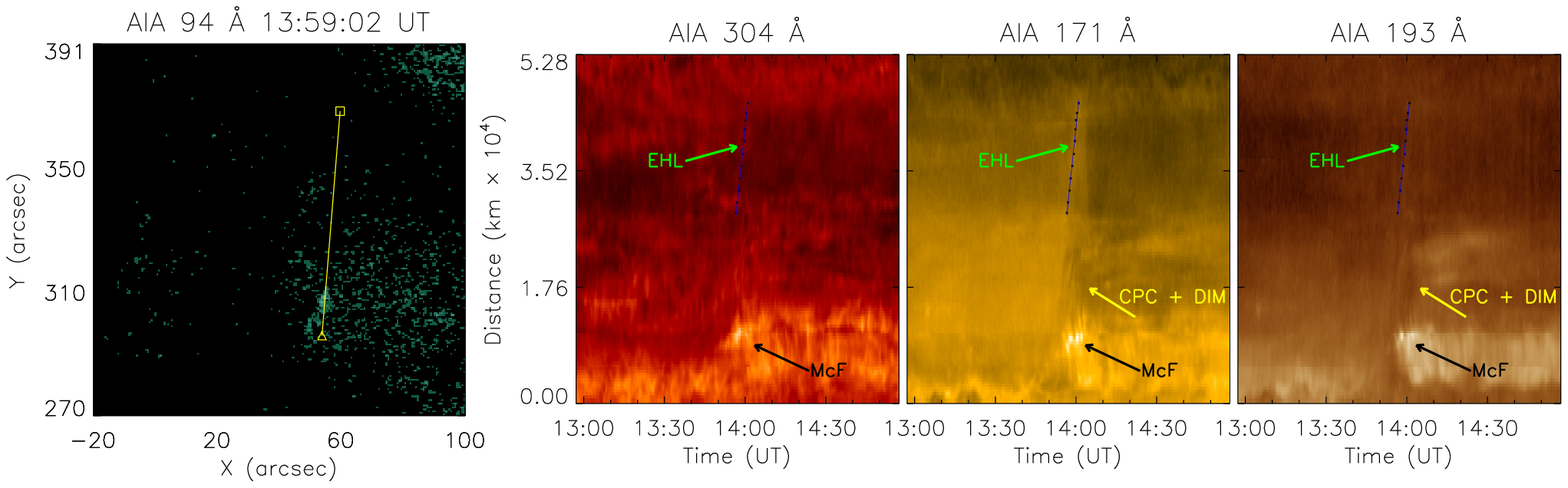}
\caption{The same as in Figure~\ref{fig:bp2_1} for BP10. The following abbreviations are marked on the panels: BP -- bright point, EHL -- eruptive hot loops, CPC -- cool plasma cloud, DIM -- dimming, and McF -- micro-flare.}
\label{fig:bp10_1}
\end{figure*}

\begin{figure*}[!ht]
\centering
\includegraphics[scale=0.30]{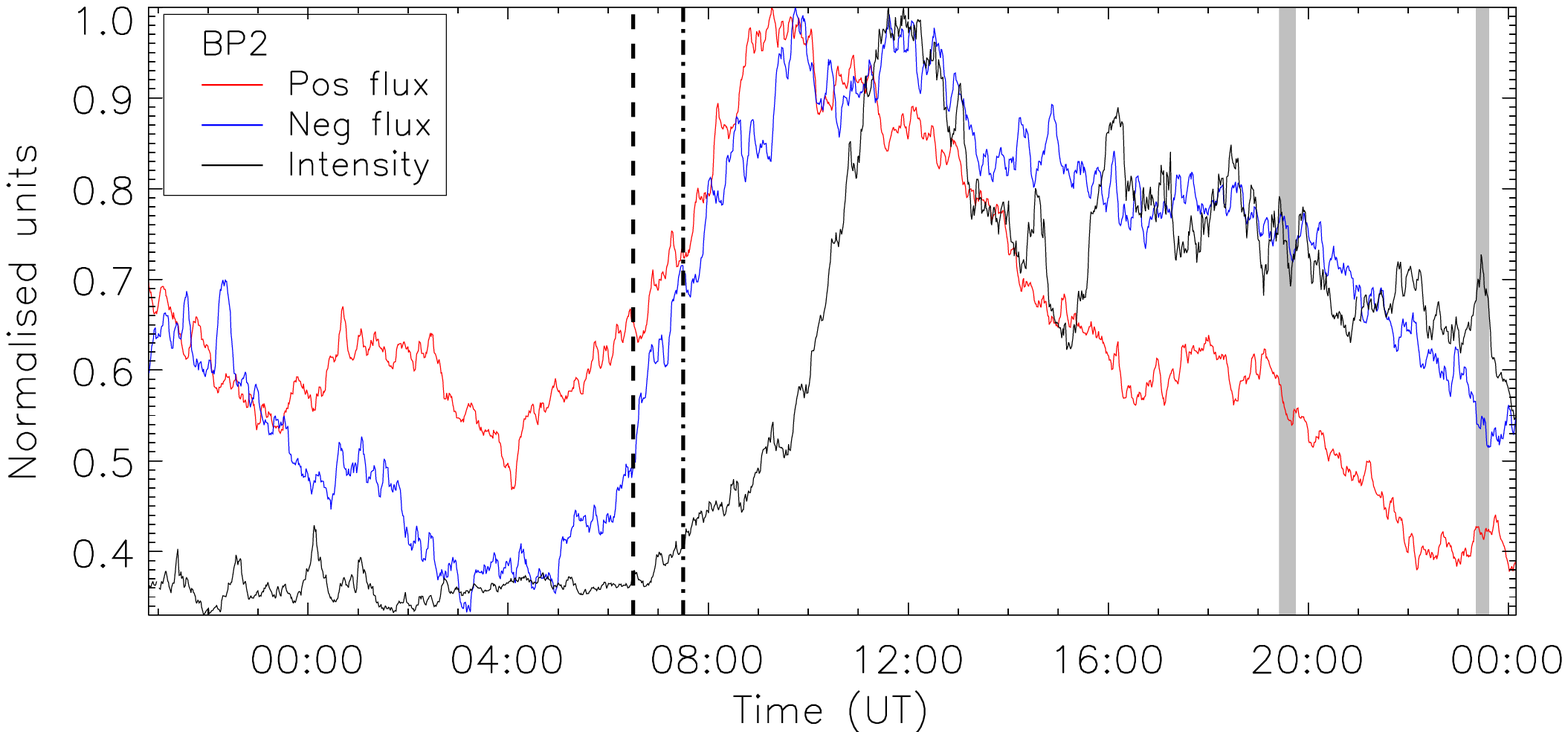}
\includegraphics[scale=0.30]{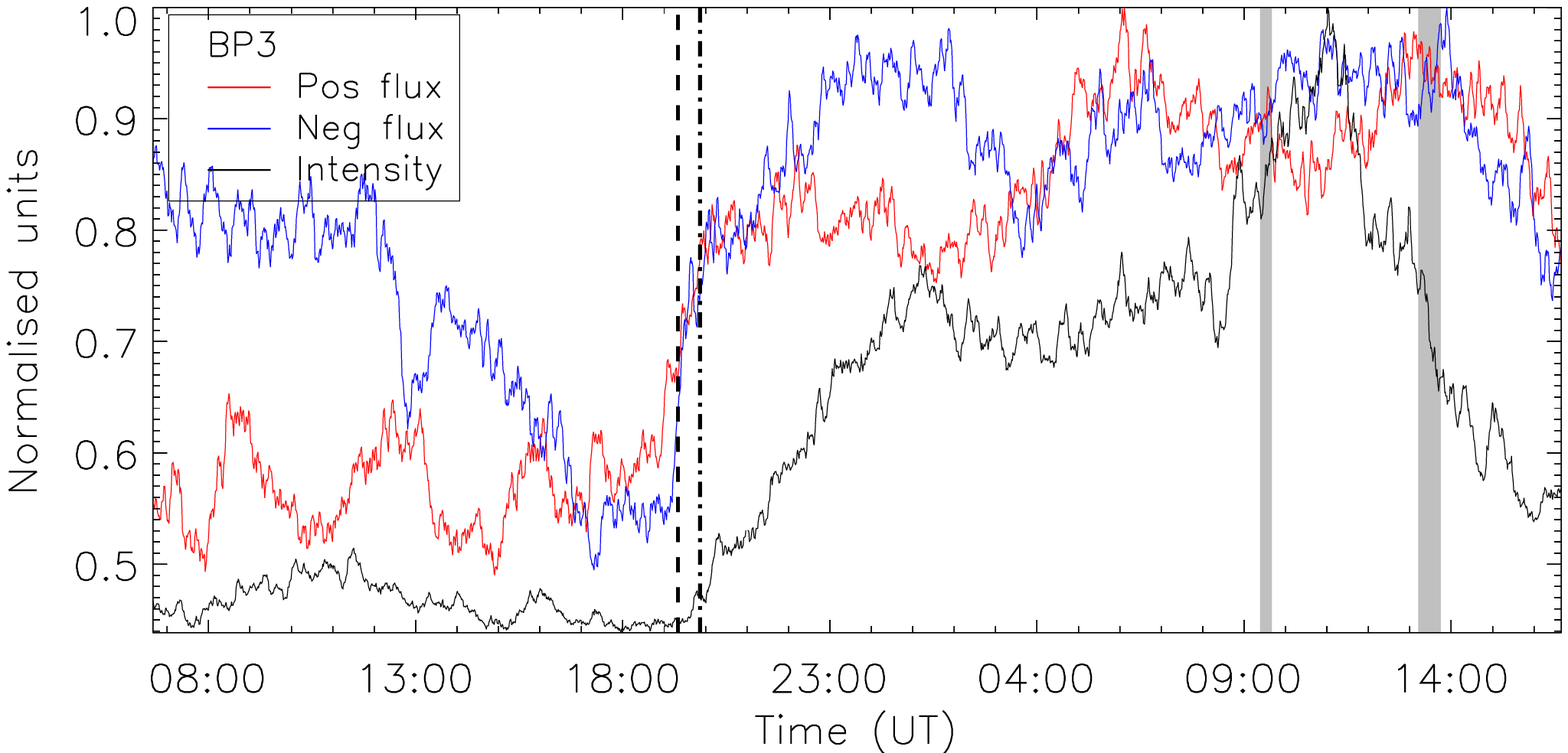}\\
\includegraphics[scale=0.30]{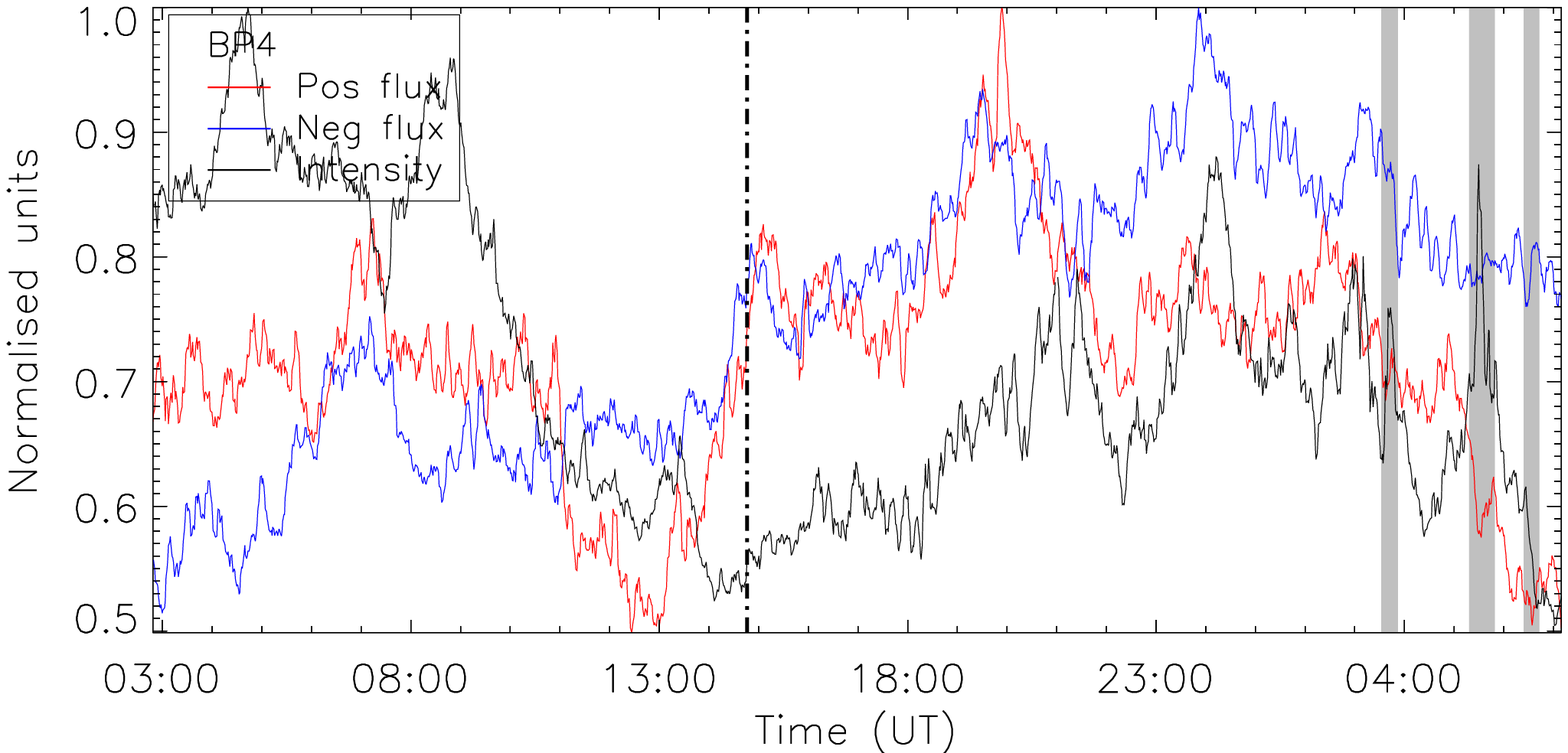}
\includegraphics[scale=0.30]{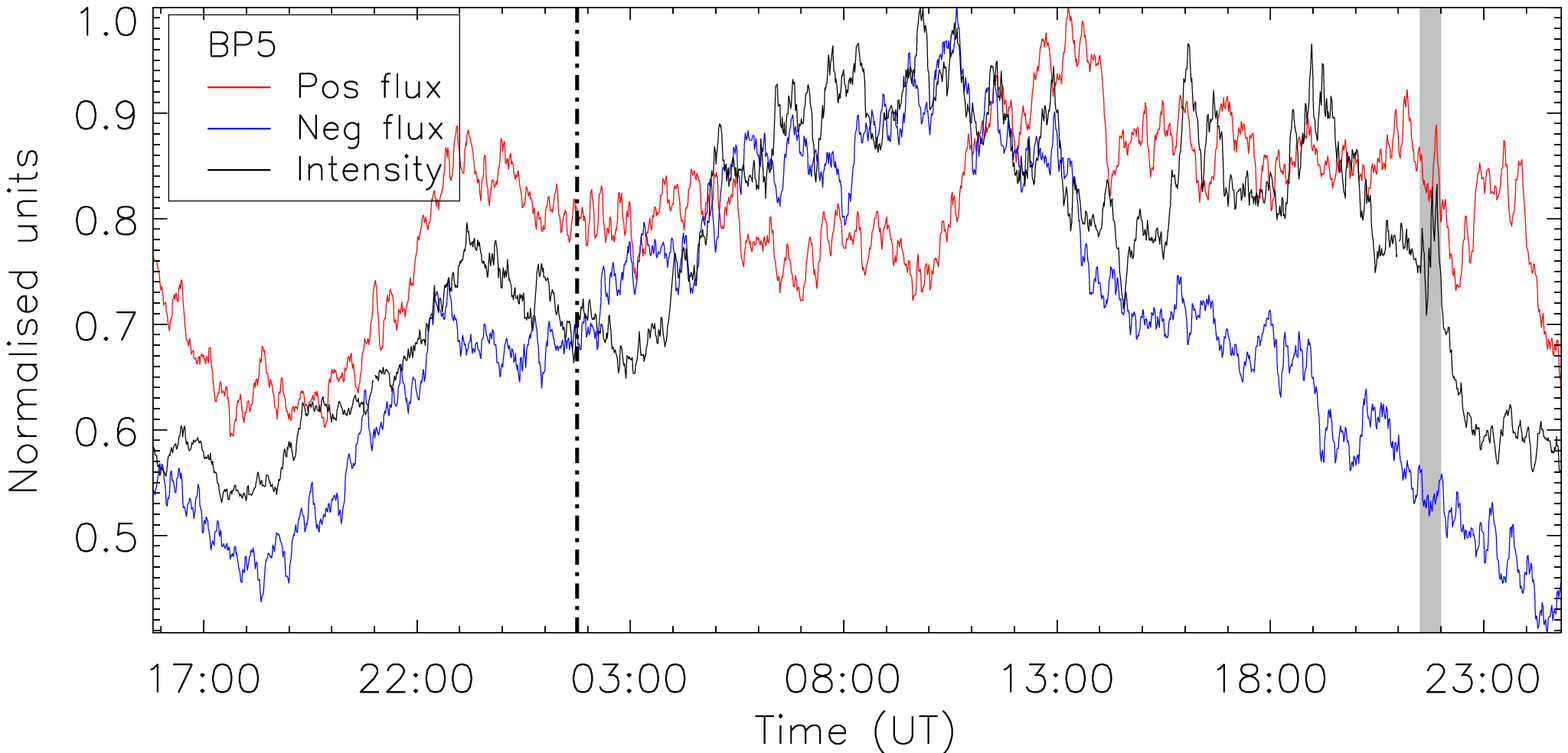}\\
\includegraphics[scale=0.30]{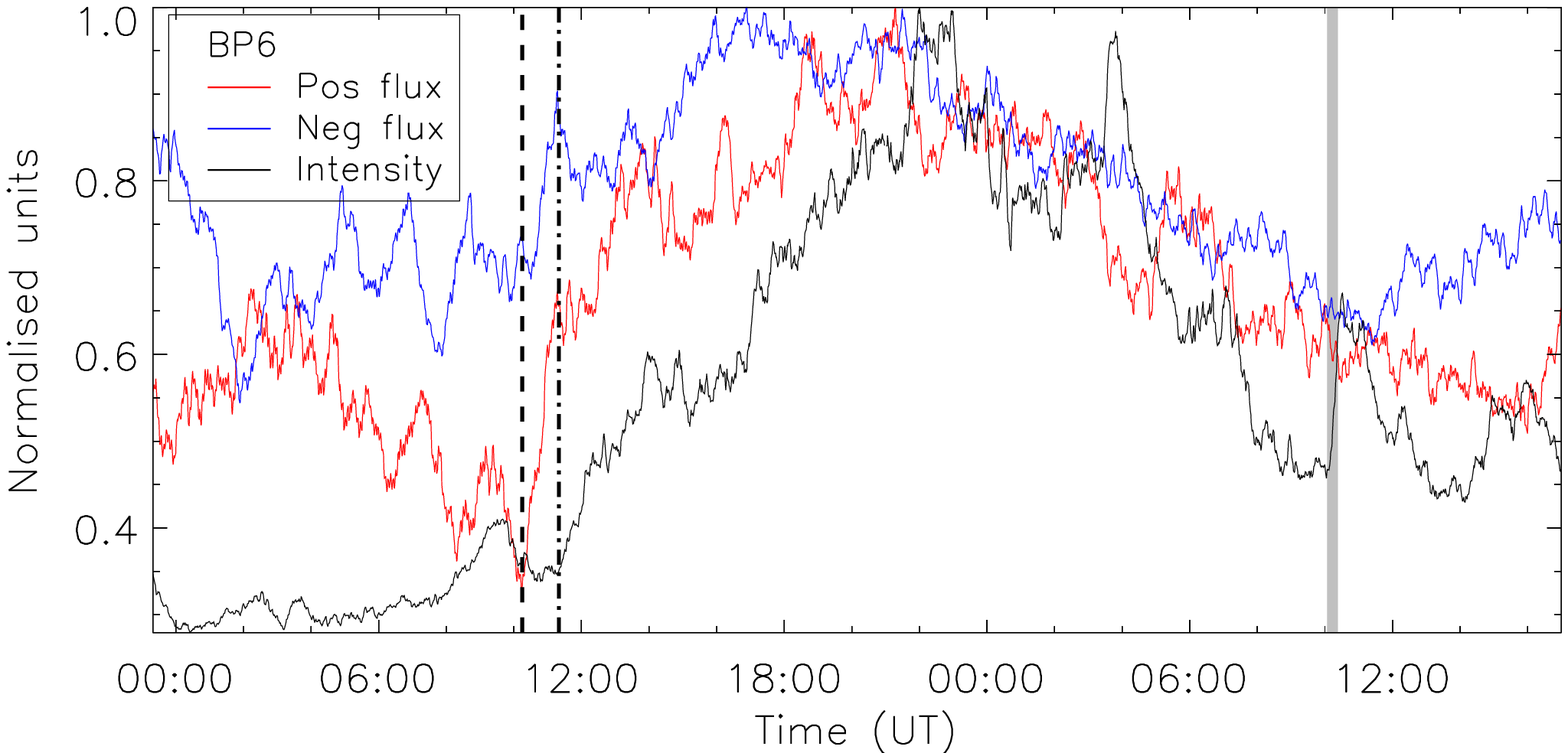}
\includegraphics[scale=0.30]{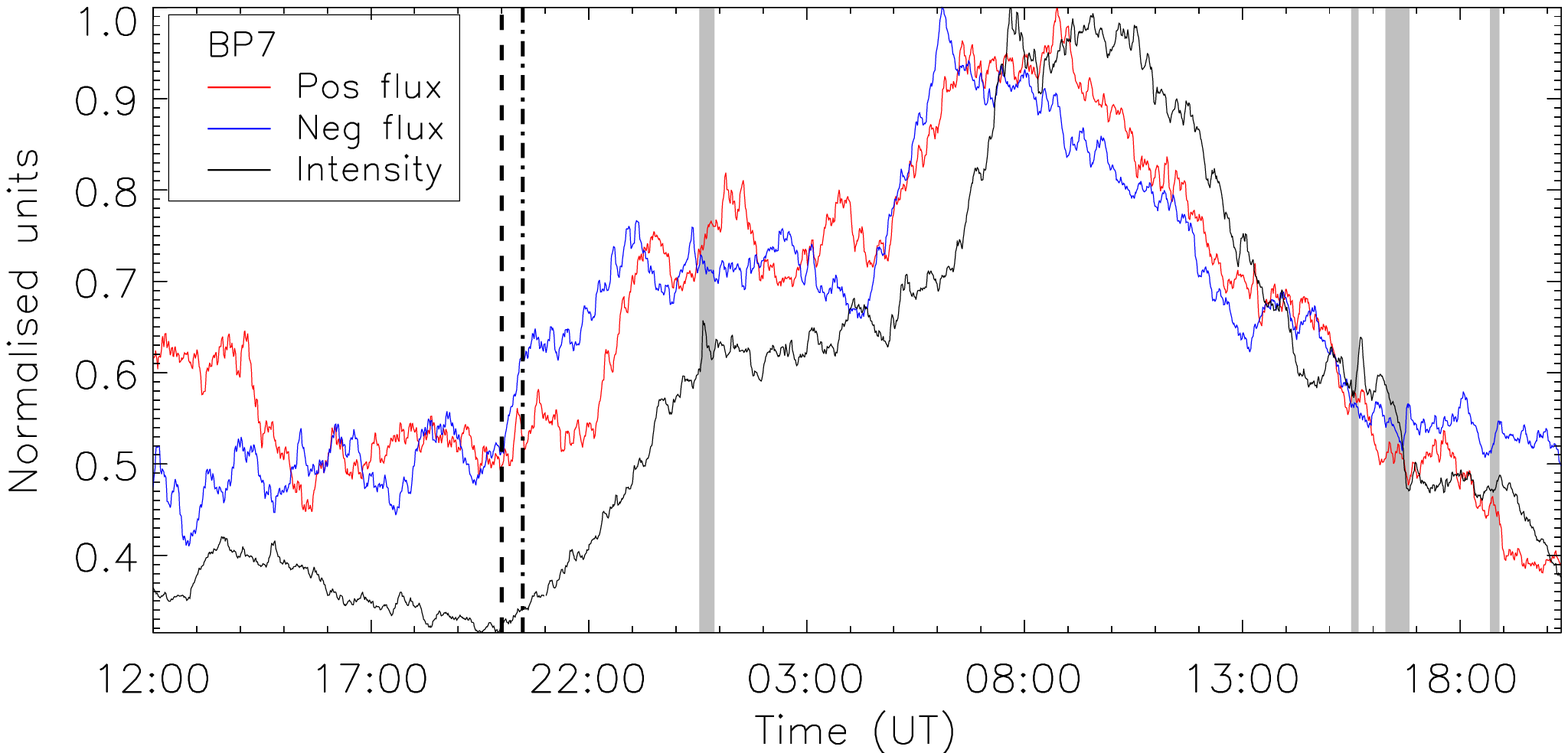}\\
\includegraphics[scale=0.30]{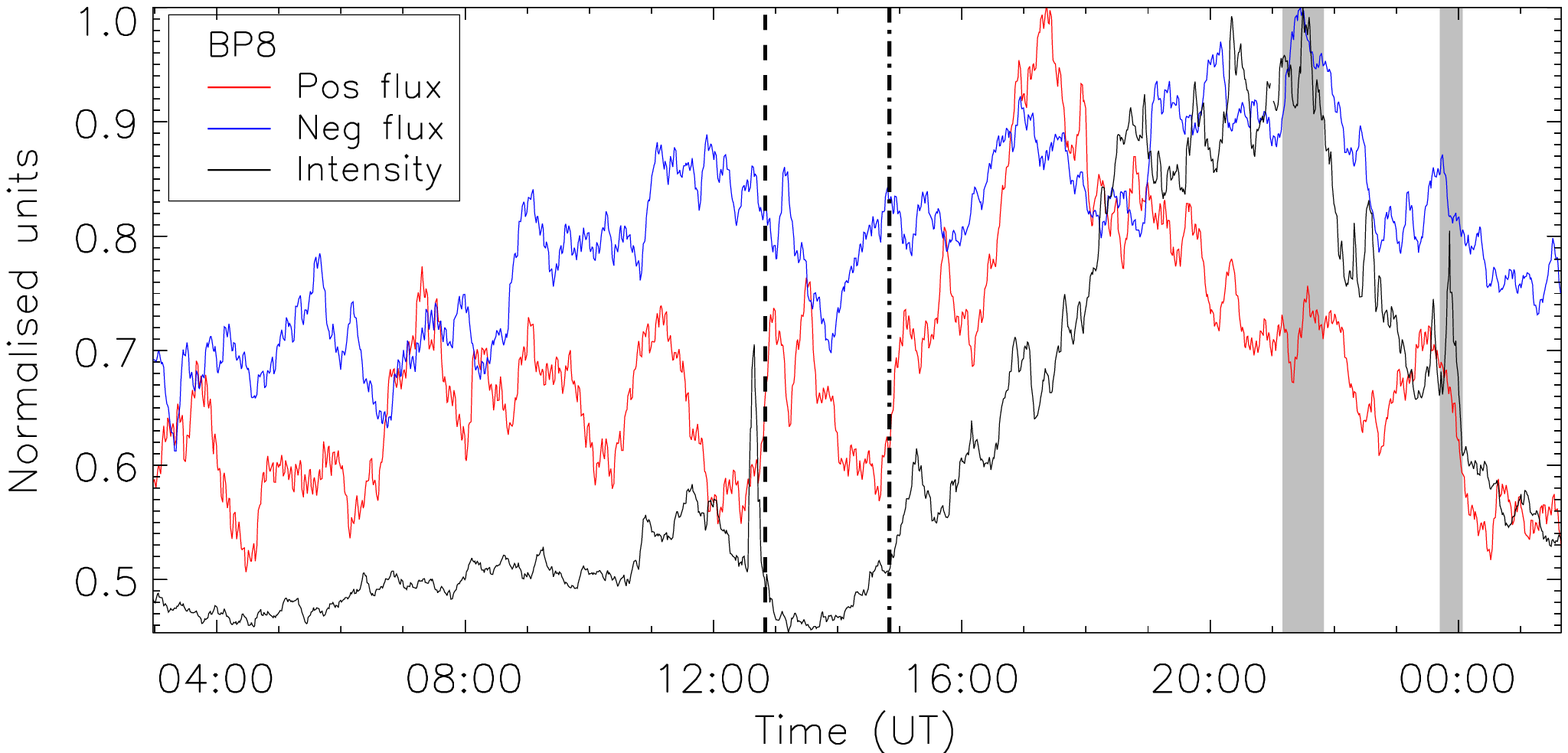}
\includegraphics[scale=0.30]{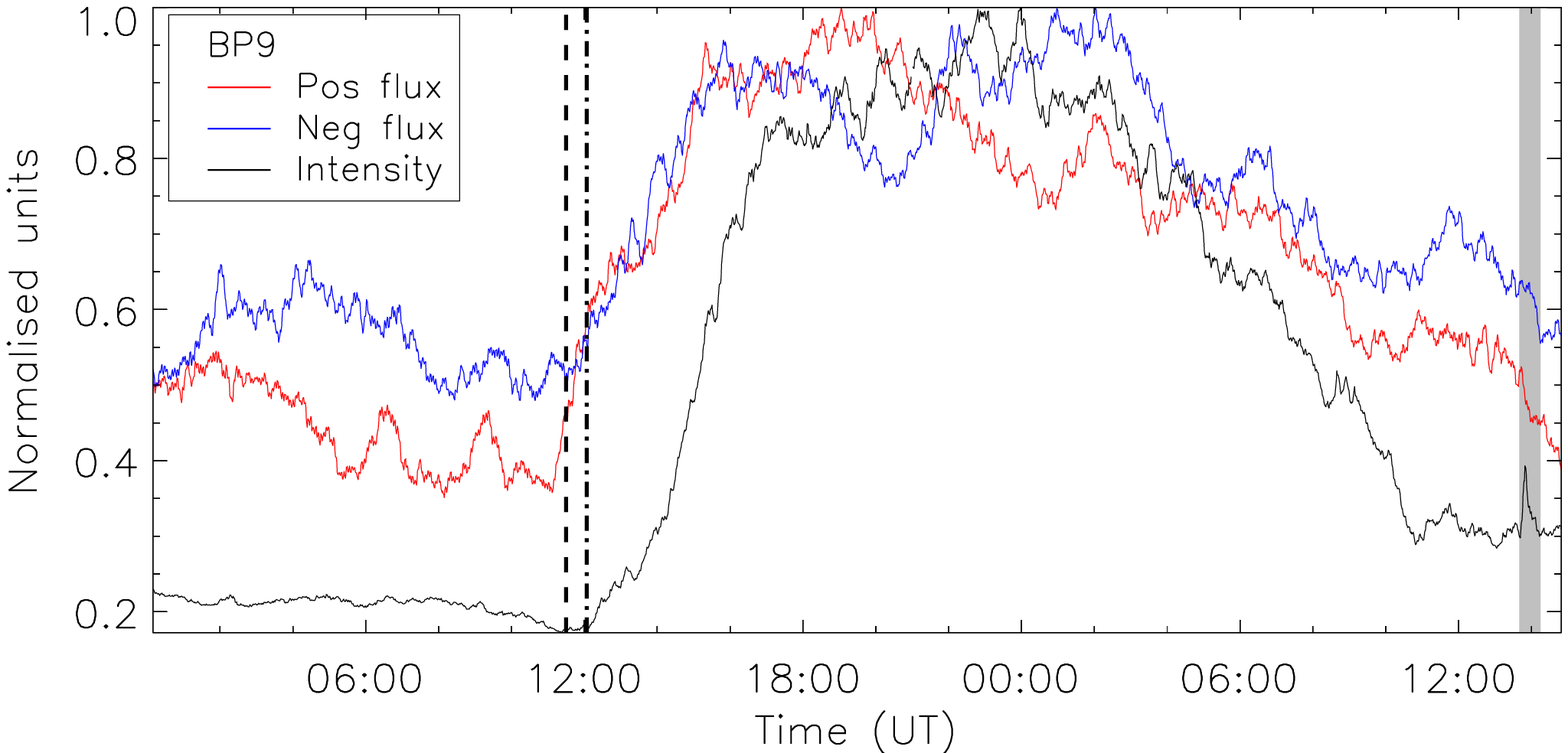}\\
\includegraphics[scale=0.30]{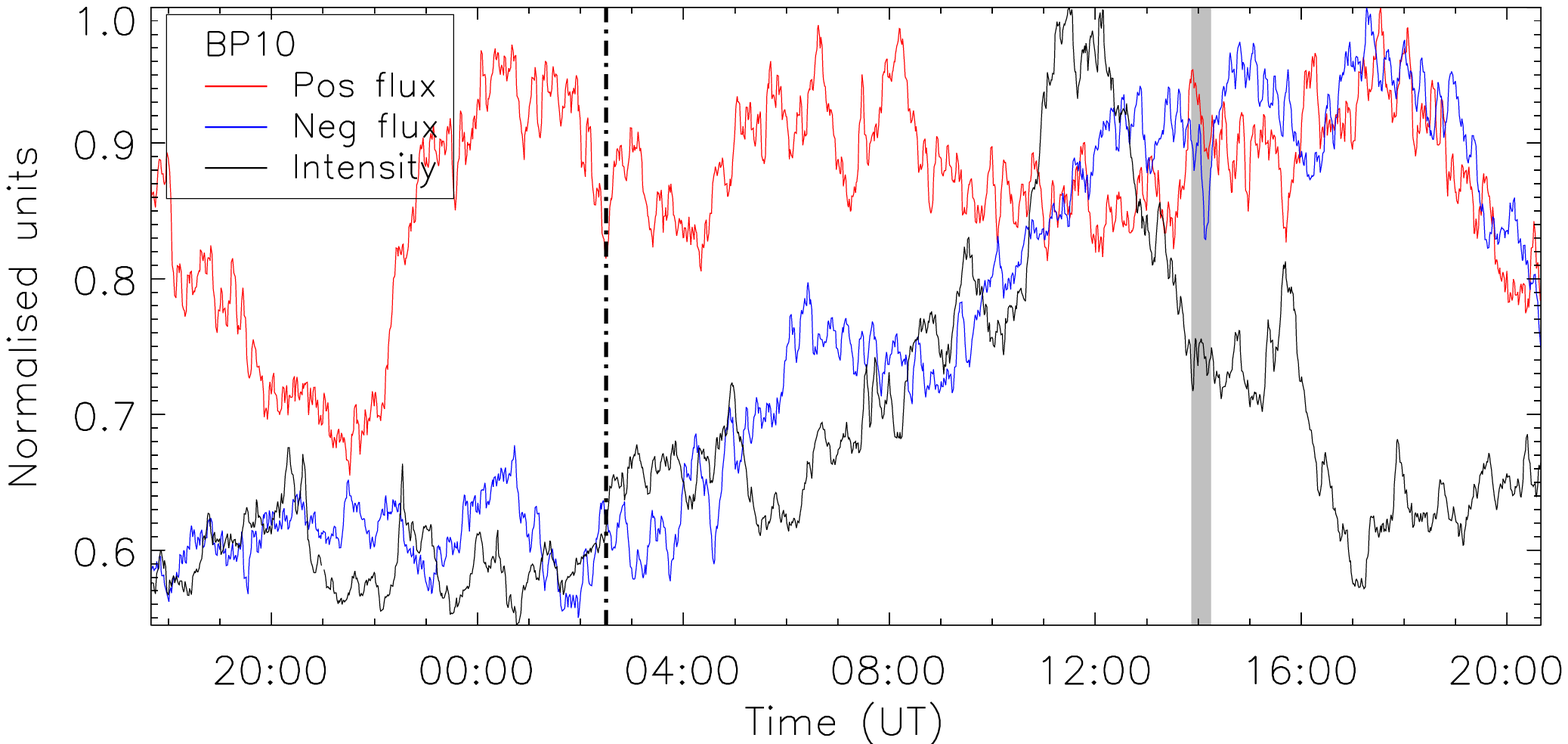}
\caption{Lightcurves of the normalized total magnetic positive (red solid line) and negative (blue solid line) flux from the region outlined with a box in the last panel of the top row of each of the BP Figures, e.g. for BP2 -- Fig.~\ref{fig:bp2_1}. The solid black line shows the normalized intensity in the AIA~193~\AA\ channel from the region outlined with a box in the third panel of the top row in Fig.~\ref{fig:bp1_1}. The grey areas show the eruption time, the dashed line the start of the flux emergence and the dashed-dotted line, the time of the BP appearance in the AIA~193~\AA\ images. }
\label{fig:bps_m}
\end{figure*}

\clearpage

\begin{figure*}[!ht]
\centering
\includegraphics[scale=0.8]{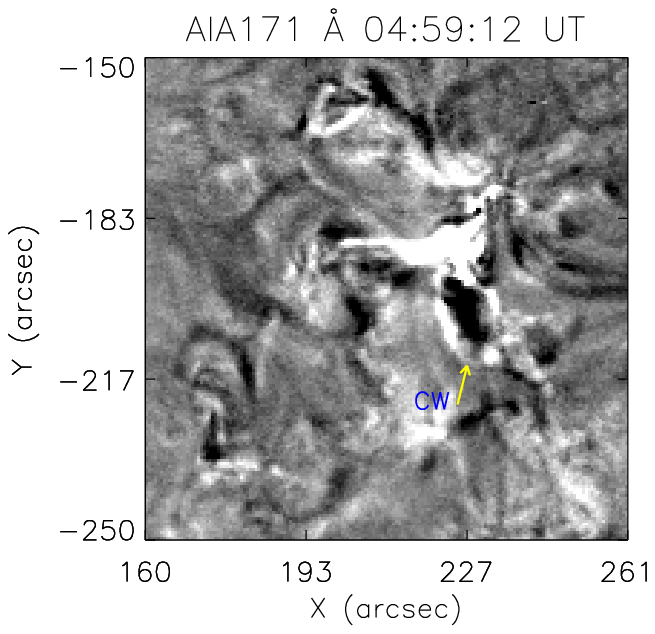}
\includegraphics[scale=0.8]{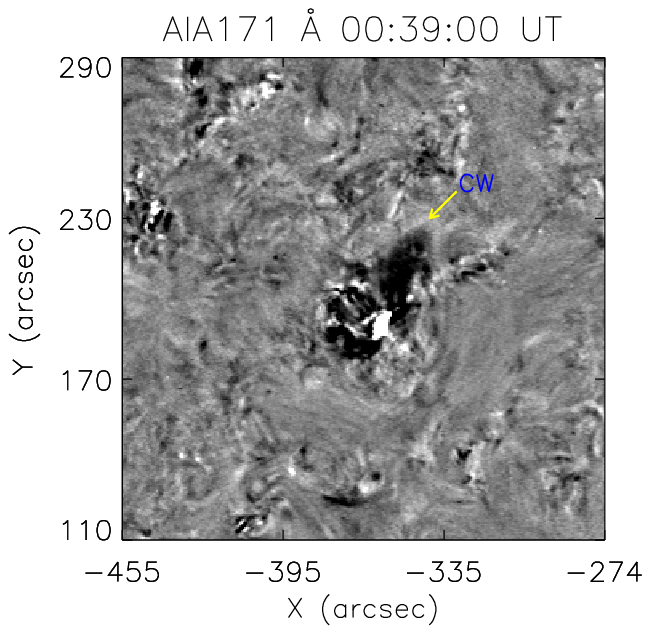}
\includegraphics[scale=0.8]{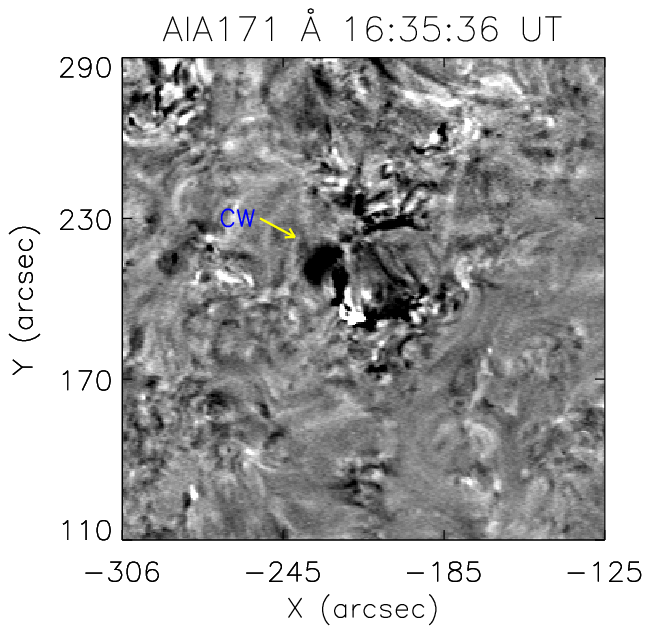}
\caption{From the left to right: The base-difference images of BP1 ER2, BP7 ER1, BP7 ER3 in AIA 171 \AA. The yellow arrows point at the coronal wave in these cases. The CW is the abbreviation of the coronal wave.}
\label{fig:base_diff}
\end{figure*}

\end{appendix}

\end{document}